\documentclass[aps,pra,superscriptaddress,12pt, longbibliography,reprint,hyphens]{revtex4-2}

\makeatletter
\renewcommand{\p@subsection}{}
\renewcommand{\p@subsubsection}{}
\makeatother

\usepackage{amsmath,amsfonts,amssymb,amsthm}
\allowdisplaybreaks

\usepackage{mathtools}
\usepackage{bm}
\usepackage{setspace}
\usepackage[normalem]{ulem} 
\usepackage{stmaryrd}       
\usepackage{dsfont}         
\usepackage{expdlist}
\usepackage{physics}
\usepackage{verbatim}
\usepackage{xspace}
\usepackage{subfigure}
\usepackage{scalerel,stackengine}

\newcommand{\RNum}[1]{\uppercase\expandafter{\romannumeral #1\relax}}

 \newcommand{\nord}[1]{\mathopen{:}#1\mathclose{:}}

\newcommand{\anormord}[1]{\mathrel{\vdots}\mathrel{#1}\mathrel{\vdots}}

\stackMath
\newcommand\reallywidehat[1]{%
\savestack{\tmpbox}{\stretchto{%
  \scaleto{%
    \scalerel*[\widthof{\ensuremath{#1}}]{\kern.1pt\mathchar"0362\kern.1pt}%
    {\rule{0ex}{\textheight}}
  }{\textheight}%
}{2.4ex}}%
\stackon[-6.9pt]{#1}{\tmpbox}%
}
\usepackage[breaklinks=true]{hyperref}

\usepackage{graphicx,xcolor}
\hypersetup{pdfstartview=}

\usepackage{verbatim}
\usepackage{alltt}
\usepackage{moreverb}

\usepackage{xr}

\providecommand{\ignore}[1]{}

\newif\ifcmnt
\cmnttrue
\ifdefined\cmntsoff\cmntfalse\fi

\ifcmnt
    \providecommand{\aucmnt}[1]{#1}

\else
    \providecommand{\aucmnt}[1]{}

\fi

\newcommand{\cB}{\mathcal{B}}
\newcommand{\cC}{\mathcal{C}}

\newcommand{\cE}{\mathcal{E}}

\newcommand{\cH}{\mathcal{H}}
\newcommand{\cI}{\mathcal{I}}
\newcommand{\cJ}{\mathcal{J}}

\newcommand{\cM}{\mathcal{M}}

\newcommand{\cP}{\mathcal{P}}

\newcommand{\cR}{\mathcal{R}}
\newcommand{\cS}{\mathcal{S}}
\newcommand{\cT}{\mathcal{T}}
\newcommand{\cU}{\mathcal{U}}

\newcommand{\cW}{\mathcal{W}}
\newcommand{\cX}{\mathcal{X}}
\newcommand{\cY}{\mathcal{Y}}

\newtheorem{theorem}{Theorem}[section]
\newtheorem*{theorem*}{Theorem}

\newtheorem*{problem*}{Problem}
\newtheorem{lemma}[theorem]{Lemma}
\newtheorem*{lemma*}{Lemma}
\newtheorem{corollary}[theorem]{Corollary}

\newtheorem{prop}[theorem]{Proposition}
\newtheorem*{prop*}{Proposition}

\begin{document}

\title{Analysis of a Homodyne-Like Measurement Scheme Using Photon Counters and Weak Local Oscillator}

\author{Arik Avagyan}
\altaffiliation{Current address: Kestrel Robotics, Inc., Champaign, Illinois 61822, USA}
\affiliation{National Institute of Standards and Technology, Boulder, Colorado 80305, USA}
\affiliation{Department of Physics, University of Colorado, Boulder, Colorado, 80309, USA}
\author{Emanuel Knill}
\affiliation{National Institute of Standards and Technology, Boulder, Colorado 80305, USA}
\affiliation{Center for Theory of Quantum Matter, University of Colorado, Boulder, Colorado 80309, USA}
\author{Scott Glancy}
\affiliation{National Institute of Standards and Technology, Boulder, Colorado 80305, USA}

\begin{abstract} 
In standard homodyne detection, an unknown optical signal state is combined with a local oscillator (LO) at a beam splitter (BS), and the appropriately integrated intensities at the two outputs of the BS are subtracted to obtain an approximate quadrature measurement. Accurate quadrature measurements require a high-amplitude LO and  high-efficiency detectors. By varying the LO phase, the optical state in the mode matching the LO can be reconstructed. In many practical settings, a high-intensity LO is not desirable, which has led to the exploration of homodyne-like measurement schemes that replace photodiodes with photon-counting detectors and use an LO with a low average photon number. In this paper, we investigate what features of the unknown state can be determined by counting the total-photon number in one or both outgoing paths after the BS, keeping the mode of the LO fixed but adjusting its phase and  magnitude. We find that it is possible to determine the unknown optical state in the mode matching the LO conditional on the number of photons in the orthogonal modes on the same path. If both the phase and the intensity of the LO can be varied, the statistics of just one of the detectors is sufficient to infer these parameters, but both detectors are required when the LO has a fixed intensity. Our analysis is carried out in the infinite-data limit, assuming access to the full photon-number distribution from one or both detectors.
\end{abstract}

\maketitle


\section{introduction}\label{sec intro}

Standard homodyne detection is a key technique for characterizing
quantum states of light \cite{bachor2019guide}. In this scheme, the
unknown state is combined with a high-intensity coherent state, known as the
local oscillator (LO) or phase reference, on a balanced beam splitter
(BS) at various relative phase settings spanning the full angular
range. The output beams are detected by photodiodes, producing
electrical signals that are integrated over time and subtracted. In
the limit of infinite LO intensity, the result corresponds to a
quadrature measurement of the unknown state in the mode matching the
LO at the given phase. Here, “mode” typically refers to a normalized
spatially localized solution of Maxwell’s equations in free space or a
dielectric medium such as an optical fiber. Measuring quadratures
across all phases allows for reconstruction of the reduced density
matrix of the state in the mode matching the LO \cite{LEONHARDT2,
  MAURODARIANO2003205}.

Standard homodyne detection has some limitations - the most obvious
being the need for a high-intensity reference, which can potentially
disturb sensitive quantum information in systems such as Rydberg
clouds \cite{maxwell:qc2013a,Beck:qc2016, peyronel:qc2013a,Lukin}, in
quantum non-demolition experiments probing atomic ensembles
\cite{Kuzmich1999,Takano2009,Koschorreck2010}, and in non-destructive
cavity QED experiments \cite{Peden2009,Niederriter2020}. Another
obvious limitation is the blindness of the scheme to the modes
orthogonal to the LO. These two features of standard
homodyne detection are closely related. Even if photodiodes were ideal
intensity detectors, in the weak LO regime the final processed signal
would no longer correspond to a quadrature measurement, but would
contain information about the modes orthogonal to the LO.

Photon-number-resolving (PNR) detectors, also referred to as photon counters, serve a role similar to photodiodes but are designed for situations where, on average, only a few photons are present in the absorbed pulse. \cite{migdall2013experimental, slussarenko2019photonic}. Current state-of-the-art PNR detectors, such as transition-edge-sensors (TESs), can distinguish up to sixteen photons in a single pulse \cite{gerrits2016super}. Further, a combination of TESs in a multiplexed scheme, acting as an effective PNR detector, can resolve up to $100$ photons \cite{Eaton2023res}. These advances motivate the exploration of homodyne-like measurement schemes in which photodiodes are replaced by PNR detectors, and the LO is attenuated to contain only a few photons on average. 

These schemes, often referred to as “hybrid” or “weak-field homodyne” detectors \cite{migdall2013experimental}, have been studied both theoretically and experimentally. Early theoretical work showed that a single PNR detector and an unbalanced BS suffice to determine an unknown single-mode state \cite{wallentowitz1996unbalanced}. Experimentally, such schemes have been used to measure two-photon phase coherence \cite{Koashi}, test Bell inequalities \cite{Walmsley1}, create superpositions of vacuum and single-photon states with well-defined phases \cite{Resch}, perform state discrimination \cite{Olivares2,Olivares3,Muller}, and estimate entanglement in bipartite systems \cite{Walmsley3}. Further demonstrations include state reconstruction for single-mode \cite{Olivares1, Olivares4, Walmsley2, Walmsley4, Thekkadath} and two-mode squeezed states \cite{Walmsley5}. Other studies have experimentally characterized weak-field homodyne detectors under the assumption that there are no photons
in detectable modes other than the mode matching the LO \cite{zhang2012recursive, xu2020exp}. Interestingly, it was shown that, in principle, an arbitrary multimode state can be determined using one single-photon detector, provided the LO can be prepared in any desired mode \cite{Shadow}. Recently, there has also been interest in exploring detection schemes where the LO is not a coherent state, but rather a non-classical state such as a cat state \cite{combes2022hom}.

A common assumption across these studies of weak-field homodyne is that either there is only one, known mode containing photons
and this mode matches that of the LO, or the set of modes of
interest is known and the LO can be matched to or scanned over
them. The former does not
account for photons in other modes, while the latter has substantial
measurement overhead. For a fixed LO mode, a desirable goal of
weak-field homodyne is to not only determine the marginal state of
the mode, but also its correlation with photons in other modes,
without the need to change or scan the LO mode. Correlations with photons in other
modes can yield valuable insights, whereas neglecting them,
especially when their content is non-negligible, can lead to
significant errors in estimation. Strengthening the power of weak-field homodyne to achieve these
goals is one of the key motivations for the present work.

We theoretically investigate what can be learned about an unknown
state with an unspecified mode structure by combining it with a LO in
a weak coherent state on a BS (not necessarily balanced) and
performing photon counting on one or both output ports. We assume that
the LO is pulsed, where the LO mode is defined by the pulse shape. The
PNR detectors count photons during a time window containing the LO
pulse, and they are equally sensitive to all modes that may have
photons in this time window. For this work, we assume that the
probabilities of all photon number outcomes are learned. This is the
infinite data limit for ideal PNR detectors. Perfect PNR detectors are
not required in the infinite data limit. For example, inefficiency or
loss and added noise can be modeled as classical noise processes
following perfect photon counting, and it is often possible to invert
the classical noise process given the count
probabilities~\cite{ivan2011operator}.

Our analysis of the infinite data limit establishes what is
fundamentally inferrable about the unknown state for the experimental
configurations considered. We find that, for a fixed LO mode, the
measurement statistics yield significantly more information about the
unknown state than is possible with standard homodyne detection. In
particular, it is possible to determine the unknown state in the mode
matching the LO conditional on each number of photons present in the
orthogonal modes, and one may also determine the photon number
probabilities in those modes. We find that the measurement statistics
from a single PNR detector, collected over a sufficient range of LO
magnitudes and phases, are sufficient to extract this
information. When two PNR detectors are used, a fixed-magnitude LO is
sufficient to extract this information provided the measurement
statistics are known for all phase values.

\section{Overview}

We begin by summarizing the main theoretical results established in
this work. Our analysis considers an unknown multimode input
state $\rho$ combined with a weak LO in a fixed mode on an arbitrary
lossless BS, followed by detection with one or two noiseless and
lossless PNR detectors that do not distinguish between the modes in
their respective BS ports (Fig.~\ref{fig: meas_config}) in the
  infinite data limit. Throughout the text, such
idealized PNR detectors are referred to simply as counters.
Sec.~\ref{sec prelim} provides a detailed description of the
measurement configuration, relevant physical quantities, and the main
mathematical instruments used in deriving the results. Following this
setup, we make a key observation in Sec.~\ref{sec sym} that the
measurement statistics exhibit a nontrivial symmetry. Specifically,
Prop.~\ref{thm: PLTpresmeasstat} and Prop.~\ref{thm PLTpresnumone}
together establish that the statistics are invariant under precisely
those passive linear transformations (PLTs) on the input state that act trivially on the LO-matched mode (mode
$1$). Averaging over the symmetry group yields the so-called
twirled-state, given in Thm.~\ref{thm Tau(rho)}, with Cor.~\ref{cor
  rho_t} showing that there exists a two-mode representative $\rho_t$
that is indistinguishable from $\rho$. We refer to $\rho_t$ as the
two-mode twirled state corresponding to $\rho$.

\newcommand{\hatnpa}{\hat{n'}_a}
\newcommand{\hatnpb}{\hat{n'}_b}
\newcommand{\hatna}{\hat{n}_a}
\newcommand{\hatnb}{\hat{n}_b}

Section~\ref{sec results} then addresses the central question of
whether the measurement statistics determine $\rho_t$. The analysis is
based on representing the measurement statistics by the moments of the
total photon-number operators ($\hatnpa $ and $\hatnpb $) measured
by the two counters. This in turn imposes a requirement on $\rho_t$:
its Husimi function must be determined by its moments. We next show
with Prop.~\ref{prop: rho_t_mom} that, in order to uniquely determine
$\rho_t$, the normally ordered moments
$\langle (\hat a_1^\dagger)^k \hat a_1^\ell (\hat a_2^\dagger)^n \hat
a_2^n\rangle$ for $k \geq \ell$ and arbitrary $n$ are both sufficient
and necessary. Here, $\hat{a}_1$ and $\hat{a}_2$ denote the
annihilation operators for modes $1$ and $2$, respectively. For input
states with bounded photon number, Cor.~\ref{cor: rho_t_mom}
establishes that only finitely many of these moments are
required. These results rely on the operator-ordering relation of
Lem.~\ref{lem: norm-antinorm}, which establishes a one-to-one
correspondence between the Husimi moments of $\rho_t$ and the moments
$\langle (\hat a_1^\dagger)^k \hat a_1^\ell (\hat a_2^\dagger)^n \hat
a_2^n\rangle$.

Within this framework, we obtain detailed results for both the single- and two-counter configurations by asking whether the measurement statistics, known for a given set of probe LO states, determine the moments $\langle (\hat a_1^\dagger)^k \hat a_1^\ell (\hat a_2^\dagger)^n \hat a_2^n\rangle$. 
For a single counter, Thm.~\ref{thm: 1cntr_uncount} shows that access to statistics across a neighborhood of probe amplitudes suffices to recover all required moments, thereby determining $\rho_t$. When the maximum photon number of the input state is bounded by $N$, Thm.~\ref{thm:1CntrNbnd} and Cor.~\ref{cor: suf_set_gamma} relaxes this requirement, showing that $(N+1)^2$ probe amplitudes are sufficient. If there is no bound on the maximum number of photons, Prop.~\ref{prop: inf_N_fin_gamma} establishes that no finite set of probe amplitudes suffices. Moreover, for fixed-magnitude probes with variable phase, Prop.~\ref{prop: fix_R} shows that the single-counter scheme is limited to reconstructing states with at most one photon.

The two-counter configuration is strictly more efficient in terms of
the number of measurement settings required to obtain the same
information. While Prop.~\ref{prop: inf_N_fin_gamma_2} shows that
finitely many probes are still insufficient to determine $\rho_t$
without a photon-number bound, Thm.~\ref{thm: twodet_R_fixed}
demonstrates that collecting measurement statistics by scanning the
probe phase at fixed magnitude recovers $\rho_t$ in full. The final
result, Prop.~\ref{prop: twodet_R_fixed}, establishes a connection
between the two-counter measurement scheme and standard homodyne
detection. It shows that the reduced state of the system is fully
characterized by the expectations of the normally ordered operators
$ \nord{(\hatnpb - \xi \hatnpa )^n} $ for all $n$. Here, $\xi$ is a
parameter determined by the BS, taking the value $\xi = 1$ for a
balanced BS. Homodyne detection is recovered in the limit of a strong
LO, where
$ \langle \nord{(\hatnpb - \hatnpa )^n} \rangle \approx \langle
(\hatnpb - \hatnpa )^n \rangle $. In the appendix we provide basic
results involving normal and antinormal ordering and PLTs, and we show
the Husimi functions of Gaussian states, Fock states and their
superpositions and mixtures are determined by their moments.

\section{Preliminaries}\label{sec prelim}

\subsection{Description of the Measurement Configuration}\label{subsec descr}

\begin{figure}[t]
\includegraphics[width=\columnwidth]{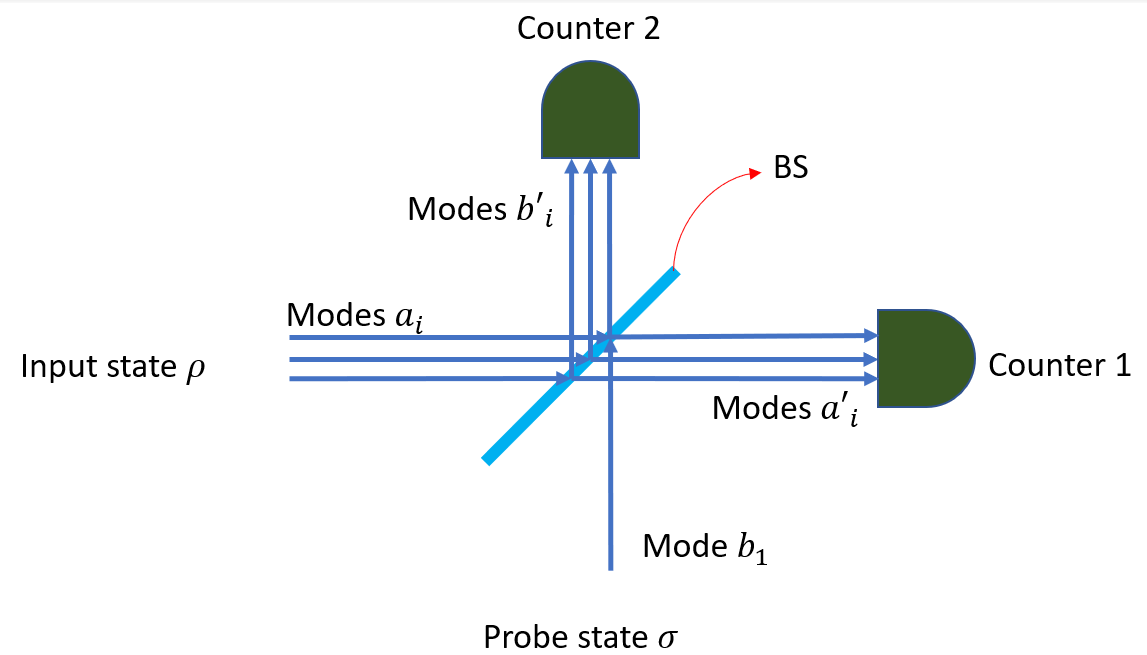}
\caption{Diagram of the measurement configuration with both PNR detectors present. The unknown input state $\rho$ and probe state $\sigma$ are directed towards the inputs of the BS, and the outputs are measured by two PNR detectors, denoted by counter 1 and counter 2. The detectors measure the total-photon-number operators \(\hatnpa \) and \(\hatnpb \) in their respective output paths of the BS. If the measurement configuration uses only one detector, then only counter 1 is present.}
    \label{fig: meas_config}
\end{figure}

The schematic of the measurement configuration is shown in
Fig.~\ref{fig: meas_config}. The unknown input state $\rho$
and the LO probe state $\sigma$ are directed toward the two
input ports of the BS. The input $\rho$ lives in $S$ mutually
orthogonal modes, characterized by $S$ annihilation operators
$\hat a_1, \hdots, \hat a_S$. We do not assume that $S$ is known, but
require that it be finite for our analysis. The BS induces
interference between modes in its two input ports that share identical
spatio‐temporal and polarization characteristics. Thus, there exist
$S$ modes in the LO input port, characterized by annihilation
operators $\hat b_1, \hdots, \hat b_S$, where each $\hat b_i$ is
matched with $\hat a_i$.

We denote the physical Hilbert space of the input state by $\cH_a$, and of the probe state by $\cH_b$. We further denote the Hilbert spaces of the output modes of the BS associated with the operators $ \hat a'_i  $ and $\hat b_i'$ by $\cH_{a'}$ and $\cH_{b'}$, respectively. We denote the vector of annihilation operators $(\hat a_1, \hdots, \hat a_S )$ by $\hat{\textbf{a}}$, and the vector of creation operators $(\hat a^\dagger_1,\hdots, \hat a^\dagger_S)$ by $\hat{ \textbf{a}}^\dagger$, and similarly for the operators in the other ports of the BS. We generally interpret vectors written inline as column vectors.
We refer to creation and annihilation operators as mode operators. Given any two vectors of mode operators, say $\hat{\mathbf{a}} = (\hat{a}_1, \dots, \hat{a}_S)$ and $\hat{\mathbf{b}} = (\hat{b}_1, \dots, \hat{b}_S)$, the direct sum $\hat{\mathbf{a}} \oplus \hat{\mathbf{b}}$ is defined as the concatenated vector $(\hat a_1, \hdots, \hat a_S, \hat b_1, \hdots, \hat b_S ) $.

For any vector of $K$ annihilation operators
\(\hat{\bf{c}} = (\hat{c}_1,\hdots, \hat{c}_K)\) with associated
Hilbert space \(\cH_{c}\), we refer to the \(K\)-dimensional
  vector space spanned by its components \(\hat{c}_{i}\) as the mode
space of \(\cH_{c}\).  We denote the Fock state with $n_i$ photons in
mode $\hat c_i$ by
$\ket{n_1,\hdots, n_{K}} = \bigotimes_{i=1}^{K} \ket{n_i}$. Similarly,
a coherent state with complex amplitude $\alpha_i$ in mode $\hat c_i$
is written as
$\ket{\vec{\alpha}} = \bigotimes_{i=1}^{K} \ket{\alpha_i}$.
We use
Greek letters in bra-ket notation to indicate coherent states and
Latin letters for Fock states. Throughout this work, the probe state
is the coherent state
$\sigma = \ket{\gamma}\bra{\gamma} \otimes
\ket{\vec{0}}\bra{\vec{0}}$, where $\ket{\vec{0}}$ denotes the vacuum
state for modes $\hat b_2,\hdots, \hat b_S$.

The transformation induced by the BS is described by a $2 \times 2$ unitary matrix
$B_1 = 
\left(
  \begin{array}{cc}
         \eta & \zeta \\ 
         \zeta' & \eta' \\
  \end{array}
\right)$ that, in the Heisenberg picture, combines each pair of
matching modes $(\hat a_i, \hat b_i)$ into two output modes
$(\hat a'_i, \hat b'_i)$ as follows:
$ (\hat a_i',\hat b_i')= B_1 \cdot (\hat a_i, \hat b_i)$. Here, the
central dot denotes the standard product between a matrix and a
vector. We assume $\eta \neq 0$ and $\zeta \neq 0$ to exclude trivial
BSs. The corresponding unitary operator on the Hilbert space of states
is denoted by $\cB$, so that $\hat a_i' = \cB^\dagger \hat a_i \cB$
and $\hat b_i' = \cB^\dagger \hat b_i \cB$.  The action of the BS on a
coherent state
$\ket{\vec{\alpha}}\ket{\vec{\beta}} \in \cH_a \otimes \cH_b$ produces
the coherent state
$\ket{\vec{\alpha'}}\ket{\vec{\beta'}} \in \cH_{a'} \otimes \cH_{b'}$,
where $(\alpha_i', \beta_i') = B_1 \cdot (\alpha_i, \beta_i)$.  The
outputs of the BS are directed toward two counters, referred to as
counter 1 and counter 2. Counter 1 measures the total-photon-number
operator $\hatnpa  = \sum_{i=1}^S (\hat a'_i)^\dagger \hat a'_i$, and
counter 2 measures
$\hatnpb  = \sum_{i=1}^S (\hat b'_i)^\dagger \hat b'_i$. If only one
of the counters is used, we assume counter 2 is absent. This entails
no loss of generality, as omitting counter 1 instead is equivalent to
omitting counter 2 up to a change of the BS coefficients.

Unitary transformations that act linearly on the annihilation
operators are called passive linear transformations (PLTs). For
example, the BS induces a PLT on $\cH_a \otimes \cH_b$. Any PLT
operator $\mathcal{U}$ acting on the input state is determined by a
unitary matrix \(U\) acting on vectors of mode operators or coherent
state amplitudes. We adopt the convention introduced implicitly for
BSs above: For PLTs, the unitary operator acting on the physical
Hilbert space is denoted by the calligraphic version of
the symbol used for the unitary matrix acting on mode
vectors. For example, if \(U\) is the
unitary matrix for a PLT transforming the vector of annihilation
operators $\hat{\textbf{c}}$ as
\(\hat{\textbf{c}} \mapsto U\cdot\hat{\textbf{c}}\), then
\(\mathcal{U}\) is the corresponding unitary operator on the Hilbert
space $\cH_c$.  In particular,
\(\mathcal{U}^{\dagger}\hat{c}_{i}\mathcal{U} =
\sum_{j}U_{ij}\hat{c}_{j}\).  We refer to the action of the unitary
matrix $U$ on $\hat{\textbf{c}}$ as the action or representation of
the PLT in the mode space of $\cH_c$.  For instance, the BS operator
$\cB$ is represented in mode-space basis
$\hat{\mathbf{a}} \oplus \hat{\mathbf{b}}$ by the unitary matrix
$B \equiv B_1 \otimes I_S$, with $I_S$ being the $S \times S$
identity. This is because $ B_1 \otimes I_S$ transforms each pair
$( \hat{a}_i, \hat{b}_i )$ into $B_1 \cdot ( \hat{a}_i, \hat{b}_i )$
in this basis.

\subsection{Operator Ordering and Correspondence with Expressions of Complex Variables}\label{subsec op_exp_cor}

The measurement statistics are the joint photon-counting probabilities
\( p_{ij} = \langle \Pi_i \otimes \Pi_j \rangle \), where \( \Pi_i \)
(\( \Pi_j \)) denotes the projector onto the subspace of
\( \mathcal{H}_{a'} \) (\( \mathcal{H}_{b'} \)) containing exactly
\( i \) (\( j \)) photons. The operators \(\Pi_{i}\otimes \Pi_{j}\)
are the measurement operators of the measurement configuration.  The
expectation defining \(p_{ij}\) is with respect to
$\rho \otimes \sigma$ since the total photon-number operators are
given in the Heisenberg picture using the evolved mode operators. From
the collection of joint probabilities \((p_{ij})_{ij}\), one can
determine the moments
\( \langle (\hatnpa )^{k} (\hatnpb )^{l} \rangle \) of the total
photon-number operators \( \hatnpa  \) and \( \hatnpb  \). For the
space of states considered in this work and introduced in
Sect.~\ref{sect:moment_problem}, these moments are finite, see Lem.~\ref{lem:finite_moments}.

In our analysis, it is convenient to work instead with the normally
ordered moments
\( \langle \nord{(\hatnpa )^{k} (\hatnpb )^{l}} \rangle \). Here,
the double colons ``\(:\,:\)'' denote normal ordering: for any
operator $\hat{P}$ that is a linear combination of mode operator
products, the operator \( \nord{\hat P} \) is obtained by rearranging
each term so that all creation operators appear to the left of
annihilation operators. It is crucial to emphasize that normal
ordering acts on the algebraic form of an operator expression, rather
than the operator itself. Because the canonical commutation relation
$[\hat{a}_i, \hat{a}_j^\dagger] = \delta_{ij}$ permits multiple
equivalent representations of the same operator, applying normal
ordering to these different representations produces distinct
operators. In accordance with standard practice, when we write a
normally ordered operator $\nord{\hat{P}}$ in closed form, we apply
the normal ordering to the fully expanded polynomial of the given
  expression for $\hat{P}$ obtained strictly without invoking any
commutation relations. For example, the normally ordered moment
$\langle \nord{(\hatnpa )^{k} (\hatnpb )^{l}} \rangle$ is obtained
by fully expanding the powers of the number operators, each expressed
in the standard form, that is
$\hatnpa  = \sum_{i=1}^S (\hat a'_i)^\dagger \hat a'_i$, and then
multiplying the two expansions and then applying normal ordering.

In addition to normal ordering, we also take advantage of
  anti-normal ordering.  For anti-normal ordering, creation operators
  are moved to the right of annihilation operators. We denote the
  anti-normally ordered operator obtained from an operator expression
  \(\hat{P}\) by \(\anormord{\hat{P}}\). The comments about normal
  ordering in the previous paragraph also apply to anti-normal
  ordering.

Each \( \nord{ (\hatnpa )^{k} (\hatnpb )^{l} } \) can be expressed
as a finite linear combination of the ordinary moments
\( (\hatnpa )^n (\hatnpb )^m \) and vice-versa, see
Appendix~\ref{app nor_n^k}. Thus, the ordinary moments of the
  measurement statistics are finite if and only if the
  normally-ordered moments
  $\langle \nord{ (\hatnpa )^{k} (\hatnpb )^{l} } \rangle$ are
  finite.

We use two key properties of normally ordered operators. First, given
any $K$ mode operators
\(\hat{\bf{c}} = (\hat{c}_1,\hdots, \hat{c}_K)\), the expectation of a
normally-ordered product
$ \qty(\otimes_{i=1}^K
  (\hat{c}_i^{\dagger})^{n_i})\qty(\otimes_{j=1}^K \hat{c}_j^{m_j}) $
with a coherent state $\ket{\vec{\delta}}$, where $\ket{\delta_i}$ is
the coherent state in mode $\hat c_i$, is the monomial
$ \prod_{i=1}^K (\delta^*_i)^{n_i} \prod_{j=1}^K
\delta_j^{m_j}$. Thus, for every normally ordered operator $\hat P$ we
have a corresponding polynomial expression $\cE[ \hat P ] $ in the
variables $\delta_i$ and $\delta_i^*$ with values
\(\cE[\hat P](\vec{\delta},\vec{\delta^{*}}) = \bra{\vec{\delta}}\hat
P\ket{\vec{\delta}}\) . We refer to $\cE[ \hat P ] $ as the amplitude
polynomial of $\hat P$. Conversely, given such a polynomial
expression we can compute the normally ordered operator $\hat P$ from
which it is obtained. For a polynomial \(p\) in the variables
\(\vec{\delta},\vec{\delta^{*}}\), we define \(\cE^{-1}[p]\) to be the
normally ordered operator $\hat P$ for which
\(\cE[\cE^{-1}[p]]=p\). The maps \(\hat P\mapsto \cE[\hat P]\) and
\(p\mapsto \cE^{-1}[p]\) are linear and identify the vector space
spanned by normally ordered operators with the vector space of
polynomials in \(\vec{\delta}\) and \(\vec{\delta^{*}}\). We refer to
this identification of vector spaces as the operator-polynomial
correspondence.
As operators on the Hilbert space, the vector
space spanned by normally ordered operators is the same as the vector
space spanned by general products of mode operators and their
adjoints.

The second key property is that PLTs map
normally ordered operators to normally ordered operators. This follows from the fact that PLTs transform creation operators into linear combinations of creation operators, and annihilation operators into linear combinations of annihilation operators. More generally, PLTs commute with the normal-ordering operation, see Appendix \ref{app nor_PLR} for a formal proof. For instance, \( \nord{ (\hatnpa )^{k}  (\hatnpb )^{l} } \) can be expressed as 
\begin{align}\label{eq: PLT_n_i^k}
& \nord{ (\hatnpa )^{k}  (\hatnpb )^{l}}  =  \nord{ \left( \sum_{i=1}^S (\hat a'_i)^\dagger \hat a'_i  \right)^k \left( \sum_{j=1}^S (\hat b'_j)^\dagger \hat b'_j  \right)^l} \nonumber \\
& =  \nord{ \left( \sum_{i=1}^S \cB^\dagger (\hat a_i)^\dagger  \cB  \cB^\dagger \hat a_i  \cB \right)^k   \left(  \sum_{j=1}^S \cB^\dagger (\hat b_j)^\dagger \cB \cB^\dagger \hat b_j \cB \right)^l  } \nonumber \\
& =  \nord{ \cB^\dagger \left( \sum_{i=1}^S  (\hat a_i)^\dagger  \hat a_i  \right)^k \cB \cB^\dagger   \left(  \sum_{j=1}^S  (\hat b_j)^\dagger  \hat b_j  \right)^l \cB } \nonumber \\
& =  \cB^\dagger \nord{ \left( \sum_{i=1}^S (\hat a_i)^\dagger \hat a_i \right)^k  \left( \sum_{j=1}^S (\hat b_j)^\dagger \hat b_j \right)^l } \cB \nonumber \\
& = \cB^\dagger \nord{ \hatna^k \hatnb ^l } \cB
\end{align}
where $\hatna$ and $\hatnb $ are the total photon-number
operators on $\cH_a $ and $\cH_b$, respectively. Here, we have used
the commutativity of $\cB$ with the normal-ordering operation in the
penultimate step.

The properties described in the previous paragraphs can be used to describe the transformation induced by a PLT on a normally-ordered operator $\hat P$ by a change in the polynomial $\cE[\hat P]$. In particular, consider the action of a PLT operator $\cU$ on $  \otimes_{i=1}^S  (\hat{a}_i^\dagger)^{n_i}  \otimes_{j=1}^S \hat{a}_j^{m_j}  $:
\begin{align}
   &\cU^\dagger \left[ \otimes_{i=1}^S  (\hat{a}_i^\dagger)^{n_i}  \right] \left[  \otimes_{j=1}^S \hat{a}_j^{m_j} \right]  \cU \hspace*{-1in}&\nonumber \\
  &= \otimes_{i=1}^S (\cU^\dagger \hat{a}^\dagger_i \cU)^{n_i} \otimes_{j=1}^S (\cU^\dagger \hat{a}_j \cU)^{m_j} \nonumber \\
  & = \otimes_{i=1}^S ( U^* \cdot  \hat{\textbf{a}}^\dagger)_i^{n_i} \otimes_{j=1}^S (U  \cdot  \hat{\textbf{a}})_j^{m_j},
\end{align}
where $( U^* \cdot \hat{\textbf{a}}^\dagger)_i$ is $i$th element of
the vector $U^* \cdot \hat{\textbf{a}}^\dagger$ with \(U^{*}\) the
matrix obtained by conjugating each entry of \(U\), and similarly for
$(U \cdot \hat{\textbf{a}})_j$. Therefore, the transformation of the
corresponding amplitude polynomial is
\begin{align}\label{eq: mon_trans_PLT}
 \prod_{i=1}^S (\alpha^*_i)^{n_i} \prod_{j=1}^S \alpha_j^{m_j} \to \prod_{i=1}^S (U^* \cdot \vec{\alpha }^*)_i^{n_i} \prod_{j=1}^S (U \cdot \vec{\alpha })_j^{m_j}.
\end{align} 
We use this correspondence under PLT action to describe Eq.~\ref{eq: PLT_n_i^k} as a transformation of expressions. $\cE\left[: (\hatnpa )^{k}  (\hatnpb )^{l} :\right]$ is given by
\begin{align}\label{eq: BS_trans_exp}
& \bra{\vec{\alpha'}}\bra{\vec{\beta'}} \nord{ (\hatnpa )^{k} (\hatnpb )^{l} }  \ket{\vec{\alpha'}}\ket{ \vec{\beta'} }  = \abs{\vec{\alpha'}}^{2k} \abs{\vec{\beta'}}^{2l},
\end{align}
where $\ket{\vec{\alpha'}}\ket{  \vec{\beta'} } \in \cH_{a'} \otimes \cH_{b'}$, see Eq.~\ref{eq: :n^k: = |a|^2k} in Appendix~\ref{app nor_n^k} for a derivation. Denoting a coherent state in the inputs of the BS by $\ket{\vec{\alpha}}\ket{  \vec{\beta} } \in \cH_{a} \otimes \cH_{b}$, Eq.~\ref{eq: PLT_n_i^k} implies that $\cB$ induces the following transformation of the corresponding amplitude polynomials:
\begin{align}\label{eq: ab -> a'b'}
\abs{\vec{\alpha}}^{2k} \abs{\vec{\beta}}^{2l} \to \abs{\vec{\alpha'}}^{2k} \abs{\vec{\beta'}}^{2l},
\end{align}
where, according to Eq.~\ref{eq: mon_trans_PLT}, $\vec{\alpha'} \oplus \vec{\beta'}  =  B \cdot (\vec{\alpha} \oplus  \vec{\beta}) $, or, equivalently, $(\alpha_i',\beta_i') = B_1 \cdot (\alpha_i, \beta_i)$ for each $i$. 

%

\subsection{Generating Functionology}\label{subsec genfun}

We use operator-valued generating functions, which are formal
  power series with operators as coefficients of the monomials. As we explain below, the
generating functions are chosen so that for the family of states
considered in this work and introduced in
Sect.~\ref{sect:moment_problem}, the expectations of the coefficients
can be obtained from the collection of joint photon-counting
probabilities \((p_{ij})_{ij}\), which makes it possible to work with
more convenient families of operators than the photon-number
projectors.

For a systematic treatment of generating functions and their
properties, see Ref.~\cite{wilf2005generating}. The measurement
configuration involving the joint counters can be represented by
$G(u,v) = \sum_{k=0}^\infty \sum_{l=0}^\infty \frac{1}{k!l!} \Pi_k
\otimes \Pi_l u^k v^l$, which should be interpreted as a formal power
series in the variables $u,v$.  The coefficient $\Pi_k \otimes \Pi_l $
can be extracted by taking the formal derivative
$ \partial_u^k \partial_v^l G(u,v) \mid_{u,v = 0} $, evaluated at
\(u,v=0\) to obtain the constant term of the resulting formal power
series. The generating function \(G(u,v)\) represents the measurement
configuration in the sense that for any given state, the expectations
of the operator coefficients are the probabilities of the measurement
outcomes. For our analysis, it is more convenient to represent the measurement
configuration by the generating function whose coefficients are
normally ordered moments:
\begin{align}\label{eq: H'}
H'(u,v) = \sum_{k=0}^\infty \sum_{l=0}^\infty \frac{1}{k!l!}  \nord{ (\hatnpa )^{k} (\hatnpb )^{l} } u^k v^l.
\end{align}
If the measurement configuration involves a single counter, the
corresponding generating function is obtained by setting $v=0$ in
$H'(u,v)$, resulting in a generating function whose coefficients are
the operators $\nord{ (\hatnpa )^{k} }\,$.

The coefficients of \(H'(u,v)\) are unbounded operators. It is
therefore necessary to restrict the state space in order for the
expectations of the operator coefficients to be finite. The
restriction to a dense set of states introduced in
Sect.~\ref{sect:moment_problem} suffices for this purpose.

Since the operator-valued coefficients of $H'(u,v)$ are normally ordered, we extend the operator-polynomial correspondence to $H'(u,v)$ by taking the expectation of all of its coefficients in a coherent state $\ket{\vec{\alpha'}}\ket{  \vec{\beta'} } \in \cH_{a'} \otimes \cH_{b'}$. Using Eq.~\ref{eq: BS_trans_exp}, we obtain
\begin{align}\label{eq: H'_expr}
& H'(u,v;\vec{\alpha'},\vec{\beta'})  \equiv \bra{\vec{\alpha'}}\bra{\vec{\beta'}} H'(u,v) \ket{\vec{\alpha'}}\ket{ \vec{\beta'} } \nonumber \\
&  \equiv  \sum_{k=0}^\infty \sum_{l=0}^\infty \frac{u^k v^l}{k!l!}  \abs{\vec{\alpha'}}^{2k} \abs{\vec{\beta'}}^{2l}  = e^{ \abs{\vec{\alpha'}}^2u + \abs{\vec{\beta'}}^2 v}.
\end{align}
$H'(u,v;\vec{\alpha'},\vec{\beta'}) $ is a convergent power series for
all values of the variables \(u\), \(v\), and of $\alpha'_i$,
$\beta_i'$ and their complex conjugates. The operator coefficients of
\(H'(u,v)\) can be recovered from the function
\(H'(u,v;\vec{\alpha'}\vec{\beta'})\) by evaluating the derivative
$ \partial_u^k \partial_v^l e^{ \abs{\vec{\alpha'}}^2u +
  \abs{\vec{\beta'}}^2 v} \mid_{u,v = 0} $ as usual, and applying the
operator-polynomial correspondence to the resulting polynomial of
$\alpha'_i$, $\beta_i'$ and their complex conjugates. We refer to
$H'(u,v;\vec{\alpha'}\vec{\beta'})$ as the amplitude generating
function of $H'(u,v)$, and we maintain this terminology whenever the
operator-polynomial correspondence is applied to an operator-valued
generating function with coefficients that are polynomials of
  mode operators and their adjoints.

The main advantage of working with $H'(u,v;\vec{\alpha'},\vec{\beta'}) $ lies in how it relates operators on $\cH_a \otimes \cH_b$ to operators on $\cH_{a'} \otimes \cH_{b'}$. Consider the operator-valued generating function
\begin{align}\label{eq: H}
H(u,v) = \sum_{k=0}^\infty \sum_{l=0}^\infty \frac{1}{k!l!}  \nord{ \hatna^k \hatnb ^l } \; u^k v^l,
\end{align}
and the corresponding amplitude generating function
\begin{align}
& H(u,v;\vec{\alpha},\vec{\beta}) \equiv \bra{\vec{\alpha}}\bra{\vec{\beta}} H(u,v) \ket{\vec{\alpha}}\ket{ \vec{\beta} } \nonumber \\ 
&=  e^{ \abs{\vec{\alpha}}^2u + \abs{\vec{\beta}}^2 v},
\end{align}
where $\ket{\vec{\alpha}}\ket{  \vec{\beta} } \in \cH_{a} \otimes \cH_{b}$. Using Eq.~\ref{eq: PLT_n_i^k}, $H'(u,v) = \cB^\dagger H(u,v) \cB$, where the transformation acts independently on each coefficient of the power series. The corresponding polynomials are related by the transformation of the complex variables, $(\alpha_i',\beta_i') = B_1 \cdot (\alpha_i, \beta_i)$ for each $i$, by applying Eq.~\ref{eq: ab -> a'b'} to each coefficient of $ H(u,v;\vec{\alpha},\vec{\beta})$. Thus, it is possible to express $H'$ as a function of $\vec{\alpha}$ and $\vec{\beta}$: 
\begin{align}\label{eq: genfun_main}
& H'(u,v;\vec{\alpha'},\vec{\beta'})  = e^{ \sum_{i=1}^S \abs{\alpha'_i}^2   u + \sum_{i=1}^S \abs{\beta'_i}^2   v} \nonumber \\
& =  e^{ \sum_{i=1}^S \abs{\eta \alpha_i + \zeta \beta_i}^2   u + \sum_{i=1}^S \abs{\zeta' \alpha_i + \eta' \beta_i}^2   v}.
\end{align}
Then, by again applying the operator-polynomial correspondence on $\cH_a \otimes \cH_b$ to each coefficient of $H'(u,v;\vec{\alpha'},\vec{\beta'})$, we obtain an expression for $H'(u,v)$ in terms of the mode operators $\hat{a}_i, \hat{a}_i^\dagger, \hat{b}_i$ and $ \hat{b}_i^\dagger$:
\begin{flalign}\label{eq: H'(u,v) in ab}
  H'(u,v) =   \hspace*{-.5in}&
\notag\\
  &\sum_{k=0}^\infty \sum_{l=0}^\infty \frac{1}{k!l!}
\nord{  \left[ \sum_{i=1}^S ( \eta^* \hat{a}_i^\dagger + \zeta^* \hat{b}_i^\dagger )  ( \eta \hat{a}_i + \zeta \hat{b}_i )    \right]^k \nonumber \\
 &\hphantom{\sum_{k=0}^\infty \sum_{l=0}^\infty \frac{1}{k!l!}:\,}\hspace*{-.15in}\times \left[ \sum_{j=1}^S  ( \zeta'^* \hat{a}_j^\dagger + \eta'^* \hat{b}_j^\dagger ) ( \zeta' \hat{a}_j + \eta' \hat{b}_j )    \right]^l  }\,  u^k v^l.
\end{flalign}
We could have obtained Eq.~\ref{eq: H'(u,v) in ab} more directly from Eq.~\ref{eq: H'} by expanding the coefficients $\nord{(\hatnpa )^{k} (\hatnpb )^{l} }$ in terms of the mode operators $\hat{a}_i, \hat{a}_i^\dagger, \hat{b}_i$ and $ \hat{b}_i^\dagger$. However, carrying out the derivation in this way provides a useful exercise in manipulating generating functions.

We are interested in using generating functions to represent the
linear spaces of operators on \(\cH_{a}\) whose expectations are
determined by measurements with different probe states. For a given
probe state
$\sigma = \ket{\gamma}\bra{\gamma} \otimes
\ket{\vec{0}}\bra{\vec{0}}$, the immediately available expectations
are the photon-counting probabilities \(p_{kl}\). These probabilities
are the expectations, with respect to the input state \(\rho\), of the
operators on \(\cH_{a}\) obtained by contracting the measurement
operators \(\Pi_{k}\otimes \Pi_{l}\) with \(\sigma\) by taking the
partial trace of \(\sigma\qty(\Pi_{k}\otimes \Pi_{l})\) over \(\cH_{b}\):
\begin{align}
\hat{E}_{kl}(\gamma) & \equiv \tr_{b} \left( \ket{\gamma}\bra{\gamma} \otimes
\ket{\vec{0}}\bra{\vec{0}} \Pi_{k}\otimes \Pi_{l} \right) \nonumber \\
&=  \bra{\gamma}\bra{\vec{0}} \Pi_{k}\otimes \Pi_{l}  \ket{\gamma}\ket{\vec{0}}.
\end{align}
The photon number probabilities are then given by
$p_{kl} = \tr (\rho \hat{E}_{kl}(\gamma) )$. Because the states
  under consideration have finite photon number moments
  (Lem.~\ref{lem:finite_moments}), the normally ordered moments
  $\langle \nord{(\hatnpa )^{k} (\hatnpb )^{l}} \rangle$ are
  linearly determined by the photon number probabilities \(p_{kl}\).
  These expectations are also obtained as expectations of operators on \(\cH_{a}\) with respect to
  \(\rho\), where these operators are
  obtained in the same way as \(\hat{E}_{kl}(\gamma)\) by contracting the $\nord{(\hatnpa )^{k} (\hatnpb )^{l}}$ with \(\sigma\).  By
  taking advantage of the identity of the coefficients of \(H'(u,v)\)
  as expressed in Eq.~\ref{eq: H'} with those expressed in
  Eq.~\ref{eq: H'(u,v) in ab}, the contraction is obtained as follows:
\begin{align}\label{eq: enduced_coef}
&  \bra{\gamma}\bra{\vec{0}} \nord{(\hatnpa )^{k} (\hatnpb )^{l}} \ket{\gamma}\ket{\vec{0}} \nonumber \\
& =  \bra{\gamma}\bra{\vec{0}}   \nord{ \left[\sum_{i=1}^S ( \eta^* \hat{a}_i^\dagger + \zeta^* \hat{b}_i^\dagger ) ( \eta \hat{a}_i + \zeta \hat{b}_i )   \right] ^k \nonumber \\
  &  \times  \left[ \sum_{j=1}^S ( \zeta'^* \hat{a}_j^\dagger + \eta'^* \hat{b}_j^\dagger ) ( \zeta' \hat{a}_j + \eta' \hat{b}_j ) \right]^l  }   \ket{\gamma}\ket{\vec{0}} \nonumber \\
  & = \nord{ \left[\sum_{i=1}^S ( \eta^* \hat{a}_i^\dagger + \zeta^* \gamma^* \delta_{1i} )  ( \eta \hat{a}_i + \zeta \gamma  \delta_{i1} ) \right]^{k}  \nonumber \\
  & \times \left[ \sum_{j=1}^S ( \zeta'^* \hat{a}_j^\dagger + \eta'^*\gamma^*  \delta_{j1} ) ( \zeta' \hat{a}_j + \eta' \gamma \delta_{j1} ) \right]^l    },
\end{align}
where $\delta_{ij}$ denotes the Kronecker delta.  In the last
  step, we applied the identities
  $\bra{\gamma}\bra{\vec{0}}\hat{b}_i^\dagger=\delta_{1i}\gamma^*\bra{\gamma}\bra{\vec{0}}$
  and
  $\hat{b}_j\ket{\gamma}\ket{\vec{0}}=\ket{\gamma}\ket{\vec{0}}\delta_{j1}\gamma$.
The expectation
$\langle \nord{(\hatnpa )^{k} (\hatnpb )^{l}} \rangle$ is then
obtained by taking the trace of the final result in Eq.~\ref{eq:
  enduced_coef} with $\rho$.

For a given $\gamma$, $\cC_2(\gamma)$ denotes the set of operators obtained by fully expanding Eq.~\ref{eq: enduced_coef} into a sum of monomials and collecting the operator-valued coefficients for all \(k,l\geq 0\), meaning that both counters are used. When one detector is used, $\cC_1(\gamma)$ similarly denotes the set obtained with all \(k\geq 0\) and $l=0$. We denote by
$\cE[\cC_i(\gamma)]$ the set of amplitude polynomials corresponding to
the operators in \(\cC_i(\gamma)\), obtained via the
operator–polynomial correspondence.

The derivation in Eq.~\ref{eq: enduced_coef} of the operators in
  \(\cC_{2}(\gamma)\) makes it clear that \(\cC_{2}(\gamma)\) is the
  collection of coefficients of the operator-valued generating
  function obtained by contracting \(H'(u,v)\) with \(\sigma\)
  term-by-term, that is by calculating
  \(F_\gamma(u,v) \equiv \bra{\vec{\alpha}} \bra{\gamma}\bra{\vec{0}}
  H'(u,v) \ket{\vec{\alpha}} \ket{\gamma}\ket{\vec{0}}\).
  \(\cC_{1}(\gamma)\) is therefore the collection of coefficients
  obtained by specializing \(F_{\gamma}(u,v)\) by setting \(v=0\). The
  generating function \(F_{\gamma}(u,v)\) has the following analytic
  expression:
\begin{align}\label{eq: F(u,v,gamma)_gen}
& F_\gamma(u,v) \equiv   \bra{\vec{\alpha}} \bra{\gamma}\bra{\vec{0}} H'(u,v) \ket{\vec{\alpha}} \ket{\gamma}\ket{\vec{0}} \nonumber \\
& \equiv  \sum_{k=0}^\infty \sum_{l=0}^\infty \frac{u^k v^l}{k!l!}  \bra{\vec{\alpha}}   \nord{ \left[\sum_{i=1}^S ( \eta^* \hat{a}_i^\dagger + \zeta^* \gamma^* \delta_{i1} )  ( \eta \hat{a}_i + \zeta \gamma  \delta_{i1} ) \right]^{k} \nonumber \\
&\times \left[ \sum_{j=1}^S ( \zeta'^* \hat{a}_j^\dagger + \eta'^*\gamma^*  \delta_{j1} ) ( \zeta' \hat{a}_j + \eta' \gamma \delta_{j1} ) \right]^l  } \ket{\vec{\alpha}} \nonumber \\
  & = \sum_{k=0}^\infty \sum_{l=0}^\infty \frac{u^k v^l}{k!l!} \left[\sum_{i=1}^S  ( \eta^* \alpha_i^*+ \zeta^* \gamma^* \delta_{i1} )  ( \eta \alpha_i + \zeta \gamma  \delta_{i1} ) \right]^k  \nonumber \\
  &\times \left[\sum_{j=1}^S ( \zeta'^* \alpha_j^* + \eta'^*\gamma^*  \delta_{j1} )  ( \zeta' \alpha_j + \eta' \gamma \delta_{j1}) \right]^l   \nonumber \\
  & =  e^{ \sum_{i=1}^S \abs{\eta \alpha_i + \zeta \gamma \delta_{i1}}^2   u + \sum_{j=1}^S \abs{\zeta' \alpha_j + \eta' \gamma \delta_{j1}}^2   v}.
\end{align}
The operator in Eq.~\ref{eq: enduced_coef} is
recovered by evaluating the derivative
$ \partial_u^k \partial_v^l F_\gamma(u,v) \mid_{u,v = 0} $ and
applying the operator-polynomial correspondence on $\cH_{a}$. Note
that the expression for $F_\gamma(u,v)$ could have been obtained
directly from Eq.~\ref{eq: H'(u,v) in ab} by substituting
$\vec{\beta} = (\gamma,0,\ldots,0)$—that is, by treating the arbitrary
coherent state on $\mathcal{H}_b$ used in the operator–expression
correspondence as the actual probe state. We have chosen the longer
route to avoid potential confusion between the different
roles of the variables $\vec{\alpha}$ and $\gamma$ in $F_\gamma(u,v)$.

As a final note, we emphasize that while $F_\gamma(u,v)$ is an
analytic function in the variables $u,v, \vec{\alpha},\vec{\alpha}^*$,
its physical purpose is to encode the operators $\cC_2(\gamma)$
through the operator-amplitude correspondence. We treat $u$ and $v$
strictly as generating variables and suppress the explicit
dependence on $\vec{\alpha}$ and $\vec{\alpha}^*$ in the notation for
$F_\gamma(u,v)$ to reflect this physical purpose.

\subsection{States with Moment-Determinate Husimi Functions}
\label{sect:moment_problem}

Because our analysis characterizes the measurement statistics by the moments of the measured photon number
distributions, these moments must be finite. A sufficient condition
for this is that the space of states under consideration has
well-defined expectations for all normally ordered products
$ \otimes_{j=1}^S (\hat{a}_j^\dagger)^{m_j} \otimes_{i=1}^S
\hat{a}_i^{n_i}$. The next lemma establishes the required connections
between the photon-number measurement statistics and normally
ordered operators.

\begin{lemma}\label{lem:finite_moments}
  If \(\rho\) has finite normally ordered moments, then for all LO
  amplitudes \(\gamma\), the photon-number measurement statistics
  have finite moments. If \(\rho\) satisfies the property that the moments of the
  photon-number measurement statistics are finite for a given LO
  amplitude \(\gamma\), then the expectations of the normally
  ordered total photon number moments of the BS outputs at
  \(\gamma\) are determined as finite linear combinations of moments of
  the photon-number measurement statistics at \(\gamma\).
\end{lemma}
\begin{proof}
  For the first part of the lemma, Eq.~\ref{eq: enduced_coef} shows
  that, for a given \(\gamma\), the normally ordered moments of the
  total photon number operators after the BS given by the
  \(\langle \nord{ (\hatnpa )^{k} (\hatnpb )^{l} }\rangle\), are
  finite linear combinations of normally ordered moment operators of
  the input states. If the latter have finite expectations, then so
  do the former. 
  Cor.~\ref{cor:usual_moments} in the appendix implies that the
  ordinary moments are finite linear combinations of normally
  ordered moments, implying that if the
  $\langle \nord{ (\hatnpa )^{k} (\hatnpb )^{l} }\rangle$ are
  finite, so are the ordinary moments, that is the moments of
  the photon-number measurement statistics.

  For the second part, 
  Appendix~\ref{app nor_n^k} shows that the
  $\langle \nord{ (\hatnpa )^{k} (\hatnpb )^{l} } \rangle$
  are finite linear combinations of the
  ordinary moments. Therefore, if the latter have finite
  expectations at \(\gamma\), we can determine the expectations of
  each of the former at \(\gamma\).
\end{proof}

Since any product of mode operators on $\cH_a$ can be expressed as a
finite linear combination of anti-normally ordered products by
repeated use of the commutation relations
$[\hat{a}_i^\dagger, \hat{a}_j] = \delta_{ij}$, the requirement that
all normally-ordered products possess well-defined expectations can
be naturally expressed in the Husimi representation of state
space. The Husimi function~\cite[Chap.~3]{leonhardt:qc1997a} of an
$S$-mode quantum state $\rho$ is defined as
\( Q(\vec{\alpha}, \vec{\alpha}^*) = \frac{1}{\pi^S}
\bra{\vec{\alpha}} \rho \ket{\vec{\alpha}}\). The definition
  extends to arbitrary bounded operators. For states, the Husimi
function is non-negative and normalized, so it can be interpreted as
a probability measure. Its complex moments -- the expectations of
monomials in the variables $\alpha_i$ and $\alpha_i^*$ with respect
to the measure \( Q(\vec{\alpha}, \vec{\alpha}^*) \) -- correspond
to expectations of anti-normally ordered mode operators for the
state $\rho$. Specifically, for every anti-normally ordered monomial
of the form
\( \otimes_{i=1}^S \hat{a}_i^{n_i} \otimes_{j=1}^S
(\hat{a}_j^\dagger)^{m_j} \), the following identity holds for
states \(\rho\) whose Husimi functions have well-defined complex
moments:
\begin{align}\label{eq: Husimi_identity}
  & \tr\left( \rho \left[\otimes_{i=1}^S \hat{a}_i^{n_i} \right] \left[ \otimes_{j=1}^S (\hat{a}_j^\dagger)^{m_j} \right] \right) \nonumber \\
  & = \int d\vec{\alpha}\, d\vec{\alpha}^*\, Q(\vec{\alpha}, \vec{\alpha}^*) \prod_{i=1}^S \alpha_i^{n_i} (\alpha_i^*)^{m_i}.  
\end{align}

For our analysis we require the stronger condition that the moments of
the Husimi function be determinate for the twirled state obtained from
the input state as described below in Sect.~\ref{subsec
  twirl}. Determinateness is a property that arises in the study of
the moment problem~\cite{schmudgen2017moment}.  The following
definition is adapted from
Ref.~\cite[Chap.~15]{schmudgen2017moment}. A multivariate distribution
over complex variables $\alpha_1, \ldots, \alpha_S$ and their complex
conjugates is said to have a determinate moment sequence if the
distribution is uniquely determined by the expectations of all
monomials of the form
$\prod_{i=1}^S \alpha_i^{n_i} (\alpha_i^*)^{m_i}$ for all non-negative
integers $n_i, m_i$. The family of expectations of these monomials is
referred to as the moment sequence. Being uniquely determined means
that there is no other multivariate distribution with the same moment
sequence. The space of states with determinate moments is defined as
the states whose Husimi function has a determinate moment sequence,
and is denoted by $\cS$. We refer to such states as states with a
determinate Husimi moment sequence. Gaussian states, states with
bounded photon number, and finite mixtures and superpositions of such
states satisfy this property, see Appendix~\ref{app: Husimi_det}. The
appendix also shows that the states of \(\cS\) have photon-number
measurement statistics whose moment sequences are determinate for all
LO amplitudes $\gamma$. 

The property of having determinate Husimi moments is preserved under PLTs, as the following lemma shows.
\begin{lemma}
  The set of states $\cS$ is closed under PLTs: if $\rho \in \cS$
  and $\cU$ is a PLT on $\cH_a$, then $\cU\rho\cU^\dagger \in
  \cS$. 
\end{lemma}
\begin{proof}
  Let $\rho' = \cU\rho\cU^\dagger$. We first show that each
  anti-normally ordered moment of $\rho'$ is a finite linear
  combination of the anti-normally ordered moments of $\rho$, and vice
  versa. To express the antinormally ordered moments of \(\rho'\) we
  compute
\begin{align}
 & \tr\left( \cU \rho \cU^\dagger \left[\otimes_{i=1}^S \hat{a}_i^{n_i} \right] \left[ \otimes_{j=1}^S (\hat{a}_j^\dagger)^{m_j} \right] \right) \nonumber \\
 &= \tr\left(  \rho \cU^\dagger \left[\otimes_{i=1}^S \hat{a}_i^{n_i} \right] \left[ \otimes_{j=1}^S (\hat{a}_j^\dagger)^{m_j} \right] \cU \right) \nonumber \\
  &= \tr\left(  \rho  \left[\otimes_{i=1}^S (\cU^\dagger \hat{a}_i  \cU)^{n_i} \right] \left[ \otimes_{j=1}^S ( \cU^\dagger  \hat{a}_j^\dagger \cU )^{m_j} \right] \right) \nonumber \\
&= \tr\left(  \rho   \otimes_{i=1}^S ( U  \cdot  \hat{\textbf{a}})_i^{n_i} \otimes_{j=1}^S (    U^* \cdot  \hat{\textbf{a}}^\dagger)_j^{m_j} \right).
\end{align}
Expanding the transformation
$(U \cdot \hat{\mathbf{a}})_i = \sum_k U_{ik}\hat{a}_k$ and pulling
the resulting sums outside the trace results in a linear
  combination of the antinormally ordered moments of \(\rho\).  To
  express the antinormally ordered moments of \(\rho\) in terms of
  those of \(\rho'\), it suffices to perform an identical derivation,
  where we start with $\rho = \cU^\dagger \rho' \cU$ and replace $U$
  with $U^\dagger$.

The previous paragraph implies that the Husimi function of
\(\rho'\) has a well-defined moment sequence. We now show that this moment sequence is determinate. Suppose that \(\mu'\)
is a probability distribution that has the same moment
sequence. Although \(\mu'\) might not be the Husimi function of a
  state, we can transform \(\mu'\) to \(\mu\) by the inverse of the
PLT in the same way that we transform Husimi functions. The
corresponding finite transform of the moment sequence of \(\mu'\)
gives the same moment sequence as that of the Husimi function of \(\rho\). Therefore, by moment determinateness, $\mu$ is the Husimi
function of $\rho$, which in turn implies that $\mu'$ is the Husimi
function of $\rho'$. By arbitrariness of \(\mu'\), we conclude that \(\rho'\in\cS\).
\end{proof}

\section{Symmetries of the Measurement Configuration}\label{sec sym}

\subsection{Equivalence Class of Input States}\label{subsec equivclas}

Because PLTs preserve total photon number,the measurement statistics
do not change if PLTs are applied on each of the output paths of the
BS. More specifically, if $\cX_{a} : \cH_{a'} \mapsto \cH_{a'}$ and
$\cX_{b} : \cH_{b'} \mapsto \cH_{b'}$ are PLT operators, then
$[\hatnpa , \cX_{a}] = 0$ and $[\hatnpb , \cX_{b}] = 0$.  The mode
space representation of the PLT $\cX_{a} \otimes \cX_{b}$ in the basis
$\hat{\textbf{a}'}\oplus \hat{\textbf{b}'}$ is the unitary matrix
$X_{a} \oplus X_{b}$. This is because $\cX_{a}$ acts by transforming
each $\hat{a}'_i$ into a linear combination of the $\hat{a}'_i$, and
analogously, $\cX_{b}$ transforms each $\hat{b}'_i$ into a linear
combination of the $\hat{b}'_i$.

Let $U(S-1)$ denote the group of PLTs on $\cH_a$ that have the form
$U = I_1 \oplus W$ in the mode-space basis $\hat{\textbf{a}}$, where $I_1$
is the $1 \times 1$ identity and $W$ is an $S-1 \times S-1$ unitary
matrix. We overload the notation and let $U(S-1)$ also denote
the corresponding group acting on the Hilbert spaces $\mathcal{H}_b$,
$\mathcal{H}_{a'}$, and $\mathcal{H}_{b'}$. We first show with two
propositions that a PLT $\cU$ on $\cH_a$ preserves the measurement
statistics for all input and probe states if and only if
$\cU \in U(S-1)$.

\begin{prop}\label{thm: PLTpresmeasstat}
Consider the measurement configuration with one or two counters. Consider a PLT $\cU \in \textrm{U}(S-1)$ acting on \(\cH_{a}\). Then, $\rho $ and $\cU^\dagger \rho \cU$ have identical measurement statistics when combined with the same probe state. 
\end{prop}

\begin{proof}
  It suffices to prove the claim for the configuration with two
  counters. We first note that we may apply the same $\cU$ on
  $\cH_b$. Doing so has no effect on mode $\hat{b}_1$ since $U$
  has the form $U = I_1 \oplus W$ in the mode-space basis
  $\hat{\textbf{b}}$. Applying the corresponding transformation to
  $\sigma = \ket{\gamma}\bra{\gamma} \otimes
  \ket{\vec{0}}\bra{\vec{0}}$, gives
\begin{align}
& \cI_1 \otimes \cW^\dagger (\ket{\gamma}\bra{\gamma} \otimes \ket{\vec{0}}\bra{\vec{0}}) \cI_1 \otimes \cW \nonumber \\
&= \ket{\gamma}\bra{\gamma}  \otimes \cW^\dagger  \ket{\vec{0}}\bra{\vec{0}} \cW  = \ket{\gamma}\bra{\gamma}  \otimes \ket{\vec{0}}\bra{\vec{0}}.
\end{align}
Consequently,
\begin{align}
  \cU \otimes \cU \left( \rho \otimes \sigma  \right) \cU^\dagger \otimes \cU^\dagger =( \cU \rho \cU^\dagger) \otimes \sigma
  \label{eq:pltstat1}
\end{align}
for all $\rho$ and $\sigma$. Hence, to establish the claim of the
  proposition it suffices to demonstrate that the action of
$\cU \otimes \cU$ on $\cH_a \otimes \cH_b$ does not affect the
measurement statistics.

Working in the mode-space representation, we next establish the fact that
  $B (U \oplus U) = (U \oplus U) B$, from which it follows that applying
  \(\cU\otimes \cU\) before the beam splitter is equivalent to
  applying it after.  This fact is established by the following identity:
\begin{align}
B (U \oplus U) B^\dagger &= \left[
\begin{array}{c|c}
\eta I_S & \zeta I_S \\
  \hline
   \zeta' I_S  &  \eta' I_S
\end{array}    \right] \left[
\begin{array}{c|c}
  U &0  \\
  \hline
   0  &  U
\end{array}    \right]  \left[
\begin{array}{c|c}
   \eta^* I_S & \zeta'^*I_S \\
  \hline
   \zeta^* I_S & \eta'^*I_S
\end{array}    \right]    \nonumber \\
& = \left[ \begin{array}{c|c}
\abs{\eta}^2U + \abs{\zeta'}^2U & \eta \zeta'^* U +\zeta \eta'^* U \\ 
  \hline
\eta^*\zeta' U +\zeta^* \eta' U & \abs{\zeta}^2 U + \abs{\eta'}^2 U \\
           \end{array} \right]
  \notag\\
  &=U \oplus U,
\end{align}
were we applied normalization and orthogonality of the columns of the
BS matrix to obtain the last line.
\end{proof}

Next, we show that any PLT on \(\mathcal{H}_a\) that preserves the
measurement statistics for all input and probe states must lie in
$U(S-1)$. If all measurement statistics are preserved, then
$\langle \hatnpa \rangle$ is preserved, so it suffices to prove the
proposition for the case where only \(\langle\hatnpa\rangle\) is
  assumed to be preserved.

\begin{prop}\label{thm PLTpresnumone}
  Consider the measurement configuration with one or two counters. Consider a PLT $\cU$ on $\cH_a$. Suppose that for every $\rho$ and every probe state \(\sigma= \ket{\gamma}\bra{\gamma}  \otimes \ket{\vec{0}}\bra{\vec{0}}\), the expectation $\langle \hatnpa  \rangle$ is the same for the states $\rho$ and $\cU^\dagger \rho \cU$. Then, $\cU \in U(S-1)$.
\end{prop}
\begin{proof}
The assumption in the proposition statement implies that for all \(\rho\) and \(\sigma= \ket{\gamma}\bra{\gamma}  \otimes \ket{\vec{0}}\bra{\vec{0}} \) we have
\begin{align}
\tr( \left( \rho \otimes \sigma \right) \hatnpa ) = \tr( \left( \cU^\dagger \rho \cU \otimes \sigma \right) \hatnpa ).
\end{align}
 We define $N_\sigma = \tr_b(\sigma \hatnpa )$, where the
index $b$ implies a partial trace over $\cH_b$. Then, the above
is equivalent to
 \begin{align}
\tr( \rho N_\sigma) &= \tr( \cU^\dagger \rho \cU N_\sigma) \nonumber \\
& = \tr(\rho  \cU N_\sigma \cU^\dagger)
\end{align}
for all $\rho$ and $\sigma$. Since $\rho$ is arbitrary, this is equivalent to requiring that $ \cU N_\sigma \cU^\dagger = N_\sigma $ for all $\sigma$. We show next that this condition holds if and only if $U$ has the form $U = I_1 \oplus W$.

To evaluate the partial trace defining $N_\sigma$ we use the relations $\hat{a'}_i = \eta \hat{a}_i + \zeta \hat{b}_i$ and $\hat{a'}^\dagger_i = \eta^* \hat{a}^\dagger_i + \zeta^* \hat{b}^\dagger_i$:
\begin{align}
& N_\sigma = \tr_{b} \left(  \ket{\gamma}\bra{\gamma}  \otimes \ket{\vec{0}}\bra{\vec{0}} \sum_{i=1}^S (\eta^* \hat{a}^\dagger_i + \zeta^* \hat{b}^\dagger_i) ( \eta \hat{a}_i + \zeta \hat{b}_i) \right) \nonumber \\
& = \sum_{i=1}^S\bra{\gamma}\bra{\vec{0}}  (\eta^* \hat{a}^\dagger_i + \zeta^* \hat{b}^\dagger_i) ( \eta \hat{a}_i + \zeta \hat{b}_i) \ket{\gamma}\ket{\vec{0}} \nonumber \\
&= \sum_{i=1}^S \abs{\eta}^2 \hat{a}^\dagger_i \hat{a}_i + \eta^* \zeta \gamma \hat{a}^\dagger_1 + \zeta^* \eta \gamma^* \hat{a}_1 + \abs{\zeta}^2 \abs{\gamma}^2 \nonumber \\
& = \abs{\eta}^2 \hatna +\abs{\zeta}^2 \abs{\gamma}^2 +  \eta^* \zeta \gamma \hat{a}^\dagger_1 + \zeta^* \eta \gamma^* \hat{a}_1.
\end{align}
The first two terms on the last line commute with every PLT on
$\mathcal{H}_a$. The remaining two terms are proportional to
$\hat{a}^\dagger_1$ and $\hat{a}_1$, where the two coefficients are
nonzero for nonzero \(\gamma\). Since a PLT transforms annihilation
operators into linear combinations of annihilation operators (and
likewise for creation operators), to commute with $N_\sigma$, $\cU$
must preserve $\hat{a}_1$ and $\hat{a}_1^\dagger$. Consequently, $U$
must be of the form $U = I_1 \oplus W$.
\end{proof}

\subsection{The Twirled State}\label{subsec twirl}

The set of input states can be partitioned into equivalence classes such that the states within each class are indistinguishable by the measurement configuration for all probe states. In particular, for any given $\rho$, the states of the form $\mathcal{U}^\dagger \rho \mathcal{U}$ with $\cU \in \textrm{U}(S-1)$ belong to the same equivalence class as $\rho$. Moreover, any convex mixture of these states is also in the same class. Each equivalence class contains a particularly simple representative, obtained by the following integral:
\begin{equation} \label{eq: Tau}
\mathcal{T}(\rho) \equiv \int_{\cX \in \mathrm{U}(S-1)} d\mu(X)\, \cX \rho \cX^\dagger.
\end{equation}
Here, the integration is with respect to the normalized Haar measure $\mu(X)$ on $\mathrm{U}(S-1)$, which is invariant under left and right multiplication by elements of $\mathrm{U}(S-1)$. $\mathcal{T}(\rho)$ is in the same equivalence class as $\rho$ since it is the uniform mixture of states in the equivalence class of $\rho$. Next, we evaluate this integral for an arbitrary $\rho$.

 In the theorem
below, $\cH_{a_{\bar 1}}$ denotes the subsystem Hilbert space of
$\cH_a$ corresponding to the modes $\hat{a}_2,\hdots, \hat{a}_S$, and
$\cH_{a_1}$ denotes the Hilbert space corresponding to mode
$\hat{a}_1$. We use Schur's lemma together with the representation
theory of PLTs on bosonic Fock space for the proof of the
theorem.  We let
$\cH_{a_{\bar 1}}^{(n)}$ denote the $n$-photon subspace of
$\cH_{a_{\bar 1}}$, with $\cH_{a_{\bar 1}}^{(0)}$ being the
one-dimensional space spanned by the vacuum state
$\ket{\vec{0}}$. Then,
$\cH_{a_{\bar 1}} = \oplus_{n = 0}^\infty \cH_{a_{\bar 1}}^{(n)}$.
$\cH_{a_{\bar 1}}^{(n)}$ can be identified with the symmetric subspace
of $\left(\cH_{a_{\bar 1}}^{(1)} \right)^{\otimes n}$ and therefore
has dimension $d(n) = \binom{n+S-2}{n}$. We denote the unitary group
of PLTs on $\cH_{a_{\bar 1}}$ by $\textrm{U}_F(S-1)$, so that
$\textrm{U}(S-1) = \cI_1 \otimes \textrm{U}_F(S-1)$.  For each photon
number $n$, the action of $\textrm{U}_F(S-1)$ leaves the subspace
$\cH_{a_{\bar 1}}^{(n)}$ invariant and constitutes a distinct,
irreducible representation of the group. Irreducibility follows by
  Schur-Weyl duality for the unitary group, since in the identification with the symmetric
  subspace of $\left(\cH_{a_{\bar 1}}^{(1)} \right)^{\otimes n}$,
  \(\textrm{U}_{F}(S-1)\) acts as the unitary group, identically on
  each factor.  For a given $\cW \in \textrm{U}_F(S-1)$, we denote
its restriction to the $n$-photon subspace by $W^{(n)}$. Accordingly,
with respect to the direct sum decomposition
$\cH_{a_{\bar 1}} = \bigoplus_{n = 0}^\infty \cH_{a_{\bar 1}}^{(n)}$,
the operator $\cW$ takes the block-diagonal form
$\cW = \bigoplus_{n=0}^\infty W^{(n)}$, where \(W^{(0)} \equiv I_1\).

\begin{theorem} \label{thm Tau(rho)}
Consider an arbitrary input state $\rho$. Then $\cT(\rho)$ has the form $\cT(\rho) = \sum_{n} \chi_n \otimes \frac{1}{d(n)} \Pi_n^{S-1}$, where $\chi_n $ are positive-semidefinite operators on $\cH_{a_1}$ and $\Pi_n^{S-1}$ is the projector onto the subspace of $\cH_{a_{\bar 1}}$ containing $n$ photons.
\end{theorem}

\begin{proof}
  Schur's lemma states that if a group representation decomposes into mutually inequivalent irreducible representations, its commutant consists precisely of those operators that act as scalar multiples of the identity on each irreducible subspace~\cite{fulton:qc1991a}. Because the representations $W^{(n)}$ are irreducible and mutually inequivalent, Schur's lemma restricts any operator commuting with $\textrm{U}_F(S-1)$ on $\cH_{a_{\bar 1}}$ to the form $\sum_{n=0}^\infty k_n \Pi_n^{S-1}$, where $k_n$ are scalars. The extension to the total Fock space $\cH_a = \cH_{a_1} \otimes \cH_{a_{\bar 1}}$ follows from the algebraic structure of tensor-product representations. Since the group acts on the joint space as $\cI_1 \otimes \cW$, an operator commutes with this total action if and only if it belongs to the tensor product of the respective commutants. The commutant of the trivial action $\cI_1$ is the set of all bounded operators on $\cH_{a_1}$, while the commutant of $\textrm{U}_F(S-1)$ is spanned by the projectors $\Pi_n^{S-1}$. Thus, the overall commutant consists precisely of operators taking the form $\sum_{n=0}^\infty A_n \otimes \Pi_n^{S-1}$, where $A_n$ are arbitrary bounded operators on $\cH_{a_1}$.  
  
 We now show that $\cT(\rho)$ commutes with $\textrm{U}(S-1)$. For any $\cU \in \textrm{U}(S-1)$,
\begin{align}
&  \cU \int_{\cX \in \mathrm{U}(S-1)} d\mu(X)\, \cX \rho \cX^\dagger  =  \int_{\cX \in \mathrm{U}(S-1)} d\mu(X)\, (\cU \cX) \rho \cX^\dagger \nonumber \\
& =  \int_{\cY \in \mathrm{U}(S-1)} d\mu(Y)\, \cY \rho (\cU^\dagger \cY)^\dagger \nonumber \\
& = \int_{\cY \in \mathrm{U}(S-1)} d\mu(Y)\, \cY \rho  \cY^\dagger \cU,
\end{align}
where in the second line we performed the variable substitution $\cY = \cU \cX  $, so that $\cX = \cU^\dagger \cY$, and used the invariance of $\mu(X)$ under left multiplication by elements of $\textrm{U}(S-1)$. It follows that $\cT(\rho)$ must be an operator of the form $\sum_{n=0}^\infty A_n \otimes  \Pi_n^{S-1}$. In order for $\cT(\rho)$ to be a quantum state, the corresponding $A_n$ must be positive-semidefinite operators. Defining $\chi_n$ such that $A_n = \frac{1}{d(n)} \chi_n$ produces the form of $\cT(\rho)$ in the theorem statement.
\end{proof}

The expression for $\mathcal{T}(\rho)$ in Theorem~\ref{thm Tau(rho)} has a natural interpretation: the uniform averaging over all unitary transformations that respect the symmetries effectively eliminates all components of the density matrix that are unobservable by the measurement configuration.
The form of $\cT(\rho)$ also suggests that the number of modes $S$ is unobservable, as long as $S>1$. Thus, there should exist a state in the equivalence class of each $\rho$ that occupies only modes $\hat a_1$ and $\hat a_2$.  In what follows, $\ket{n,\vec{0}} \in \cH_{a_{\bar 1}}$ refers to the Fock state with $n$ photons in mode $\hat a_2$, and no photons in the other modes. 
Furthermore, we define $\mathcal{T}_F$ as the averaging map for operators on the subsystem $\mathcal{H}_{a_{\bar 1}}$. This map is defined by integrating over $\mathrm{U}_F(S-1)$ with respect to its Haar measure:
\begin{equation}
\mathcal{T}_F(\cdot) \equiv \int_{\cX \in \mathrm{U}_F(S-1)} d\mu(X)\, \cX (\cdot) \cX^\dagger.
\end{equation}
\begin{corollary}\label{cor rho_t}
Consider an arbitrary input state $\rho$. The two-mode state $\rho_t = \sum_n \chi_n \otimes \ket{n,\vec{0}} \bra{n,\vec{0}}$, where the $\chi_n$ are the operators in $\cT(\rho) = \sum_{n} \chi_n \otimes \frac{1}{d(n)} \Pi_n^{S-1}$, is indistinguishable from $\rho$ by the measurement configuration. 
\end{corollary}
\begin{proof}
It suffices to show that $\cT(\rho_t) = \cT(\rho)$. We first argue that $\cT_F( \ket{n,\vec{0}} \bra{n,\vec{0}}) = \frac{1}{d(n)} \Pi_n^{S-1}$. Observe that $\cT_F( \ket{n,\vec{0}} \bra{n,\vec{0}})$ is in the commutant of the representation of $\textrm{U}_F(S-1)$ on $\cH_{a_{\bar 1}}$, and has total photon number $n$. Therefore $\cT_F( \ket{n,\vec{0}} \bra{n,\vec{0}})$ is in the commutant of the irreducible representation of $\textrm{U}_F(S-1)$ on $\cH_{a_{\bar 1}}^{(n)}$. By Schur's lemma, $\cT_F( \ket{n,\vec{0}} \bra{n,\vec{0}}) = \lambda \Pi_n^{S-1}$ for some constant $\lambda$. Imposing the trace condition $\tr \left( \cT_F( \ket{n,\vec{0}} \bra{n,\vec{0}}) \right) = 1$ gives \(\lambda =  \frac{1}{d(n)}\).

Then,
\begin{align}
& \cT(\rho_t)  = \int_{\cX \in \mathrm{U}(S-1)} d\mu(X)\, \cX \left( \sum_n \chi_n \otimes \ket{n,\vec{0}} \bra{n,\vec{0}} \right) \cX^\dagger   \nonumber \\
& = \sum_n \chi_n  \otimes \int_{\cX \in \mathrm{U}_F(S-1)} d\mu(X)\, \cX \ket{n,\vec{0}} \bra{n,\vec{0}}  \cX^\dagger   \nonumber \\
& =  \sum_n \chi_n \otimes  \frac{1}{d(n)} \Pi_n^{S-1} = \cT(\rho).
\end{align}
\end{proof}
We refer to $\rho_t$ as the ``two-mode twirled state" corresponding to $\rho$. Cor.~\ref{cor rho_t} implies that we may, without loss of generality, assume $S = 2$ in our analysis. Then, $\rho_t$ has the form $\rho_t = \sum_n \chi_n \otimes \ket{n}\bra{n}$. When referring to $\rho_t$ in the next section, we mean this explicit form. The trace $\tr(\chi_n)$ is the probability that $\cH_{a_{\bar 1}}$ contains $n$ photons, so $\rho_t$ assigns a state, $\chi_n/\tr(\chi_n)$, to $\cH_{a_1}$ conditional on each number of photons in $\cH_{a_{\bar 1}}$.

\section{Measurement Statistics Determine the Twirled State}\label{sec results}

As argued at the end of the last section, it is sufficient to analyze the case $S=2$ as each equivalence class of states has a representative $\rho_t$ that is in vacuum in all modes except $\hat{a}_1$, $\hat{a}_2$. Thus, to simplify our analysis, we assume $S=2$ throughout this section.

\subsection{Informationally Complete Set of Operators}\label{subsec infocomp}

In this section we explore the conditions under which the measurement
configuration can determine $\rho_t$. More specifically, we consider
the measurement statistics obtained with one or two counters for
different families of probe states. We first identify a minimal
informationally complete set of products of mode operators on $\cH_a$,
whose expectations fully determine $\rho_t$ for \(\rho_{t}\in\cS\). In what follows,
we say that ``$\rho_t$ cannot be determined in general" by the
expectations of a given set of operators, if there exists \(\rho'_{t}\) such that
  \(\rho'_{t}\ne \rho_{t}\), but the expectation of every operator in
  the set is the same on the two states. We need the following
lemma.
\begin{lemma}\label{lem: norm-antinorm}
The anti-normally ordered monomial $ \hat{a}_1^l (\hat a_1^\dagger)^k  \hat{a}_2^n ( \hat{a}_2^\dagger)^n$, where $k \geq l $, can be expressed as a finite linear combination of normally ordered monomials of the form $(\hat a_1^\dagger)^{k'} \hat{a}_1^{l'} ( \hat{a}_2^\dagger)^{n'} \hat{a}_2^{n'}$ with $0\leq k' \leq k$, $0 \leq l' \leq l$, $0 \leq n' \leq n$ and $k'-l' = k-l$. Conversely, any normally ordered monomial $(\hat a_1^\dagger)^{k} \hat{a}_1^{l} ( \hat{a}_2^\dagger)^{n} \hat{a}_2^{n}$ with $k \geq l$ can be expressed as a finite linear combination of anti-normally ordered monomials $ \hat{a}_1^{l'} (\hat a_1^\dagger)^{k'} \hat{a}_2^{n' }  ( \hat{a}_2^\dagger)^{n'}$ with $0\leq k' \leq k$, $0 \leq l' \leq l$, $0 \leq n' \leq n$ and $k'-l' = k-l$. 
\end{lemma}
\begin{proof}
    To establish the first relation, we first consider a general single-mode anti-normally ordered monomial $\hat{a}^l (\hat{a}^\dagger)^k$. Each application of the commutation rule $\hat{a} \hat{a}^\dagger = \hat{a}^\dagger \hat{a} + 1$ to a given monomial produces two terms: one in which the exponents of $\hat{a}$ and $\hat{a}^\dagger$ are each decreased by one, and another with the same exponents but with one fewer instance of a creation operator appearing to the right of an annihilation operator. Hence, iterating this substitution converts any anti-normally ordered monomial to a finite sum of normally ordered monomials of the form $(\hat{a}^\dagger)^{k'} \hat{a}^{l'}$, where in this sum, the exponents are bounded by the original ones, and the difference between creation and annihilation exponents is preserved ($k' - l' = k - l$). Applying this single-mode result independently to each mode in the product $\hat{a}_1^l (\hat{a}_1^\dagger)^k \hat{a}_2^n (\hat{a}_2^\dagger)^n$ yields the first relation. Following the same procedure, the second relation is obtained by repeatedly applying the commutation rule $\hat{a}^\dagger \hat{a} = \hat{a} \hat{a}^\dagger - 1$ to the normally ordered monomial $(\hat{a}_1^\dagger)^k \hat{a}_1^l (\hat{a}_2^\dagger)^n \hat{a}_2^n$.
\end{proof}

\begin{prop}\label{prop: rho_t_mom}
  Consider an arbitrary two-mode twirled state $\rho_t$ with
  \(\rho_{t}\in\cS\).  Then, \(\rho_{t}\) is determined by the family
  of expectations \(r(k,l,n)\) of the normally-ordered operators
  $(\hat a_1^\dagger)^k \hat{a}_1^l ( \hat{a}_2^\dagger)^n
  \hat{a}_2^n$ over non-negative integers \(k,l,n\) satisfying
  \(k\geq l\) with \(k,l,n\) not all zero.  No proper subset of this
    family of expectations suffices for determining \(\rho_{t}\).
\end{prop}

\begin{proof}
  To prove the first claim, assume the expectations $r(k,l,n)$
  are known for all $k \geq l$ and all $n$. It follows from
  Lem.~\ref{lem: norm-antinorm} that for all $k \geq l$ and all
    $n$, the expectations
    \(\langle \hat{a}_1^l (\hat a_1^\dagger)^k \hat{a}_2^n
    (\hat{a}_2^\dagger)^n \rangle\) of the anti-normally ordered
    operators are also known. The expectations for \(k<l\) can be
  determined from the expectations for \(k\geq l\).  Specifically,
  using Eq.~\ref{eq: Husimi_identity} with $S=2$, for each $k<l$,
  \begin{align}
    & r(l,k,n) = \int d\vec{\alpha}  d\vec{\alpha}^*  Q(\vec{\alpha}, \vec{\alpha}^*)  \alpha_1^{l} (\alpha_1^*)^{k} |\alpha_2|^{2n}  \nonumber \\
    &= \left( \int d\vec{\alpha}  d\vec{\alpha}^*  Q(\vec{\alpha}, \vec{\alpha}^*)  (\alpha_1^*)^{l} \alpha_1^{k} |\alpha_2|^{2n}  \right)^* \nonumber \\
    & = \langle \hat{a}_1^k (\hat{a}_1^\dagger)^l \hat{a}_2^n (\hat{a}_2^\dagger)^n \rangle^* = r(k,l,n)^{*}.
  \end{align}

  The collection of moments of the Husimi function of $\rho_t$
  consists of the expectations of antinormally ordered operators
  $ \langle \hat{a}_1^l (\hat a_1^\dagger)^k \hat{a}_2^n (
  \hat{a}_2^\dagger)^m \rangle$. However, due to the structure of
  $\rho_t$, these vanish when $n\neq m$. To see this, observe that
  \begin{align}
    \langle \hat{a}_1^l (\hat a_1^\dagger)^k  \hat{a}_2^n ( \hat{a}_2^\dagger)^m \rangle = \sum_{i} \tr (\chi_i  \hat{a}_1^l (\hat a_1^\dagger)^k ) \bra{i} \hat{a}_2^n ( \hat{a}_2^\dagger)^m \ket{i}.
  \end{align}
  But for each $\ket{i}$,
  $\bra{i} \hat{a}_2^n ( \hat{a}_2^\dagger)^m \ket{i} \propto
  \bra{i+n}\ket{i+m} = 0$, unless $n = m$. Therefore, knowing all
  expectations indicated in the lemma statement implies that the
  Husimi function is fully determined, and hence so is $\rho_t$.

  To prove the second claim, consider the set \(\cS_{N}\) of
    twirled states with at most $N$ photons. By the results of App.~\ref{app: Husimi_det},
    \(\cS_{N}\subseteq\cS\).
    \(\cS_{N}\) consists of
  the states of the form
  $\rho_t = \sum_{i=0}^N \chi_i \otimes \ket{i}\bra{i}$, where
  $\chi_i$ has at most $N-i$ photons. Then, for all $k\geq l, n$ we
  have
  \begin{align}
   r(k,l,n)  = \sum_{i = 0}^N \tr\left( \chi_i  (\hat a_1^\dagger)^{k} \hat{a}_1^{l}   \right) \bra{i} ( \hat{a}_2^\dagger)^{n} \hat{a}_2^{n} \ket{i}.
    \label{eq:nkl_exp}
  \end{align}
  Observe that
  $\bra{i} ( \hat{a}_2^\dagger)^{n} \hat{a}_2^{n} \ket{i} = 0$ for all
  $i <n$, and
  $\tr\left( \chi_i (\hat a_1^\dagger)^{k} \hat{a}_1^{l} \right) = 0$
  whenever $k>N-i$. Thus, the expectations \(r(k,l,n)\) for $n + k >N$
  are zero for all states in \(\cS_{N}\). We have \(r(0,0,0)=1\)
  since \(\rho_{t}\) has trace \(1\).  For \(k=l\), \(r(k,l,n)=r(k,k,n)\)
    is the expectation of a Hermitian operator and therefore real. For
    \(l<k\), \(r(k,l,n)\) may be complex. We count the number of real
    values not yet constrained to be \(0\) or \(1\) provided by the
    expectations \(r(k,l,n)\).  These are provided by the family
    of expectations \(r(k,l,n)\) with \(0\leq l\leq k\leq N-n\)
    excluding \(l=k=n=0\).  First fix the value of \(n\).  There are
    \(N-n+1\) real values contributed by \(r(k,k,n)\) with
    \(n+k\leq N\) and \((N-n)(N-n+1)/2\) complex values for \(r(k,l,n)\)
    with \(0\leq l<k\leq N-n\).  This gives a total of
    \((N-n+1) + (N-n)(N-n+1) = (N-n+1)^2\) real values at fixed
    \(n\). Summing over \(n\) and excluding \(l=k=n=0\) gives
    \begin{align}
      \kappa_{r}(N) = -1+\sum_{n=0}^{N}(N-n+1)^{2}.
    \end{align}

    We compare \(\kappa_{r}(N)\) to the affine dimension of
    \(\cS_{N}\), which is the number of independent real values
    required to
    specify a generic $\rho_t$ in \(\cS_{N}\). The operator $\chi_i$
    of \(\rho_{t}\) is an arbitrary positive semidefinite Hermitian
    operator on a Hilbert space of dimension $N+1-i$, and is therefore
    determined by $(N+1-i)^2$ independent real
    parameters. Consequently, the affine dimension of $\cS_{N}$ is
    \begin{align}\label{eq: ind_par_num}
      \kappa(N)  &= -1+ \sum_{i=0}^N (N-i+1  )^2,
    \end{align}
    where the summand $-1$ accounts for the unit trace condition.
    By comparing the expressions, we find that \(\kappa_{r}(N)=\kappa(N)\).

    Consider now a strict subfamily of the family of expectations
    \(r(k,l,n)\) with \(k\geq l\). Suppose this subfamily does not
    contain \(r(k_{0},l_{0},n_{0})\) with \(k_{0}\geq l_{0}\) and
    \(k_{0},l_{0},n_{0}\) not all zero. Let \(N=n_{0}+k_{0}\) and
    consider states in \(\cS_{N}\).  Because
    \(0\leq l_{0}\leq k_{0}\leq N-n_{0}\), the number of real
    parameters provided by this subfamily from expectations
    \(r(k,l,n)\) with \(0\leq l\leq k\leq N-n\) with \(k,l,n\) not all
    zero is strictly less than the affine dimension of
    \(\cS_{N}\). Since expectations \(r(k,l,n)\) for other triples
    \(k,l,n\) are constant, either \(0\) or \(1\), on \(\cS_{N}\),
    there are two distinct states in \(\cS_{N}\) which have the same
    values for all expectations in the subfamily, which establishes
    the second claim. 
\end{proof}

We also consider input states with bounded photon number. If $\rho$
contains at most $N$ photons, then $\rho_t$ does as well, since the
map $\cT(\cdot)$, being mixture of PLT operations, cannot increase the
photon number. In particular, by the results of App.~\ref{app:
  Husimi_det}, \(\rho_{t}\in\cS\). The proof of Prop.~\ref{prop:
  rho_t_mom} implies the following corollary.

\begin{corollary}\label{cor: rho_t_mom}
As in the proof of Prop.~\ref{prop: rho_t_mom}, let \(\cS_{N}\)
  be the set of two-mode twirled states $\rho_t$ with at most $N$
  photons. Consider the space \(\cR\) of
complex vectors \(\bm{r} = (\bm{r}_{\bm{j}})_{\bm{j}\in\cJ}\), where
  \(\cJ\) is the family of triples \(k,l,n\) of non-negative
  integers where $k \geq l$ and $k+n \leq N$, with \(k,l,n\) not all
  zero. Then the linear map
  \(\bm{r}:\rho_{t}\in\cS_{N}\mapsto \bm{r}(\rho_{t}) \in \cR\)
  with
  \(\bm{r}_{k,l,n}(\rho_{t})=r(k,l,n)=\langle (\hat a_1^\dagger)^k
  \hat{a}_1^l ( \hat{a}_2^\dagger)^n \hat{a}_2^n \rangle_{\rho_{t}} \)
  is one-to-one and its range is a convex set with non-empty interior.
\end{corollary}

\begin{proof}
  The result follows from the first part of Prop.~\ref{prop:
    rho_t_mom} and the argument given to establish the second part of
  that proposition.  That the map is one-to-one follows from the
  observation that the values of \(r(k,l,n)\) for \(k,l,n\) not in
  \(\cJ\) are constant on \(\cS_{N}\) or determined by
    \(\bm{r}(\rho_{t})\). That the range of the map is convex follows
  from the convexity of its domain, which the range inherits because
  the map is linear. That the interior of the range is non-empty
  follows because the real dimension of \(\cR\) is the same
    as the affine dimension of \(\cS_{N}\).
\end{proof}

Before proceeding to the next section, we present the generating functions for both the two-counter and single-counter configurations adapted to the $S=2$ case. In particular, when two counters are used, Eq.~\ref{eq: F(u,v,gamma)_gen} becomes
\begin{align}\label{eq: F(u,v,gamma)}
F_\gamma(u,v) & \equiv e^{  \abs{\eta \alpha_1 + \zeta \gamma}^2   u + \abs{\zeta' \alpha_1 + \eta' \gamma}^2   v +  \abs{\eta \alpha_2}^2 u +\abs{\zeta' \alpha_2}^2 v }.
\end{align}
If the measurement configuration involves a single counter, we denote the corresponding generating function by $F_{1,\gamma}(u)$:
\begin{align}
F_{1,\gamma}(u) &=  e^{  \abs{\eta \alpha_1 + \zeta \gamma}^2   u  +  \abs{\eta \alpha_2}^2 u } \nonumber \\
& =  e^{  ( \abs{ \eta \vec{\alpha}}^2 +  \eta\zeta^* \gamma^*  \alpha_1 + \zeta\eta^*\gamma  \alpha_1^* +  \abs{\zeta \gamma}^2)u}.
\end{align}
We consider this measurement configuration first.

\subsection{Results With One Counter}\label{subsec res_one}

In this section we present our results for the measurement configuration with one counter. We first introduce several definitions and notational conventions that will be used throughout the statements and proofs of the subsequent results. For all non-negative integers \(k, l, n\), we define the following amplitude polynomials:
\begin{align}
& m(k,l,n) \equiv \cE[\nord{ (\hat a_1^\dagger)^k \hat{a}_1^l  \hatna^n }  ] = (\alpha_1^*)^{k} \alpha_1^{l}  |\vec{\alpha}|^{2n} \nonumber \\
& m'(k,l,n) \equiv  \cE[ (\hat a_1^\dagger)^k \hat{a}_1^l ( \hat{a}_2^\dagger)^n \hat{a}_2^n ]= (\alpha_1^*)^{k} \alpha_1^{l}  |\alpha_2|^{2n}. \nonumber
\end{align}
For each non-negative integer \(N\), we define the sets of these
expressions restricted by their degrees as
\(\cM_N = \{m(k,l,n) : n + \max(k,l) \leq N\}\) and
\(\cM'_N = \{m'(k,l,n) : n + \max(k,l) \leq N\}\). For
  convenience, we define \(\cM_{-1}=\cM'_{-1}=\varnothing\). We have
  \(\cM_{N}\subseteq\cM_{N'}\) and \(\cM'_{N} \subseteq\cM'_{N'}\) for
  \(N\le N'\). 

We adopt the notation of expectation brackets
  $\langle \cdot \rangle$ around an amplitude polynomial $P$ to denote
  the expectation value of the corresponding operator $\cE^{-1}[P]$
  with respect to the twirled state \(\rho_{t}\) under
  consideration. We refer to this expectation simply as the
  expectation of \(P\).  For example,
  \(\langle m'(k,l,n) \rangle \equiv \tr(\rho_t (\hat a_1^\dagger)^k
  \hat{a}_1^l ( \hat{a}_2^\dagger)^n \hat{a}_2^n)\).  In the proof of
  Prop.~\ref{prop: rho_t_mom} we analyzed the set of expectations
  \(r(k,l,n)\) with \(0\leq l\leq k\leq N-n\). Because
  \(r(k,l,n)=r(l,k,n)^{*}\), the set of expectations \(r(k,l,n)\) with
  \(0\leq \max(k,l)\leq N-n\) may be identified with the set of
  expectations of members of \(\cM'_{N}\).  Furthermore, for
  \(\rho_{t}\in\cS_{N}\) the expectations for the \(m'(k,l,n)\) not in
  \(\cM'_{N}\) are zero and \(\rho_{t}\) is determined by the
  expectations of the members of \(\cM'_{N}\).  

\begin{lemma}\label{lem: M_N_equiv_M'_N}
Every element of $\cM_N\setminus\cM_{N-1}$ can be expressed as a linear combination of elements in $\cM'_N\setminus\cM'_{N-1}$, and conversely, every element of $\cM'_N\setminus\cM'_{N-1}$ can be expressed as a linear combination of elements in $\cM_N\setminus\cM_{N-1}$.
\end{lemma}
\begin{proof}
\(\cM_{N}\setminus\cM_{N-1}\) consists of the \(m(k,l,n)\) with
  \(\max(k,l)=N-n\) and similarly for
  \(\cM'_{N}\setminus\cM'_{N-1}\).  The result follows directly from
  expanding the magnitude terms $|\vec{\alpha}|^{2n}$ via the
  binomial theorem, utilizing the relation
  $|\vec{\alpha}|^2 = |\alpha_1|^2 + |\alpha_2|^2$. We
  can express the elements of $\cM'_N$ in terms of $\cM_N$ as
  \begin{align}\label{eq:m'_from_m}
    m'(k,l,n) & =     ( \alpha_1^*)^k  \alpha_1^l (|\vec{\alpha}|^{2}-|\alpha_1|^{2})^{n}\nonumber\\      
              & =  \sum_{j=0}^{n}\binom{n}{j} (-1)^j ( \alpha_1^*)^k  \alpha_1^l |\alpha_1|^{2j} |\vec{\alpha}|^{2(n-j)} \nonumber\\
              &= \sum_{j=0}^{n}\binom{n}{j}  (-1)^j m(k+j,l+j,n-j).
  \end{align}
  For the converse, we can express the elements of $\cM_N$ in terms of
  $\cM'_N$ as
  \begin{align}
    m(k,l,n) &=     ( \alpha_1^*)^k  \alpha_1^l (|\alpha_1|^{2}+|\alpha_2|^{2})^{n}\nonumber\\      
             &=
               \sum_{j=0}^{n}\binom{n}{j} 
               ( \alpha_1^*)^k  \alpha_1^l |\alpha_1|^{2j} |\alpha_2|^{2(n-j)} \nonumber\\
             &= \sum_{j=0}^{n}\binom{n}{j}   m'(k+j,l+j,n-j).
               \label{eq:m_from_m'}
  \end{align}
  The lemma follows because \(\max(k,l)+n=\max(k+j,l+j)+n-j\).
\end{proof}

We previously showed that the expectations of
\(m'(k,l,n)\notin\cM'_{N}\) are zero when \(\rho_{t}\in\cS_{N}\).
Observe that as a consequence of the above lemma, the expectations of
the \(m(k,l,n)\notin\cM_{N}\) are also zero when
\(\rho_{t}\in\cS_{N}\).  Cor.~\ref{cor: rho_t_mom} can be phrased in
terms of the expectations of the \(m'(k,l,n)\). Since the above lemma
implies that there is an invertible linear relationship between
\(\cM_{N}\) and \(\cM'_{N}\), this observation enables replacing
\(m'(k,l,n)\) in the proof of this corollary with \(m(k,l,n)\), which
yields the following result.

  \begin{corollary}
    \label{cor: rho_t_momn}
  The conclusion of Cor.~\ref{cor: rho_t_mom} holds if the linear map \(\bm{r}(\rho_{t})\) is replaced by the map \(\bm{s}:\rho_{t}\mapsto (\langle m(k,l,n)\rangle_{\rho_{t}})_{k,l,n:l\leq k\leq N-k, (k,l,n)\ne (0,0,0)}\).
\end{corollary}

 In the following proofs, we make use of the fact that the product of two generating functions yields another generating function, with coefficients given by finite linear combinations of the products of the original coefficients. In particular, we find it useful to work with the generating function 
\begin{align}
F'_{1,\gamma}(u) &= e^{ -  \abs{\zeta \gamma}^2 u} F_{1,\gamma}(u) \nonumber \\
& = e^{  ( \abs{ \eta \vec{\alpha}}^2 +  \eta\zeta^* \gamma^*  \alpha_1 + \zeta\eta^*\gamma  \alpha_1^* )u}.\label{eq:F'1,gamma}
\end{align}
The coefficient of $u^n$ in $F'_{1,\gamma}(u)$ is a discrete
convolution of the first $n+1$ coefficients of $F_{1,\gamma}(u)$ and
$e^{- \abs{\zeta \gamma}^2 u}$, as can be seen by writing
  \(F_{1,\gamma}(u)\) and \(e^{- \abs{\zeta \gamma}^2 u}\) in terms of
  their explicit formal power series, formally multiplying, and adding
  terms of equal degree in \(u\). The coefficients of $F_{1,\gamma}(u)$ encode the operators'
polynomial correspondence. Because
the coefficients of $e^{- \abs{\zeta \gamma}^2 u}$ are simply scalar
constants proportional to $(-\abs{\zeta \gamma}^2)^k$, it follows that
the operators corresponding to the coefficients of $F'_{1,\gamma}(u)$
are strictly finite linear combinations of the operators in
$\cC_1(\gamma)$. Conversely, because the exponential factor
$e^{ - \abs{\zeta \gamma}^2 u} $ is invertible, the coefficients of
$F_{1,\gamma}(u)$ can similarly be expressed as linear combinations of
the coefficients of $F'_{1,\gamma}(u)$.

This relationship between the coefficients of $F_{1,\gamma}(u)$ and $F'_{1,\gamma}(u)$ motivates a formal definition: we consider two generating functions to be equivalent if their respective sets of coefficients algebraically span the same linear space. 
\begin{lemma}\label{lem: F_1g_equiv}
The generating functions $F_{1,\gamma}(u)$ and $F'_{1,\gamma}(u)$ are equivalent.
\end{lemma}
\begin{proof}
This follows from the preceding discussion.\end{proof}

We begin by assuming the measurement statistics are known for a finite number of probe amplitudes. For each non-negative integer $N$, let $\cP_{N}$ denote the space of polynomials spanned by all monomials of the form $\gamma^k (\gamma^*)^l$, where $k,l \leq N$.

\begin{theorem}\label{thm:1CntrNbnd}
  Consider a measurement configuration with a single counter. Suppose the input state, and hence the two-mode twirled state \( \rho_t \), contains at most \( N \) photons. Let \( \Gamma = \{\gamma_1,\hdots, \gamma_K\} \subset \mathbb{C} \) be a finite set satisfying the following condition: if a polynomial \( p \in \mathcal{P}_N \)  satisfies \( p(\gamma_i) = 0 \) for all \( \gamma_i \in \Gamma \), then \( p =0\). Then, $\rho_t$ is determined by the collection of the measurement statistics for each probe amplitude $\gamma \in \Gamma $.
\end{theorem}
\begin{proof}
We show that the expectations of all operators $(\hat a_1^\dagger)^k \hat{a}_1^l (\hat a_2^\dagger)^n \hat{a}_2^n$ with $n+\max(k,l) \le N$ are determined by the measurement statistics. The theorem statement then follows from Cor.~\ref{cor: rho_t_mom}. Note that the set of $ \cE[(\hat a_1^\dagger)^k \hat{a}_1^l (\hat a_2^\dagger)^n \hat{a}_2^n]$ with $n+\max(k,l) \le N$ is exactly $\cM'_N$. Lem.~\ref{lem: M_N_equiv_M'_N} shows that $\cM'_N$ is in the span
of $\cM_N$, so it suffices to show that the expectations of the operators in $\cM_N $ can be extracted from the generating functions $F'_{1,\gamma}(u)$ for the $\gamma \in \Gamma$, which we proceed to do. By ``extracted" we mean that they can be obtained from the expectations of the coefficients of the generating functions by taking their linear combinations.

The assumption on $\Gamma$ in the theorem statement implies that the
map defined by \(p\in\cP_{N}\mapsto (p(\gamma_i))_{i=1}^K\) is linear
and injective. We can express this map in the basis of monomials
\((\gamma^*)^{l}\gamma^{k}\in\cP_{N}\) as a matrix with independent
columns, which implies that the row span of the matrix is complete. In
particular, for every basis vector, there is a linear combination of
the rows identical to this basis vector. The matrix entry
corresponding to monomial \(p\) and row \(i\) is \(p(\gamma_{i})\). It follows that for any given monomial $p\in\cP_{N}$,
there is a tuple of coefficients $(c_{i})_{i=1}^K$ such that for all
monomials $q\in\cP_{N}$ different from $p$,
\begin{align}
  \sum_{i=1}^K c_{i} q(\gamma_i)&=0,\nonumber\\
  \sum_{i=1}^K c_{i}p(\gamma_i)&= 1.
\end{align}
We define $S_{p}(u)=\sum_{i=1}^K c_{i}F'_{1,\gamma_i}(u) $. As a
generating function, the coefficients of $S_{p}(u)$ are linear
combinations of the coefficients of the $F'_{1,\gamma_i}(u) $. The
coefficient of $u^{n}$ in $S_{p}(u)$ is given by
\begin{align}\label{eq: S_p_coeff_n}
   &  (n!) \mathrm{coeff}_{u^{n}}S_{p} =  \sum_{i=1}^K c_{i} ( \abs{ \eta \vec{\alpha}}^2 +  \eta\zeta^* \gamma_i^*  \alpha_1 + \zeta\eta^*\gamma_i  \alpha_1^* )^n \nonumber \\
          & =  \sum_{i=1}^K c_{i} \sum_{r+s+t = n} \binom{n}{r,s,t}  (\eta\zeta^*\alpha_1)^{r}(\gamma_i^*)^r (\zeta\eta^*\alpha_1^*)^{s} \gamma_i^s |\eta \vec{\alpha}|^{2t}   \nonumber \\
          & = \sum_{r+s \leq n}\binom{n}{r,s,n-r-s}   (\zeta\eta^*)^{s}(\eta\zeta^*)^r |\eta|^{2(n-r-s)}   \nonumber\\
          & \hphantom{==} \times \left( \sum_{i=1}^K c_{i} (\gamma_i^*)^r \gamma_i^s  \right)  m(s,r,n-r-s).
\end{align}
By construction, when $r, s \leq N$, the sum $\sum_{i=1}^K c_i (\gamma_i^*)^r \gamma_i^s = 0$ unless $r = l$ and $s = k$. Consequently, if $n < k+l$, the entire expression vanishes. For $n \geq k+l$, we consider the expectation of the coefficient. Because $\rho_t$ contains at most $N$ photons, the expectations $\langle m(s,r,t) \rangle$ strictly vanish whenever the relevant degree exceeds this bound, namely when $r+t > N$ or $s+t > N$. This truncation ensures that the value of $\sum_{i=1}^K c_i (\gamma_i^*)^r \gamma_i^s$ is irrelevant whenever $r > N$ or $s > N$, as it multiplies a zero expectation. Taking the expectation on both sides of Eq.~\ref{eq: S_p_coeff_n} thus yields
\begin{flalign*}
 &n! \langle \mathrm{coeff}_{u^{n}}S_{p}  \rangle  = \binom{n}{l,k,n-k-l} \nonumber \\     
 &\times (\zeta\eta^*)^{k}(\eta\zeta^*)^l |\eta|^{2(n-k-l)} \langle m(k,l,n-k-l) \rangle 
\end{flalign*} 
when $n \geq k+l$ such that $n-l \leq N$ and $n-k \leq N$, and
$ \langle \mathrm{coeff}_{u^{n}}S_{p} \rangle =0$ otherwise, so that
\(\langle m(k,l,n-k-l)\rangle\) can be determined from the expectation
of \(\mathrm{coeff}_{u^{n}}S_{p}\) of \(S_{p}(u)\).  Performing this
procedure for all monomials $\gamma^l (\gamma^*)^k \in \cP_N$ extracts
the expectation of every member of $\cM_N$, which completes the proof.
  \end{proof}

Thm.~\ref{thm:1CntrNbnd} indicates that, for $\rho_t$ with at most $N$ photons, a minimal number $h_1(N)$ of probe amplitudes exists that are necessary and sufficient to determine $\rho_t$.
\begin{corollary}\label{cor: suf_set_gamma}
Consider a measurement configuration with a single counter. Suppose the input state, and hence the two-mode twirled state \( \rho_t \), contains at most \( N \) photons. 
Suppose the conditions of Thm.~\ref{thm:1CntrNbnd} are satisfied. Then, there exist \( (N+1)^2 \) distinct probe amplitudes that determine \( \rho_t \). Moreover, almost every set of \( (N+1)^2 \) distinct probe amplitudes possesses this property.
\end{corollary}
\begin{proof}
The claim implies that one can take $K = (N+1)^2$ in Thm.~\ref{thm:1CntrNbnd}. Namely, there exist $(N+1)^2$ distinct $\gamma_i \in \mathbb{C}$ such that the only polynomial $p \in \cP_N$ that satisfies $p(\gamma_i) = 0$ for all $i =1,\hdots, (N+1)^2$ is the zero polynomial. This is a standard result in Lagrange interpolation theory on a real-analytic polynomial space over $\mathbb{C}$, where $\gamma$ and $\gamma^*$ are treated as independent variables \cite{lorentz1992multi}. The last claim follows from the fact that the Lagrange interpolation map $p \to  (p(\gamma_i))_{i=1}^K$ fails to be injective only for \((\gamma_{i})_{i=1}^{K}\) in a Lebesgue-null subset of $\mathbb{C}^{K}$ (see Cor.~5.6.3 in Ref.~\cite{lorentz1992multi}).  
\end{proof}

Cor.~\ref{cor: suf_set_gamma} implies that $h_1(N) \leq (N+1)^2$ since Thm.~\ref{thm:1CntrNbnd} shows only that the stated number of probe amplitudes are sufficient for determining $\rho_t$.  

Next, we consider the case where measurement statistics are known for an infinite number of probe amplitudes and we do not assume an upper bound on the number of photons. To that end, for each non-negative integer $N$, we introduce the set $\cM_{=N} = \{m(k,l,n) : k+l+n = N \}$. We further denote by $\cP_{N,>}$ the polynomial space spanned by all monomials of the form $(\gamma^*)^l \gamma^k$ satisfying $k+l \leq N$. 

\begin{theorem}\label{thm: 1cntr_uncount}
  Consider a measurement configuration with a single counter. Consider
  an arbitrary two-mode twirled state $\rho_t$ with
  \(\rho_{t}\in\cS\). If the measurement statistics are known for all
  probe amplitudes in a set \(\Gamma\) of positive Lebesgue measure, then $\rho_t$ is
  uniquely determined.
\end{theorem}

\begin{proof}
  We show that the
  expectations of the operators
  $(\hat a_1^\dagger)^k \hat{a}_1^l (\hat a_2^\dagger)^n \hat{a}_2^n$
  for all non-negative integers $k,l,n$ are determined by the
  measurement statistics. The theorem statement then follows from
  Prop.~\ref{prop: rho_t_mom}.

  Note that $(n!) \mathrm{coeff}_{u^{n}}S_{p}(u)$ in Eq.~\eqref{eq:
    S_p_coeff_n} depends only on the terms $m(r,s,t)$ with
  $r+s+t = n$, which exactly constitute the set $\cM_{=n}$. Since
  $\bigcup_N \cM_N = \bigcup_N \cM_{=N}$, it suffices to show that
  for every $N$, $\cM_{=N}$ can be extracted from the coefficients of
  the generating functions $F'_{1,\gamma}(u)$ for probe amplitudes
  $\gamma$ in the given neighborhood.

  As stated in the proof of Cor.~\ref{cor: suf_set_gamma}, the
  polynomial evaluation map
  $p \in \cP_{N,>} \mapsto (p(\gamma_i))_{i=1}^{K}$, where $K$ is a
  function of $N$ satisfying the conditions of
  Thm.~\ref{thm:1CntrNbnd}, fails to be injective only for
    \((\gamma_{i})_{i=1}^{K}\) in a measure-zero subset of
    $\mathbb{C}^{K}$. Since \(\gamma\) has positive measure, so does
    \(\Gamma^{K}\subseteq\mathbb{C}^{K}\) for all \(K\).
  Consequently, for any integer $N$, we can
  always choose a tuple of \(K(N)\) probe amplitudes
  $(\gamma_i)_{i=1}^{K(N)}\in\Gamma^{K}$
  such that the map
  $p \in \cP_{N,>} \mapsto (p(\gamma_i))_{i=1}^{K(N)}$ is injective.
  The number of required amplitudes, $K(N)$, can be chosen to match
  the dimension of $\cP_{N,>}$, which is exactly the number of pairs
  $(k,l)$ satisfying $k+l \leq N$, namely
  $K(N) = \frac{1}{2}(N+1)(N+2)$.

  With $N$ fixed, for each target monomial $p = (\gamma^*)^l \gamma^k$
  satisfying $k+l \leq N$, we construct the linear combination
  $S_p(u) = \sum_{i=1}^{K(N)} c_i F'_{1,\gamma_i}(u)$ as in the proof
  of Thm.~\ref{thm:1CntrNbnd}, choosing the coefficients $c_i$ such
  that $\sum c_i q(\gamma_i) = \delta_{qp}$ for
  \(q\in\cP_{N,>}\). Then, the $N$-th order term in the expansion of
  $S_p(u)$ in Eq.~\eqref{eq: S_p_coeff_n}, is proportional to
  $m(k,l,N-k-l)$. Because the expansion is strictly restricted to
  terms \(m(r,s,t))\)where $r+s+t = N$, no terms of degree higher than
  $N$ appear in the summation, rendering the physical photon-number
  truncation step from Thm.~\ref{thm:1CntrNbnd}
  unnecessary. Performing this procedure for each value of $N$
  extracts all monomials in $\bigcup_N \cM_{=N}$, which completes the
  proof.
\end{proof}

In the remainder of this section we establish two no-go results for
general states in \(\cS\). The first establishes that
$\lim_{N \to \infty}h_1(N) $ diverges. Namely, no finite collection of
probe amplitudes are sufficient to obtain $\rho_t$ in the absence of a
bound on total photon number. The second shows that if only one
counter is used, measuring at just one probe magnitude is
insufficient. The proofs of the results first establish that there are
states with bounded photon numbers that cannot be distinguished based
on knowledge of the photon-number moment sequences associated with the
observed photon number distributions at the set of probe amplitudes
under consideration. States with bounded photon numbers have
determinate photon-number moment sequences (see
Cor.~\ref{cor:pnr_moment_determinate} in Appendix \ref{app:
  Husimi_det}), which implies that knowledge of the photon-number
moment sequence is equivalent to knowledge of the photon-number
probability distribution.  The no-go results therefore imply that
there are states that cannot be distinguished based on the measurement
statistics, that is, the observed photon number distributions at the
given set of probe amplitudes.  A subtle point is that with this proof
strategy, the two states that cannot be distinguished based on the
photon-number moment sequence must also be indistinguishable by their
photon-number distributions to complete the no-go result. To ensure
the latter indistinguishability, we need the photon-number moment
sequence to be determinate in the sense that there is a unique
photon-number probability distribution with this moment sequence. We
do not know whether all states in \(\cS\) have determinate
photon-number moment sequences.

\begin{prop}\label{prop: inf_N_fin_gamma}
  Consider a measurement configuration with a single counter. Consider
  an arbitrary two-mode twirled state $\rho_t$ with
  \(\rho_{t}\in\cS\). Suppose the measurement statistics are known for
  only a finite set of probe amplitudes $\Gamma$. Then, in
    general, the two-mode twirled state $\rho_t$ is not determined by
  the collection of measurement statistics for the
  $\gamma \in \Gamma$.
\end{prop}

\begin{proof}
  Recall that for each $\gamma \in \Gamma$, the coefficients of the
  generating function $F_{1,\gamma}(u)$ correspond to the operators in
  $\cC_1(\gamma)$. The set of expectations of these operators are
  strictly equivalent to the set of all normally ordered moments
  $\langle \nord{ (\hatnpa )^k } \rangle$, which in turn contain the
  exact same information as the set of all expectations
  $\langle (\hatnpa)^k \rangle$ of photon-number moments, as shown in
  Appendix~\ref{app nor_n^k}. Furthermore, because $F_{1,\gamma}(u)$
  and $F'_{1,\gamma}(u)$ are equivalent generating functions
  (Lem.~\ref{lem: F_1g_equiv}), the information accessible from the
  measurement statistics is completely characterized by the linear
  span of the coefficients of the $F'_{1,\gamma}(u)$ collected over
  the probe amplitudes \(\gamma\in\Gamma\). 

  The coefficient of \(u^{j}\) in \(F'_{1,\gamma}(u)\) is
  \( ( \abs{ \eta \vec{\alpha}}^2 + \eta \zeta^* \gamma^* \alpha_1 +
  \eta^* \zeta \gamma \alpha_1^* )^j/j!\), which expands as a linear
  combination of the monomials $m(k,l,n)$ with $k+l+n = j$.  Thus, the
  sets \(\cM_{=j} \;=\;\bigl\{\,m(k,l,n)\mid k + l + n = j\bigr\}\)
  partition the monomials \(m(k,l,n)\) according to the degree at
  which they occur in the coefficients of \(F'_{1,\gamma}(u)\).  Each
  probe amplitude yields the expectation of one complex linear
  combination of monomials in \(\cM_{=j}\) for each \(j\). For
  \(K=|\Gamma|\) probe amplitudes and a given \(N'\), this yields at
  most \(N'K\) such linear combinations of members of
  \(\cM_{\leq N'}=\bigcup_{1\leq j\leq N'} \cM_{=j}\).  We now
  restrict the twirled states to \(\cS_{N}\). Because the
    expectations for \(m(k,l,n)\) not in \(\cM_{N}\) are zero for
    states in \(\cS_{N}\), they do not contribute to the expectations
    of linear combinations of monomials. Since
\(\cM_{N}\subseteq\cM_{\leq 2N}\), the monomials outside \(\cM_{\leq 2N}\)
    do not contribute and it suffices to consider \(N'=2N\). The
    information available from the measurement statistics is 
    therefore contained in expectations of linear combinations of
    monomials in \(\cM_{\leq 2N}\), which gives at most \(2NK\)
    independent complex or \(4NK\) independent real values.  Since
    \(K\) is constant, this grows much more slowly with \(N\) than the
    \(\kappa(N)=\Omega(N^{3})\) independent such values required to
    identify the states of \(\cS_{N}\) according to Eq.~\eqref{eq:
      ind_par_num}.
\end{proof}

Our final result in this subsection addresses the case where measurement statistics are known only for probe states with a fixed magnitude $|\gamma|$. This scenario is experimentally relevant, since it is typically much easier to vary the phase of the LO and collect data than to change its magnitude.
\begin{prop}\label{prop: fix_R}
  Consider a measurement configuration with a single counter. Suppose
  that, for a fixed probe magnitude \(R = |\gamma|\), the measurement
  statistics are known for all phase values $\theta$ of
  \(\gamma = R e^{i\theta}\). Then, the two-mode twirled state
  $\rho_t$ is determined by this collection of measurement statistics
  if and only if the number of photons is bounded by $N \leq 1$.
\end{prop}
\begin{proof}
The proof follows from similar considerations as the proof of Prop.~\ref{prop: inf_N_fin_gamma}, allowing us to omit some of the intermediate logical steps. By hypothesis, the expectations of the coefficients of $F'_{1,\gamma}(u)$ are available for all $\theta \in [0, 2\pi)$. We first demonstrate that the individual monomials in $\cM_{=n}$ for $n \geq 2$ cannot be fully extracted from the $\mathrm{coeff}_{u^{n}} F'_{1,\gamma}(u)$. Plugging $\gamma = R e^{i\theta}$ into the $n$-th coefficient, we obtain
\begin{align} 
  &  (n!) \mathrm{coeff}_{u^{n}}F'_{1,\gamma} =  ( \abs{ \eta \vec{\alpha}}^2 + \eta \zeta^*  \alpha_1 R e^{- i\theta} + \eta^* \zeta \alpha_1^* R e^{i\theta} )^n \nonumber \\
  &= \sum_{r+s+t = n} \binom{n}{r,s,t} (\eta \zeta^* \alpha_1)^{r}  R^r e^{-i r \theta} (\eta^* \zeta \alpha_1^*)^{s} R^s e^{i s \theta} |\eta \vec{\alpha}|^{2t}   \nonumber \\
  & = \sum_{r+s+t = n}\binom{n}{r,s,t}  |\eta|^{2t}  (\eta \zeta^* \alpha_1)^{r} (\eta^* \zeta \alpha_1^*)^{s} |\vec{\alpha}|^{2t}  R^{s+r} e^{i \theta (s-r)}.
\end{align}

Let us define the scaled monomial
\[m_R(s,r,t) = \binom{n}{r,s,t} |\eta|^{2t} R^{s+r} (\eta \zeta^*)^r
  (\eta^* \zeta)^s (\alpha_1^*)^s \alpha_1^r |\vec{\alpha}|^{2t}, \]
which is proportional to $m(s,r,t)$ up to a fixed constant factor. The
expansion then simplifies to the phase-dependent sum
$\sum_{r+s+t = n} e^{i\theta (s-r)} m_R(s,r,t)$, which are the linear
combinations of the \(m_{R}(s,r,t)\) whose expectations can be
determined from the measurement statistics. These linear combinations
satisfy that for \(n\) fixed, the \(m_{R}(s+s',s,n-2s-s')\) always
occur with the same coefficient of \(e^{i\theta s'}\). For
\(N\geq 2\), consider the set \(\cM_{2}\) with respect to states in
\(\cS_{2}\).  The span of \(\cM_{2}\) contains \(m_{R}(0,0,2)\) and
\(m_{R}(1,1,0)\).  The expectations that can be determined from the
measurement statistics are therefore already determined by the
subspace of linear combinations of members of \(\cM_{2}\) with
identical coefficients on \(m_{R}(0,0,2)\) and \(m_{R}(1,1,0)\). Since
this is a codimension \(1\) subspace of the span of \(\cM_{2}\) and by
Cor.~\ref{cor: rho_t_mom}~and~\ref{cor: rho_t_momn}, this is
insufficient information to identify members of \(\cS_{2}\).

If $\rho_t$ has at most one photon, the expectations of monomials in $\cM_{=j}$ with $j\geq 2$ vanish, with the exception of $\langle \alpha_1^* \alpha_1 \rangle$.  Consequently, excluding the trivial trace condition $\langle F'_{1,\gamma}(0) \rangle = 1$, only the expectations $\langle \mathrm{coeff}_{u}F'_{1,\gamma}(u) \rangle$ and $\langle \mathrm{coeff}_{u^2}F'_{1,\gamma}(u) \rangle$ can be non-zero. Because the higher-order monomials in $\cM_{=2}$ vanish, $\langle \mathrm{coeff}_{u^2}F'_{1,\gamma}(u) \rangle$ is directly proportional to $\langle \alpha_1^* \alpha_1 \rangle$.  For the first-order coefficient, we have
 \begin{align}
 \mathrm{coeff}_{u}F'_{1,\gamma}  = \abs{ \eta \vec{\alpha}}^2 + \alpha_1 R e^{- i\theta} +  \alpha_1^* R e^{i\theta}. \nonumber
 \end{align} 
Evaluating this expectation at three distinct phases — for example, $\theta = 0, \pi/2, \pi$ — produces three linearly independent equations in the unknowns $\langle |\eta \vec{\alpha}|^2 \rangle$, $R\langle \alpha_1 \rangle$, and $R\langle \alpha_1^* \rangle$. Inverting the system we obtain the monomials in $\cM_{=1}$. Since $\abs{\vec{\alpha}}^2 = \alpha_1^* \alpha_1 + \alpha_2^* \alpha_2$, knowing $\langle |\eta \vec{\alpha}|^2 \rangle$ and $\langle \alpha_1^* \alpha_1 \rangle$ allows us to immediately deduce $\langle \alpha_2^* \alpha_2 \rangle$. By Cor.~\ref{cor: rho_t_mom}, these extracted expectations ($\langle \alpha_1 \rangle$, $\langle \alpha_1^* \rangle$, $\langle \alpha_1^* \alpha_1 \rangle$, and $\langle \alpha_2^* \alpha_2 \rangle$) are completely sufficient to uniquely determine $\rho_t$.
\end{proof}

\subsection{Results With Both Counters}\label{subsec res_two}

For a measurement configuration with both counters it is convenient to redefine the variables $u$ and $v$ of the generating function $F_{\gamma}(u,v)$ in Eq.~\ref{eq: F(u,v,gamma)} by making the substitutions  $ u \to u/|\eta|^2$ and $ v \to v/|\zeta'|^2$. Note that this variable substitution merely rescales the coefficients of the original generating function by scalar factors, and thus yields an equivalent generating function that encodes the operators in \(\mathcal{C}_2(\gamma)\). By an abuse of notation, we retain the symbol $F_{\gamma}(u,v)$ for the rescaled generating function throughout the remainder of this section. Since $B_1$ is unitary, we have the identities $|\eta|^2 = |\eta'|^2$, $|\zeta|^2 = |\zeta'|^2$ and $\eta'/\zeta' = -\eta^*/\zeta^*$, which we use to write the rescaled $F_{\gamma}(u,v)$ as
\begin{align}
 & F_{\gamma}(u,v) = \nonumber \\
 & \exp \left( \abs{ \alpha_1 + \frac{\zeta}{\eta} \gamma}^2   u + \abs{ \alpha_1 - \frac{\eta^*}{\zeta^*}  \gamma}^2 v +  \abs{ \alpha_2}^2 u +\abs{ \alpha_2}^2 v \right). \nonumber
\end{align}

We further denote $\xi = \eta/\zeta$ so that
\begin{align}
& F_{\gamma}(u,v) =  e^{\abs{ \alpha_1 + \xi^{-1} \gamma}^2   u + \abs{ \alpha_1 -\xi^* \gamma}^2 v +  \abs{ \alpha_2}^2 u +\abs{ \alpha_2}^2 v} \nonumber \\
& = \mathrm{exp} \Big( \abs{ \alpha_1}^2 u + ( \alpha_1 \xi^{-1*} \gamma^* + \alpha_1^* \xi^{-1} \gamma + |\xi^{-1} \gamma |^2 ) u + \nonumber \\
&  \abs{ \alpha_1}^2 v + ( - \alpha_1 \xi \gamma^* - \alpha_1^* \xi^* \gamma + |\xi \gamma |^2 ) v  +\abs{ \alpha_2}^2 u +\abs{ \alpha_2}^2 v \Big) \nonumber \\
& = \mathrm{exp} \Big( ( \alpha_1 \xi^{-1*} \gamma^* + \alpha_1^* \xi^{-1} \gamma + |\xi^{-1} \gamma |^2 ) u + \nonumber \\
&  ( - \alpha_1 \xi \gamma^* - \alpha_1^* \xi^* \gamma + |\xi \gamma |^2 ) v  + \abs{ \vec{\alpha}}^2 u + \abs{ \vec{\alpha}}^2  v \Big)
\end{align}
and introduce the generating function 
\begin{align}\label{eq: F_g'(u,v)}
& F'_{\gamma}(u,v) = e^{- |\xi^{-1} \gamma |^2  u -  |\xi \gamma |^2  v } F_{\gamma}(u,v) \nonumber \\
& = e^{( \alpha_1 \xi^{-1*} \gamma^* + \alpha_1^* \xi^{-1} \gamma  + \abs{ \vec{\alpha}}^2 ) u  + ( - \alpha_1 \xi \gamma^* - \alpha_1^* \xi^* \gamma + \abs{ \vec{\alpha}}^2  )v },
\end{align}
which is equivalent to $F_{\gamma}(u,v)$ as the following lemma shows.
\begin{lemma}\label{lem: F_g_equiv}
The generating functions $F_{\gamma}(u,v)$ and $F'_{\gamma}(u,v)$ are equivalent.
\end{lemma}
\begin{proof}
  The argument is identical to the one used for the proof
  of Lem.~\ref{lem: F_1g_equiv}. In particular, the coefficient of
  $u^n v^m$ in $F'_{\gamma}(u,v)$ is a discrete convolution of the
  coefficients of $F_{\gamma}(u,v)$ and of
  $e^{- |\xi^{-1} \gamma |^2 u - |\xi \gamma |^2 v }$ that go up to
  degree $u^n v^m$, and due to the multiplicative invertibility of the
  exponential factor, the same is true in the reverse direction.
\end{proof}

Lem.~\ref{lem: F_g_equiv} allows us to conduct our analysis using $F'_{\gamma}(u,v)$.

If the measurement statistics from one counter determine $\rho_t$ for
a given set of probe states, then the statistics from both counters
for the same set of probe states do as well, since the former are a
marginal distribution of the latter. Consequently, Thms.~\ref{thm:
  1cntr_uncount} and~\ref{thm:1CntrNbnd}, as well as Cor.~\ref{cor:
  suf_set_gamma} are true for the configuration with two counters. The
relevant question is whether there are interesting cases where the
measurement statistics for a restricted set of probe states fails to
determine $\rho_t$ with one counter but succeeds with two counters. We
first show that the no-go result of Prop.~\ref{prop: inf_N_fin_gamma},
namely, that a finite number of probe amplitudes does not determine $\rho_t$,
also holds for the two-counter configuration.

\begin{prop}\label{prop: inf_N_fin_gamma_2}
  Consider a measurement configuration with a two
  counters. Consider an arbitrary two-mode twirled state $\rho_t$
    with \(\rho_{t}\in\cS\). Suppose the measurement statistics are
    known for only a finite set of probe amplitudes $\Gamma$. Then, in
    general, the two-mode twirled state $\rho_t$ is not determined by
    the collection of measurement statistics for the
    $\gamma \in \Gamma$.
\end{prop}
\begin{proof}
  The proof closely follows the proof of Prop.~\ref{prop:
    inf_N_fin_gamma}, replacing the generating function
  \(F'_{\gamma}(u)\) there with \(F'_{\gamma}(u,v)\) here. We refer to
  that proof for details omitted here.
  
  The coefficient of $u^i v^j$ in $ F'_{\gamma}(u,v)$ is $\frac{1}{i!j!}$ times 
  \begin{align}
    (\abs{ \vec{\alpha}}^2  + \alpha_1 \xi^{-1*}\gamma^* +  \alpha_1^* \xi^{-1}\gamma   )^i  ( \abs{\vec{\alpha}}^2   - \alpha_1 \xi \gamma^* - \alpha_1^* \xi^* \gamma  )^j, \nonumber
  \end{align}
  which expands as a linear combination of monomials $m(k,l,n)$ with
  $k+l+n = i+j$. Thus, for each $N$ and for a given $j'$, the
  elements of $\cM_{=j'}$ appear only among the coefficients of
  $u^i v^j$ with $i+j = j'$, and there are exactly $j'+1$ such
  coefficients. Consequently, for each probe amplitude \(\gamma\),
  there are at most \(j'+1\) complex linear combinations of members of
  \(\cM_{=j'}\) whose expectations can be determined from the
  measurement statistics at this probe amplitude. For \(K=|\Gamma|\)
  probe amplitudes and a given \(N'\), this yields at most
  \((N'+1)(N'+2)K/2\) such linear combinations of members of
  \(\cM_{\leq N'}=\bigcup_{1\leq j'\leq N'} \cM_{=j'}\).  Following
  the proof of Prop.~\ref{prop: inf_N_fin_gamma} 
  we restrict the twirled states to \(\cS_{N}\), and because
  \(\cM_{N}\subseteq \cM_{\leq 2N}\), we find that  the
  number of independent real values obtained for these states is now
  at most \((2N+1)(2N+2)K\), which is \(O(N^{2})\) and 
  thus grows much more slowly than the \(\Omega(N^{3})\) required
  to identify the states of \(\cS_{N}\).
\end{proof}

It follows from Prop.~\ref{prop: inf_N_fin_gamma_2} that for $\rho_t$ with at most $N$ photons, there is some minimal number of probe amplitude $h_2(N) \leq (N+1)^2$ that are necessary and sufficient to determine $\rho_t$ with this measurement configuration. Since there is significantly more information with two counters than with one counter, we expect that $h_2(N) < h_1(N)$. We leave finding $h_1(N)$ and $h_2(N)$ as open problems. The final result in our analysis of the single-counter configuration concerned the case where the magnitude of the probe is restricted to a single value, while allowing its phase to vary freely, as is done in standard homodyne measurements. We now demonstrate that, unlike with the single-counter configuration, $\rho_t$ can be determined under these constraints if both counters are used.

\begin{theorem}\label{thm: twodet_R_fixed}
  Consider a measurement configuration with two counters. Consider an
  arbitrary two-mode twirled state $\rho_t$ with
  \(\rho_{t}\in\cS\). Suppose that, for a fixed probe magnitude
  \(R = |\gamma|\), the measurement statistics are known for all phase
  values $\theta$ of \(\gamma = R e^{i\theta}\). Then, $\rho_t$ is
  determined by this collection of measurement statistics.
\end{theorem}
\begin{proof}
We treat $u$ and $v$ as real variables in $F'_{\gamma}(u,v)$, and show that with suitable variable transformations involving $u,v,$ we can extract every monomial $m(k,l,n)$ from the $F'_{\gamma}(u,v)$ for all values of the phase $\theta$. Note that $\xi^*$ and $\xi^{-1}$ have the same complex argument, which we denote by $\phi$; consequently, $\xi$ and $\xi^{-1*}$ share the argument $-\phi$. This allows us to rewrite the exponent of $F'_{\gamma}(u,v)  $ as
\begin{align}\label{eq:exp F'(u,v)}
& \alpha_1 \gamma^* ( \xi^{-1*} u - \xi v )  + \alpha_1^*  \gamma (\xi^{-1} u - \xi^* v) + \abs{ \vec{\alpha}}^2 (u+v) \nonumber \\
 & =  \alpha_1 R  e^{-i\theta} ( \xi^{-1*} u - \xi v )  + \alpha_1^*  R  e^{i\theta} e^{i2\phi} (\xi^{-1*} u - \xi v)  \nonumber \\
 &+ \abs{ \vec{\alpha}}^2 (u+v) \nonumber \\
 & =( \alpha_1  e^{-i\theta} + \alpha_1^*  e^{i2\phi} e^{i\theta} ) R( \xi^{-1*} u - \xi v ) + \abs{ \vec{\alpha}}^2 (u+v)
\end{align}
We now change variables from $(u,v)$ to the independent variables $t = u+v$ and $s = R( \xi^{-1*} u - \xi v )$, in terms of which the exponent of $F'$ takes the form
\begin{align}
 \abs{\vec{\alpha}}^2 t  + ( \alpha_1  e^{-i\theta} + \alpha_1^*  e^{i2\phi} e^{i\theta} ).
\end{align}
The coefficient of $t^n s^r$ is given by
\begin{align}
 & n!r! \textrm{coeff}_{t^n s^r} F' = \abs{\vec{\alpha}}^{2n} ( \alpha_1  e^{-i\theta} + \alpha_1^*  e^{i2\phi} e^{i\theta})^r \nonumber \\
& = \sum_{j=0}^r \binom{r}{j} \abs{\vec{\alpha}}^{2n}  (\alpha_1^*)^j e^{i2\phi j} \alpha_1^{r-j} e^{i\theta (2j-r)} \nonumber \\
& = e^{-i \theta r}  \sum_{j=0}^r \binom{r}{j} e^{i2\phi j} m(j,r-j,n) e^{i 2 \theta j}.
\end{align}
Since this expression is available for all values of $\theta$, the sum can be inverted to obtain the $m(j,r-j,n)$ using Fourier analysis. For every $k = 0,\hdots, r$, the following integration isolates $m(k, r-k, n)$:
\begin{align}
& \int_{0}^{2\pi} d\theta e^{- i 2 \theta k} \left[ e^{i \theta r} n!r! \textrm{coeff}_{t^n s^r} F' \right] \nonumber \\
& = \int_{0}^{2\pi} d\theta e^{- i 2 \theta k} \sum_{j=0}^r \binom{r}{j} e^{i2\phi j} m(j,r-j,n)e^{i 2 \theta j} \nonumber \\
& = \sum_{j=0}^r \binom{r}{j}e^{i2\phi j} m(j,r-j,n) \int_{0}^{2\pi} d\theta e^{- i 2 \theta (k-j) } \nonumber \\
 &= 4 \pi  \binom{r}{k} e^{i2\phi k} m(k,r-k,n).
\end{align}
Performing this procedure for every pair $(r,n)$ with $r + n = N$ yields all monomials in $\cM_N$, and iterating over all $N$ recovers every monomial $m(k,l,n)$.
\end{proof}

Our final result aims to clarify the connection between this measurement scheme and standard homodyne detection. Let \(\rho_r\) denote the reduced state in mode \(\hat{a}_1\). In experimental implementations of homodyne detection, the probe amplitude \(|\gamma|\) is effectively fixed at a large value, and the photodiode signal approximates a measurement of the total photon number. Consequently, the measurement statistics are equivalent to the set of expectations \(\langle (\hatnpa  - \hatnpb )^r \rangle\) for all $r\geq 0$, sampled over a sufficiently dense set of probe phases covering the full angular range. Since our scheme does not require \( R \) to be large, it is interesting to explore whether some form of subtraction between the measurements from the two counters can directly measure \(\rho_r\).

\begin{prop}\label{prop: twodet_R_fixed}
Suppose the Husimi function of the two-mode twirled state $\rho_t$ has well-defined moments. Suppose the Husimi function of its marginal on mode 1, $\rho_r$, has a determinate moment sequence. Suppose that for a fixed probe magnitude \( R = |\gamma| \), the normally ordered moments \( \langle \nord{ ( |\eta|^2 \hatnpb  - |\zeta|^2 \hatnpa  )^n } \rangle \) are known for all integers \( n \geq 0 \) and for all phase values \( \theta \) of \( \gamma = R e^{i\theta} \). Then this set of expectations alone suffices to determine \( \rho_r \). 
\end{prop}
\begin{proof}
By Lem.~\ref{lem: norm-antinorm}, the expectations of the anti-normally ordered operators 
\(\hat{a}_1^l (\hat{a}_1^\dagger)^k\) for all \(k \geq l\) can be expressed in terms of the expectations of the 
normally ordered operators \((\hat{a}_1^\dagger)^k \hat{a}_1^l\). Since the collection 
\(\langle \hat{a}_1^l (\hat{a}_1^\dagger)^k \rangle\) forms the complete moment sequence of the Husimi 
function of \(\rho_r\), and therefore uniquely determines \(\rho_r\), it suffices to extract the 
expectations \(\langle (\hat{a}_1^\dagger)^k \hat{a}_1^l \rangle\) for all \(k \geq l\) from the set of expectations that are assumed to be known in the proposition statement.

For each value of \(\gamma\), we introduce the operator-valued generating function
\begin{align}
H_{s,\gamma}'(v) = \sum_{n=0}^\infty \frac{1}{n!} \frac{1}{|\eta|^{2n} } \, \nord{ \big( |\eta|^2 \hatnpb  - |\zeta|^2 \hatnpa  \big)^n } \, v^n,
\end{align}
whose coefficients span the measurement operators whose expectations are assumed known for
that $\gamma$. $H_{s,\gamma}'(v)$ is obtained from the two-counter generating
function $H'(u,v)$ in Eq.~\ref{eq: H'} by setting $u = - |\zeta/\eta|^2 v  = -|\xi^{-1}|^2 v$:
\begin{align}
& H'(  -|\xi^{-1}|^2 v,v)  = \sum_{k=0}^\infty \sum_{l=0}^\infty \frac{1}{k!l!}  \nord{ (\hatnpa )^{k} (\hatnpb )^{l} } ( -|\xi^{-1}|^2 v)^k v^l \nonumber  \\
& =  \sum_{n=0}^\infty \sum_{k=0}^n \frac{1}{k!(n-k)!}  \nord{ (\hatnpa )^{k} (\hatnpb )^{n-k} } ( -|\xi^{-1}|^2 v)^k v^{n-k} \nonumber  \\ 
& =  \sum_{n=0}^\infty  \frac{1}{n!} \nord{ \big( \hatnpb  - |\xi^{-1}|^2 \hatnpa  \big)^n } \, v^n, \nonumber \\
& = \sum_{n=0}^\infty \frac{1}{n!} \frac{1}{|\eta|^{2n} } \, \nord{ \big( |\eta|^2 \hatnpb  - |\zeta|^2 \hatnpa  \big)^n } \, v^n =  H_{s,\gamma}'(v),
\end{align}
where we have relabeled the indices in the second line, and then applied the binomial identity.

The generating function for the measurement operators on $\cH_a$ is therefore obtained from $F_\gamma(u,v)$ by setting $u = -|\xi^{-1}|^2 v$. Working instead with the equivalent generating function $F'_\gamma(u,v)$ in Eq.~\ref{eq: F_g'(u,v)} (where $|\eta|^2 u \to u$ and $|\zeta'|^2 v \to v$), this substitution
becomes simply $u = -v$. Reading off the exponent of $F'_\gamma(-v, v)$ from
Eq.~\ref{eq:exp F'(u,v)} with $u = -v$ gives
\begin{align}
& ( \alpha_1  e^{-i\theta} + \alpha_1^*  e^{i2\phi} e^{i\theta} ) R( - \xi^{-1*} v - \xi v ) + \abs{ \vec{\alpha}}^2 (-v+v) \nonumber \\
& = ( \alpha_1  e^{-i\theta} + \alpha_1^*  e^{i2\phi} e^{i\theta} ) R( \xi^{-1*} + \xi  ) (-v).
\end{align}
Crucially, the $|\vec{\alpha}|^2$ term vanishes, so the exponent carries no dependence on
$\alpha_2$ or $\alpha_2^*$, as intended. The coefficient of $v^n$ in $F'_\gamma(-v,v)$ is therefore
\begin{align}
\frac{\left[-R(\xi^{-1*}+\xi)\right]^n}{n!}
\left( \alpha_1 e^{-i\theta} + \alpha_1^* e^{i(\theta+2\phi)} \right)^n.
\end{align}
As in the proof of Theorem~\ref{thm: twodet_R_fixed}, the availability of this expression
for all $\theta$ allows one to isolate each monomial $(\alpha_1^*)^k \alpha_1^{n-k}$ for
$k = 0, \ldots, n$ by Fourier inversion in $\theta$. Repeating for every $n$ recovers
the full collection of monomials $(\alpha_1^*)^k \alpha_1^l$, whose expectations
$\langle (\hat{a}_1^\dagger)^k \hat{a}_1^l \rangle$ determine $\rho_r$.

\end{proof}

If the BS is balanced, as in standard homodyne detection, the
expectations $\langle \nord{ (\hatnpb - \hatnpa )^n } \rangle$ in
Prop.~\ref{prop: twodet_R_fixed} are the normally ordered versions of
the usual homodyne moments $\langle(\hatnpb - \hatnpa
)^n\rangle$. Although this establishes a neat connection with homodyne
detection, and avoids assuming a high-intensity probe, these normally
ordered moments cannot be accessed directly in experiment without
measuring the full joint total-photon‑number distribution of both
counters. In particular, unlike in standard homodyne detection, where
subtracting the photodiode signals directly measures the moments
$\langle(\hatnpb - \hatnpa)^n\rangle$ to a good approximation, there
is no straightforward method to measure the normally ordered moments
$\langle \nord{ (\hatnpb - \hatnpa )^n } \rangle$ directly for
arbitrary probe amplitudes. However, if the probe has high intensity,
the normally ordered moments
$\langle \nord{ (\hatnpb - \hatnpa )^n } \rangle$ admit a significant
simplification.  Since normally ordered products of $\hatnpa $ and
$\hatnpb $ differ from their standard counterparts by terms involving
lower-order operators, and since each such correction is suppressed by
a power of the probe intensity relative to the leading term, the
normally ordered and standard moments become asymptotically equal in
the high-intensity limit. This is a well-established observation in
standard homodyne theory \cite{leonhardt:qc1997a}. Thus, it could be
said that for large enough probe magnitudes, the
$\langle \nord{ (\hatnpb - \hatnpa )^n } \rangle$ can be approximately
measured by subtracting the outcomes of the counters in the
two-counter configuration.

\section{Discussion}

Standard homodyne detection, viewed within our framework, appears as a
limiting case of the two-counter measurement scheme in the presence of
a strong LO. Our work provides a general theoretical comparison of the
information content accessible through an idealized weak-field
homodyne scheme and that obtained from standard homodyne detection for
a fixed LO mode. Existing literature performing this comparison
assumes the unknown input state resides only in the LO-matched mode
with the other modes in their vacuum states. We relax this assumption
entirely, providing an analysis where the orthogonal modes are not
necessarily in vacuum.  Our analysis demonstrates that employing
photon counters together with a weak LO enables the measurement to
extract strictly richer information about the unknown input state,
while requiring no additional device complexity beyond a BS and the
counters. Notably, this advantage arises even in the single-counter
configuration, provided that statistics are collected for a
sufficiently large set of LO amplitudes with varying intensities.

From a practical perspective, realizing the informational advantage of
photon counters requires the characterization of detectors that can
resolve photon number sufficiently well for inferrability of the
photon number probabilities.  With the recent progress in
superconducting
detectors~\cite{gerrits2016super,lita2022development,Eaton2023res,sauer2023resolving}
and multiplexed photodetectors~\cite{cheng2023pixel,Eaton2023res}, the
difficulty of this characterization challenge has been significantly
reduced. By applying detector tomography
techniques~\cite{lundeen2009tomography,humphreys2015tomography,fiurasek2025fundamental,brida2012quantum,zhang2012recursive,morais2024precisely},
it is now feasible to infer photon number probabilities for photon
numbers below increasingly high detector saturation limits.  Thus,
while photon counters are more challenging to implement and
characterize, the advantages they offer in information gain, together
with technological advances, make them increasingly useful for state
characterization with weak probes in continuous-variable experiments.

It is worth highlighting that the sufficiency of the single-counter configuration offers a practical simplification. In standard homodyne detection, the two photodiodes may not perfectly capture the full LO mode, due to imperfect coupling, finite detector area, or spectral mismatch, which can introduce ambiguity in the measured quadrature. Similarly, with a weak-field scheme with PNR detectors, ensuring that both detectors measure exactly the same combination of modes can be technically demanding and complicate the interpretation of the resulting statistics. By contrast, a single-counter configuration reduces this source of ambiguity: although there is still uncertainty about which exact modes are measured, compared to the two-counter configuration the consequences of this uncertainty are straightforward to model and interpret. Thus, the single-counter scheme alleviates a common experimental challenge, provided that statistics are collected over a sufficient set of LO amplitudes.

Most of the results presented here were first developed in a doctoral
thesis \cite{arik_thesis}, with several extensions and refinements
introduced here. The thesis included analysis of measurement
configurations in which the BS acts differently on distinct mode sets,
as occurs in polarizing, fiber-optic beam splitters with unequal
transmission–reflection ratios for orthogonal
polarizations. Accompanying this analysis, a demonstration experiment
with such a BS was conducted, while the theoretical results showed
that two-counter measurements can reveal significantly more
information about the unknown input state than is contained in the
two-mode twirled state. The thesis also reported simulations
reconstructing the two-mode twirled state via maximum-likelihood
estimation for input states containing up to eight photons, with probe
states of order one photon in amplitude to ensure resolvability by
modern PNR detectors. Fidelities exceeding $99.6\%$ between
reconstructed and true two-mode twirled states were obtained with
$10^6$ samples distributed over fewer than ten LO amplitudes, and
remained high even at $50\%$ detector efficiency, indicating that
statistical noise and known losses are unlikely to pose significant
limitations. A first attempt to experimentally implement these
weak-field homodyne schemes is presented in the thesis of
D. S. Phillips~\cite{phillips2020advanced}.

The simulations did not consider the effect of mode-dependent loss,
which violates the assumptions of the model since the total photon
number can no longer be inferred from the detector signal. Future work
could investigate the impact of such unaccounted-for losses, which
might arise in experimental settings. Additional directions include
exploring how the information obtained from this measurement
configuration could be enhanced by employing different sets of LO
modes. For example, one might consider the situation in which, for the
same unknown state, multiple two-mode twirled states have been
constructed for multiple, potentially overlapping LO modes, and ask
what conclusions can then be drawn about the underlying unknown state.

\begin{acknowledgments}
  \label{ack}
  This work includes contributions of the National Institute of
  Standards and Technology, which are not subject to
  U.S. copyright. The U.S. Government is authorized to reproduce and
  distribute reprints for governmental purposes notwithstanding any
  copyright annotation thereon. The use of trade names and software
  does not imply endorsement by the US government, nor does it imply
  these are necessarily the best available for the purpose used
  here. This document has not been peer reviewed but has been cleared
  by NIST for release.

  Use of AI: Most results were obtained without the use of AI and can
  be found in the pre-AI thesis~\cite{arik_thesis}. When writing this
  paper, AI was used for researching the relevant literature, finding
  relevant works to reference and take advantage of in updated proofs
  and added results, and to improve the language. 
\end{acknowledgments}

\bibliography{WFH_PRA.bib}

@article{combes2022hom,
  title = {Homodyne measurement with a Schr\"odinger cat state as a local oscillator},
  author = {Combes, Joshua and Lund, Austin P.},
  journal = {Phys. Rev. A},
  volume = {106},
  issue = {6},
  pages = {063706},
  numpages = {19},
  year = {2022},
  month = {Dec},
  publisher = {American Physical Society},
  doi = {10.1103/PhysRevA.106.063706},
  url = {https://link.aps.org/doi/10.1103/PhysRevA.106.063706}
}

@Article{Eaton2023res,
author={Eaton, Miller
and Hossameldin, Amr
and Birrittella, Richard J.
and Alsing, Paul M.
and Gerry, Christopher C.
and Dong, Hai
and Cuevas, Chris
and Pfister, Olivier},
title={Resolution of 100 photons and quantum generation of unbiased random numbers},
journal={Nature Photonics},
year={2023},
month={Jan},
day={01},
volume={17},
number={1},
pages={106-111},
issn={1749-4893},
doi={10.1038/s41566-022-01105-9},
url={https://doi.org/10.1038/s41566-022-01105-9}
}

@article{Kuzmich1999,
  title     = {Atomic quantum non-demolition measurements and squeezing},
  author    = {Kuzmich, A. and Mandel, L. and Bigelow, N. P.},
  journal   = {Phys. Rev. Lett.},
  volume    = {85},
  number    = {8},
  pages     = {1594--1597},
  year      = {2000},
  publisher = {American Physical Society},
  doi       = {10.1103/PhysRevLett.85.1594}
}

@article{Takano2009,
  title     = {Spin squeezing of a cold atomic ensemble with the nuclear spin of one-half},
  author    = {Takano, T. and Fuyama, M. and Namiki, R. and Takahashi, Y.},
  journal   = {Phys. Rev. Lett.},
  volume    = {102},
  number    = {3},
  pages     = {033601},
  year      = {2009},
  publisher = {American Physical Society},
  doi       = {10.1103/PhysRevLett.102.033601},
  eprint    = {0808.2353},
  archivePrefix = {arXiv},
  primaryClass = {quant-ph}
}

@article{Koschorreck2010,
  title     = {Quantum nondemolition measurement of large-spin ensembles by dynamical decoupling},
  author    = {Koschorreck, M. and Napolitano, M. and Dubost, B. and Mitchell, M. W.},
  journal   = {Phys. Rev. Lett.},
  volume    = {105},
  number    = {9},
  pages     = {093602},
  year      = {2010},
  publisher = {American Physical Society},
  doi       = {10.1103/PhysRevLett.105.093602},
  eprint    = {1005.2807},
  archivePrefix = {arXiv},
  primaryClass = {quant-ph}
}

@article{Peden2009,
  title     = {Nondestructive cavity {QED} probe of Bloch oscillations in a gas of ultracold atoms},
  author    = {Peden, B.~M. and Meiser, D. and Chiofalo, M.~L. and Holland, M.~J.},
  journal   = {Physical Review A},
  volume    = {80},
  number    = {4},
  pages     = {043803},
  year      = {2009},
  doi       = {10.1103/PhysRevA.80.043803},
  url       = {https://arxiv.org/abs/0906.1765}
}

@article{Niederriter2020,
  title     = {Cavity probe for real-time detection of atom dynamics in an optical lattice},
  author    = {Niederriter, Robert D. and Schlupf, Chandler and Hamilton, Paul},
  journal   = {arXiv preprint},
  year      = {2020},
  eprint    = {2006.04776},
  archivePrefix = {arXiv},
  primaryClass = {quant-ph},
  url       = {https://arxiv.org/abs/2006.04776}
}

@book{lorentz1992multi,
author = {Rudolph A. Lorentz},
year = 1992,
title = "Multivariate {B}irkhoff Interpolation",
publisher = {Springer Berlin, Heidelberg},
doi = {10.1007/BFb0088788},
series = {Lecture Notes in Mathematics},
}

@unpublished{arik_thesis,
      title={Quantum State Characterization Using Measurement Configurations Inspired by Homodyne Detection}, 
      author={Arik Avagyan},
      year={2023},
      eprint={2305.19397},
      archivePrefix={arXiv},
      primaryClass={quant-ph},
      url={https://arxiv.org/abs/2305.19397}, 
}

@misc{MAURODARIANO2003205,
      title={Quantum Tomography}, 
      author={G. Mauro D'Ariano and Matteo G. A. Paris and Massimiliano F. Sacchi},
      year={2003},
      eprint={quant-ph/0302028},
      archivePrefix={arXiv},
      primaryClass={quant-ph},
}

@article{LEONHARDT2,
title = "Sampling of photon statistics and density matrix using homodyne detection",
journal = "Optics Communications",
doi = {https://doi.org/10.1016/0030-4018(96)00061-2},
volume = "127",
number = "1",
pages = "144 - 160",
year = "1996",
issn = "0030-4018",
author = "U Leonhardt and M Munroe and T Kiss and Th Richter and M.G Raymer",
}

@article{Olivares1,
doi = {10.1088/1367-2630/ab4afb},
year = {2019},
month = {oct},
publisher = {IOP Publishing},
volume = {21},
number = {10},
pages = {103045},
author = {Stefano Olivares and Alessia Allevi and Giovanni Caiazzo and Matteo G A Paris and Maria Bondani},
title = {Quantum tomography of light states by photon-number-resolving detectors},
journal = {New Journal of Physics},
}

@article{Olivares2,
author = {Matteo Bina and Alessia Allevi and Maria Bondani and Stefano Olivares},
journal = {Opt. Express},
number = {9},
pages = {10685--10692},
publisher = {OSA},
title = {Homodyne-like detection for coherent state-discrimination in the presence of phase noise},
doi = {10.1364/OE.25.010685},
volume = {25},
month = {May},
year = {2017},
}

@article{Olivares3,
author = {Bina, Matteo and Allevi, Alessia and Bondani, Maria and Olivares, Stefano},
year = {2016},
month = {05},
pages = {26025},
title = {Phase-reference monitoring in coherent-state discrimination assisted by a photon-number resolving detector},
volume = {6},
journal = {Scientific Reports},
doi = {10.1038/srep26025},
}

@article{Olivares4,
author = {Allevi, Alessia and Bina, Matteo and Olivares, Stefano and Bondani, Maria},
year = {2017},
month = {12},
pages = {1740016},
title = {Homodyne-like detection scheme based on photon-number-resolving detectors},
doi = {10.1142/S0219749917400160},
journal = {International Journal of Quantum Information},
}

@article{Muller,
  author={C R M{ü}ller and M A Usuga and C Wittmann and M Takeoka and Ch Marquardt and U L Andersen and G Leuchs},
  title={Quadrature phase shift keying coherent state discrimination via a hybrid receiver},
  journal={New Journal of Physics},
  doi = {10.1088/1367-2630/14/8/083009},
  volume={14},
  number={8},
  pages={083009},
  year={2012},
}

@article{Resch,
  title = {Quantum State Preparation and Conditional Coherence},
  author = {Resch, K. J. and Lundeen, J. S. and Steinberg, A. M.},
  journal = {Phys. Rev. Lett.},
    doi = {10.1103/PhysRevLett.88.113601},
  volume = {88},
  issue = {11},
  pages = {113601},
  numpages = {4},
  year = {2002},
  month = {Mar},
  publisher = {American Physical Society},
}

@article{wallentowitz1996unbalanced,
  title = {Unbalanced homodyning for quantum state measurements},
  author = {Wallentowitz, S. and Vogel, W.},
  journal = {Phys. Rev. A},
  volume = {53},
  issue = {6},
  pages = {4528--4533},
  numpages = {0},
  year = {1996},
  month = {Jun},
  publisher = {American Physical Society},
  doi = {10.1103/PhysRevA.53.4528},
}

@article{Walmsley1,
  title = {Violation of {B}ell's Inequality by a Generalized {E}instein-{P}odolsky-{R}osen State Using Homodyne Detection},
  author = {Kuzmich, A. and Walmsley, I. A. and Mandel, L.},
  journal = {Phys. Rev. Lett.},
  volume = {85},
  issue = {7},
  pages = {1349--1353},
  numpages = {0},
  year = {2000},
    doi = {10.1103/PhysRevLett.85.1349},
  month = {Aug},
  publisher = {American Physical Society},
 }

@article{Walmsley2,
  title = {Bridging Particle and Wave Sensitivity in a Configurable Detector of Positive Operator-Valued Measures},
    doi = {10.1103/PhysRevLett.102.080404},
  author = {Puentes, Graciana and Lundeen, Jeff S. and Branderhorst, Matthijs P. A. and Coldenstrodt-Ronge, Hendrik B. and Smith, Brian J. and Walmsley, Ian A.},
  journal = {Phys. Rev. Lett.},
  volume = {102},
  issue = {8},
  pages = {080404},
  numpages = {4},
  year = {2009},
  month = {Feb},
  publisher = {American Physical Society},
}

@article{Walmsley3,
  author={Graciana Puentes and Animesh Datta and Alvaro Feito and Jens Eisert and Martin B Plenio and Ian A Walmsley},
  title={Entanglement quantification from incomplete measurements: applications using photon-number-resolving weak homodyne detectors},
  journal={New Journal of Physics},
  doi = {10.1088/1367-2630/12/3/033042},
  volume={12},
  number={3},
  pages={033042},
  year={2010},
 }

@article{Walmsley4,
author = {Zhang, Lijian and B. Coldenstrodt-Ronge, Hendrik and Datta, Animesh and Puentes, Graciana and Lundeen, J and Jin, Xian-Min and Smith, Brian and Plenio, M and Walmsley, Ian},
year = {2012},
month = {05},
pages = {364},
title = {Mapping coherence in measurement via full quantum tomography of a hybrid optical detector},
doi = {10.1038/nphoton.2012.107},
volume = {6},
journal = {Nature Photonics},
}

@article{Walmsley5,
  title={Observing optical coherence across {F}ock layers with weak-field homodyne detectors},
  doi = {10.1038/ncomms6584},
  author={Gaia Donati and Tim J. Bartley and Xian-min Jin and M. D. Vidrighin and Animesh Datta and Marco Barbieri and Ian A. Walmsley},
  journal={Nature communications},
  year={2014},
  volume={5},
  pages={5584},
}

@article{Shadow,
  author={J Tiedau and V S Shchesnovich and D Mogilevtsev and V Ansari and G Harder and T J Bartley and N Korolkova and Ch Silberhorn},
  doi = {10.1088/1367-2630/aaad8a},
  title={Quantum state and mode profile tomography by the overlap},
  journal={New Journal of Physics},
  volume={20},
  number={3},
  pages={033003},
  year={2018},

}

@article{Koashi,
  title = {Probing the two-photon phase coherence of parametrically down-converted photons by a local oscillator},
  author = {Koashi, M. and Kono, K. and Matsuoka, M. and Hirano, T.},
  journal = {Phys. Rev. A},
  volume = {50},
  issue = {5},
  pages = {R3605--R3608},
  numpages = {0},
  year = {1994},
  month = {Nov},
  publisher = {American Physical Society},
    doi = {10.1103/PhysRevA.50.R3605},
 }

@article{Lukin,
  title={Symmetry-protected collisions between strongly interacting photons},
  author={Jeff D Thompson and Travis L Nicholson and Qi-Yu Liang and Sergio H Cantu and Aditya V Venkatramani and Soonwon Choi and Ilya A. Fedorov and Daniel Viscor and T. Pohl and Mikhail D. Lukin and Vladan Vuletic},
  journal={Nature},
  year={2017},
  volume={542},
  pages={206-209},
  doi  = {10.1038/nature20823},
}

@article{Thekkadath,
  title = {Tuning between photon-number and quadrature measurements with weak-field homodyne detection},
  author = {Thekkadath, G. S. and Phillips, D. S. and Bulmer, J. F. F. and Clements, W. R. and Eckstein, A. and Bell, B. A. and Lugani, J. and Wolterink, T. A. W. and Lita, A. and Nam, S. W. and Gerrits, T. and Wade, C. G. and Walmsley, I. A.},
  journal = {Phys. Rev. A},
  volume = {101},
  issue = {3},
  pages = {031801},
  numpages = {5},
  year = {2020},
  month = {Mar},
  publisher = {American Physical Society},
  doi = {10.1103/PhysRevA.101.031801},
}

@article {Beck:qc2016,
	author = {Beck, Kristin M. and Hosseini, Mahdi and Duan, Yiheng and Vuleti{\'c}, Vladan},
	title = {Large conditional single-photon cross-phase modulation},
	volume = {113},
	number = {35},
	pages = {9740--9744},
	year = {2016},
	doi = {10.1073/pnas.1524117113},
	publisher = {National Academy of Sciences},
	issn = {0027-8424},
	journal = {Proceedings of the National Academy of Sciences},
}

@article{maxwell:qc2013a,
  title = {Storage and Control of Optical Photons Using {R}ydberg Polaritons},
  author = {Maxwell, D. and Szwer, D. J. and Paredes-Barato, D. and Busche, H. and Pritchard, J. D. and Gauguet, A. and Weatherill, K. J. and Jones, M. P. A. and Adams, C. S.},
  journal = {Phys. Rev. Lett.},
  volume = {110},
  issue = {10},
  pages = {103001},
  numpages = {5},
  year = {2013},
  month = {Mar},
  publisher = {American Physical Society},
  doi = {10.1103/PhysRevLett.110.103001},
  url = {https://link.aps.org/doi/10.1103/PhysRevLett.110.103001},
}

@article{peyronel:qc2013a,
  author = "T. Peyronel and O. Firstenberg and Q.-Y. Liang and S. Hofferberth and A. Gorshkov and T. Pohl and M. Lukin and V. Vuletic",
  title = "Quantum Nonlinear Optics with Single Photons Enabled by Strongly Interacting Atoms",
  journal = "Nature",
  volume = "488",
  pages = "57--60",
  doi = "10.1038/nature11361",
year = {2012},
month = {July},  
}

@book{schmudgen2017moment,
  title={The Moment Problem},
  author={Schm{\"u}dgen, K.},
  isbn={9783319645469},
  series={Graduate Texts in Mathematics},
doi = {https://doi.org/10.1007/978-3-319-64546-9},
  year={2017},
  publisher={Springer International Publishing},
  
}

@article{brida2012quantum,
doi = {10.1088/1367-2630/14/8/085001},
year = {2012},
month = {aug},
publisher = {IOP Publishing},
volume = {14},
number = {8},
pages = {085001},
author = {Giorgio Brida and Luigi Ciavarella and Ivo Pietro Degiovanni and Marco Genovese and Lapo Lolli and Maria Griselda Mingolla and Fabrizio Piacentini and Mauro Rajteri and Emanuele Taralli and Matteo G A Paris},
title = {Quantum characterization of superconducting photon counters},
journal = {New Journal of Physics},

}

@article{slussarenko2019photonic,
author = {Slussarenko,Sergei  and Pryde,Geoff J. },
title = {Photonic quantum information processing: A concise review},
journal = {Applied Physics Reviews},
volume = {6},
number = {4},
pages = {041303},
year = {2019},
doi = {10.1063/1.5115814},

}

@article{zhang2012recursive,
doi = {10.1088/1367-2630/14/11/115005},
year = {2012},
month = {nov},
publisher = {IOP Publishing},
volume = {14},
number = {11},
pages = {115005},
author = {Lijian Zhang and Animesh Datta and Hendrik B Coldenstrodt-Ronge and Xian-Min Jin and Jens Eisert and Martin B Plenio and Ian A Walmsley},
title = {Recursive quantum detector tomography},
journal = {New Journal of Physics},
}

@Inbook{gerrits2016super,
author="Gerrits, Thomas
and Lita, Adriana
and Calkins, Brice
and Nam, Sae Woo",
title="Superconducting Transition Edge Sensors for Quantum Optics",
bookTitle="Superconducting Devices in Quantum Optics",
year="2016",
publisher="Springer International Publishing",
address="Cham",
pages="31--60",
isbn="978-3-319-24091-6",
doi="10.1007/978-3-319-24091-6_2",
}

@book{fulton:qc1991a,
  author = {W. Fulton and J. Harris},
  title = {Representation Theory:  A First Course},
  series = {Graduate Texts in Mathematics},
  number = {129},
  publisher = {Springer},
  address = {New York},
  year = {1991},
}

@book{leonhardt:qc1997a,
  author = "U. Leonhardt",
  title = "Measuring the Quantum State of Light",
  publisher = "Cambridge University Press",
  address = "Cambridge, UK",
  year = 1997
}

@book{wilf2005generating,
author = {Wilf, Herbert S.},
year = 2005,
title = "Generatingfunctionology",
edition = {third},
publisher = {A K Peters/CRC Press},
doi = {10.1201/b10576},
}

@book{rudin1987real,
  title={Real and Complex Analysis},
  author={Rudin, W.},
  isbn={9780070542341},
  lccn={86000007},
  series={Higher Mathematics Series},
  url={https://books.google.com/books?id=Z\_fuAAAAMAAJ},
  year={1987},
  publisher={McGraw-Hill Education},
}

@book{bachor2019guide,
author = {Hans A. Bachor and Timothy C. Ralph},
year = {2019},
month = {Oct},
edition = {third},
title = {A Guide to Experiments in Quantum Optics},
publisher = {Wiley-VCH},
doi = {10.1002/9783527695805},
}

@phdthesis{phillips2020advanced,
  publisher = {University of Oxford},
  school = {University of Oxford},
  title = {Advanced measurements for quantum photonics and quantum technologies},
  author = {Phillips, D. S.},
  year = {2020},
}

@book{migdall2013experimental,
author = {Alan Migdall and Sergey V. Polyakov and Jingyun Fan and Joshua C. Bienfang},
series = {Experimental Methods in the Physical Sciences},
publisher = {Academic Press},
volume = {45},
year = {2013},
title = {Single-Photon Generation and Detection},
issn = {1079-4042},
doi = {https://doi.org/10.1016/B978-0-12-387695-9.00015-9},
}

@article{xu2020exp,
  title = {Experimental Quantification of Coherence of a Tunable Quantum Detector},
  author = {Xu, Huichao and Xu, Feixiang and Theurer, Thomas and Egloff, Dario and Liu, Zi-Wen and Yu, Nengkun and Plenio, Martin B. and Zhang, Lijian},
  journal = {Phys. Rev. Lett.},
  volume = {125},
  issue = {6},
  pages = {060404},
  numpages = {7},
  year = {2020},
  month = {Aug},
  publisher = {American Physical Society},
  doi = {10.1103/PhysRevLett.125.060404},
  url = {https://link.aps.org/doi/10.1103/PhysRevLett.125.060404},
}

@article{ivan2011operator,
  title={Operator-sum representation for bosonic Gaussian channels},
  author={Ivan, J Solomon and Sabapathy, Krishna Kumar and Simon, Rajiah},
  journal={Physical Review A},
  volume={84},
  number={4},
  pages={042311},
  year={2011},
  publisher={APS},
  doi={10.1103/PhysRevA.84.042311},
  url={https://doi.org/10.1103/PhysRevA.84.042311}
}

@article{humphreys2015tomography,
  doi = {10.1088/1367-2630/17/10/103044},
  year = {2015},
  month = {oct},
  publisher = {IOP Publishing},
  volume = {17},
  number = {10},
  pages = {103044},
  author = {Peter C. Humphreys and Benjamin J. Metcalf and Thomas Gerrits and Thomas Hiemstra and Adriana E. Lita and Joshua Nunn and Sae Woo Nam and Animesh Datta and W. Steven Kolthammer and Ian A. Walmsley},
  title = {Tomography of photon-number resolving continuous-output detectors},
  journal = {New Journal of Physics},
}

@article{fiurasek2025fundamental,
  title = {Fundamental limits on determination of photon number statistics from measurements with multiplexed on/off detectors},
  author = {Fiur\'a\v{s}ek, Jarom\'ir},
  journal = {Phys. Rev. Research},
  volume = {7},
  issue = {3},
  pages = {033147},
  year = {2025},
  month = {Aug},
  publisher = {American Physical Society},
  eprint = {2503.17480},
  archivePrefix = {arXiv},
  primaryClass = {quant-ph},
}

@article{lita2022development,
  author = {Lita, Adriana E. and Reddy, Dileep V. and Verma, Varun B. and Mirin, Richard P. and Nam, Sae Woo},
  title = {Development of Superconducting Single-Photon and Photon-Number Resolving Detectors for Quantum Applications},
  journal = {Journal of Lightwave Technology},
  volume = {40},
  number = {23},
  pages = {7578--7597},
  year = {2022},
  doi = {10.1109/JLT.2022.3195000},
}

@misc{sauer2023resolving,
  title = {Resolving photon numbers using ultra-high-resolution timing of a single low-jitter superconducting nanowire detector},
  author = {Sauer, Gregor and Kolarczik, Mirco and Gomez, Rodrigo and Conrad, Johanna and Steinlechner, Fabian},
  year = {2023},
  eprint = {2310.12472},
  archivePrefix = {arXiv},
  primaryClass = {quant-ph},
  url = {https://arxiv.org/abs/2310.12472},
}

@Article{cheng2023pixel,
  author = {Cheng, Risheng and Zhou, Yiyu and Wang, Sihao and Shen, Mohan and Taher, Towsif and Tang, Hong X.},
  title = {A 100-pixel photon-number-resolving detector unveiling photon statistics},
  journal = {Nature Photonics},
  year = {2023},
  month = {Jan},
  volume = {17},
  number = {1},
  pages = {112--119},
  issn = {1749-4893},
  doi = {10.1038/s41566-022-01119-3},
  url = {https://doi.org/10.1038/s41566-022-01119-3},
}

@Article{lundeen2009tomography,
  author = {Lundeen, J. S. and Feito, A. and Coldenstrodt-Ronge, H. and Pregnell, K. L. and Silberhorn, Ch. and Ralph, T. C. and Eisert, J. and Plenio, M. B. and Walmsley, I. A.},
  title = {Tomography of quantum detectors},
  journal = {Nature Physics},
  year = {2009},
  month = {Jan},
  volume = {5},
  number = {1},
  pages = {27--30},
  issn = {1745-2481},
}

@article{morais2024precisely,
  doi = {10.22331/q-2024-05-23-1355},
  title = {Precisely determining photon-number in real time},
  author = {Morais, Leonardo Assis and Weinhold, Till and de Almeida, Marcelo Pereira and Combes, Joshua and Rambach, Markus and Lita, Adriana and Gerrits, Thomas and Nam, Sae Woo and White, Andrew G. and Gillett, Geoff},
  journal = {Quantum},
  volume = {8},
  pages = {1355},
  year = {2024},
  month = {May},
  publisher = {{Verein zur F{\"{o}}rderung des Open Access Publizierens in den Quantenwissenschaften}},
}

\appendix

\section*{Appendix}

In this appendix, we derive results concerning the use of normal and anti-normal orderings introduced in the main text. We consider an arbitrary collection of \(S\) orthogonal bosonic modes, denoted by \(\hat{a}_1, \dots, \hat{a}_S\), with the total number operator defined as \( \hat{n} = \sum_{i=1}^S \hat{a}_i^\dagger \hat{a}_i \, .\)

\section{Normal ordering and PLTs}\label{app nor_PLR}
We begin by establishing that the normal ordering operation commutes with PLTs. To clarify, let $\hat P$ denote an arbitrary operator that is a linear combination of products of mode operators. We want to show that for any PLT $\cU$,
\begin{align}
\cU^\dagger \nord{\hat P} \cU = \nord{\cU^\dagger \hat P \cU}.
\end{align}
By the linearity of both PLTs and the normal ordering operation, it suffices to verify this property for a single product of mode operators. Let $
M(\hat{\mathbf{a}}^\dagger, \hat{\mathbf{a}})$ denote such a product. Then
\begin{align}\label{eq: U^*MU}
\cU^\dagger M(\hat{\mathbf{a}}^\dagger, \hat{\mathbf{a}}) \cU 
= M(U^* \cdot \hat{\mathbf{a}}^\dagger, \, U \cdot \hat{\mathbf{a}}),
\end{align}
which follows by inserting the identity $\cU \cU^\dagger$ between each consecutive pair of mode operators in the product and using 
\(\cU^\dagger \hat{a}_i \cU = (U \cdot \hat{\mathbf{a}})_i\) and 
\(\cU^\dagger \hat{a}_i^\dagger \cU = (U^* \cdot \hat{\mathbf{a}}^\dagger)_i\), 
where the index $i$ denotes the $i$-th element of the corresponding vector.

Denote the transformed mode operators \((U \cdot \hat{\mathbf{a}})_i\) by \(\hat{b}_i\), so that \((U^* \cdot \hat{\mathbf{a}}^\dagger)_i = \hat{b}_i^\dagger\). We further define \( \hat{\mathbf{b}} = (\hat{b}_1, \dots, \hat{b}_S) \) and \(
\hat{\mathbf{b}}^\dagger = (\hat{b}_1^\dagger, \dots, \hat{b}_S^\dagger) \).
For a product of mode operators, we have
\begin{align}
\nord{M(\hat{\mathbf{a}}^\dagger, \hat{\mathbf{a}})} = \prod_{i=1}^S (\hat{a}_i^\dagger)^{k_i} \prod_{j=1}^S \hat{a}_j^{l_j},
\end{align}
where \(k_i\) and \(l_j\) denote the number of occurrences of \(\hat{a}_i^\dagger\) and \(\hat{a}_j\) in \(M(\hat{\mathbf{a}}^\dagger, \hat{\mathbf{a}})\), respectively. Hence,
\begin{align}
\cU^\dagger \nord{M(\hat{\mathbf{a}}^\dagger, \hat{\mathbf{a}})} \cU 
&= \prod_{i=1}^S (U^* \cdot \hat{\mathbf{a}}^\dagger)_i^{k_i} \prod_{j=1}^S (U \cdot \hat{\mathbf{a}})_j^{l_j} \nonumber\\
&= \prod_{i=1}^S (\hat{b}_i^\dagger)^{k_i} \prod_{j=1}^S \hat{b}_j^{l_j},
\end{align}
where the second line follows using the same technique as in Eq.~\ref{eq: U^*MU}. On the other hand,
\begin{align}
& \nord{\cU^\dagger M(\hat{\mathbf{a}}^\dagger, \hat{\mathbf{a}}) \cU} 
= \nord{M(U^* \cdot \hat{\mathbf{a}}^\dagger, U \cdot \hat{\mathbf{a}})} \nonumber\\
&= \nord{M(\hat{\mathbf{b}}^\dagger, \hat{\mathbf{b}})} 
= \prod_{i=1}^S (\hat{b}_i^\dagger)^{k_i} \prod_{j=1}^S \hat{b}_j^{l_j}.
\end{align}

\section{Normally ordered powers of \(\hat{n}\) }\label{app nor_n^k}  
Here, we show that the normally ordered powers of \(\hat{n}\) can be expressed as linear combinations of ordinary powers of \(\hat{n}\). We define the falling factorial of the operator \(\hat{n}\) by  
\begin{align}\label{eq: (n)_k}
(\hat{n})_{k} = \hat{n}(\hat{n}-1)\cdots(\hat{n}-k+1),
\end{align}
where each integer is understood as a scalar multiple of the identity operator on the full multimode Hilbert space.

\begin{lemma}\label{lem:ntotnum}
For every nonnegative integer \(k\), the normally ordered power of \(\hat{n}\) is given by
\begin{align}
\nord{\hat{n}^k} = (\hat{n})_k.
\end{align}
\end{lemma}

\begin{proof}
The operator $\nord{\hat{n}^k}$ is normally ordered, while the operator $(\hat{n})_k$ can be written in normally ordered form by repeated application of the commutation relations $\hat{a}_i \hat{a}_i^\dagger = 1 + \hat{a}_i^\dagger \hat{a}_i$. Hence, the operator–expression correspondence implies that the claim follows if we show that for all $\vec{\alpha}$,
\begin{align}
  \bra{\vec{\alpha}} \nord{\hat{n}^k} \ket{\vec{\alpha}} &= |\vec{\alpha}|^{2k}, \\
  \bra{\vec{\alpha}} (\hat{n})_k \ket{\vec{\alpha}} &= |\vec{\alpha}|^{2k}.
\end{align}
Let $\vec{e_1}=(1,0,\hdots,0)$ be the vector with $1$ in the first entry and $0$'s elsewhere.
Given $\vec{\alpha}$, define $\alpha=|\vec{\alpha}|$, and let $\cU$ be a PLT that transforms $\cU^\dagger \ket{\vec{\alpha}} = \ket{\alpha \vec{e_1}}$. We use the fact that the number operator commutes with PLTs, together with the fact that normal ordering also commutes with PLTs, in the following:
\begin{align}\label{eq: :n^k: = |a|^2k}
  \bra{\vec{\alpha}} \nord{\hat{n}^k}  \ket{\vec{\alpha}}
  &=\bra{\alpha\vec{e_1}}  \cU^{\dagger}
       \nord{\hat{n}^k} \cU \ket{\alpha\vec{e_1}}  \nonumber \\
  &=\bra{\alpha\vec{e_1}}    \nord{ \cU^{\dagger}  \hat{n}^k \cU }   \ket{\alpha\vec{e_1}}
       \nonumber \\
  &=\bra{\alpha \vec{e_1}}   \nord{  \hat{n}^k }   \ket{\alpha \vec{e_1}} \nonumber\\
  &=\bra{\alpha \vec{e_1}} (a_{1}^{\dagger})^{k}a_{1}^{k}   \ket{\alpha \vec{e_1}}\nonumber \\
  &=\alpha^{2k} = |\vec{\alpha}|^{2k}.
\end{align}
When replacing $ \nord{  \hat{n}^k } $ with $(a_{1}^{\dagger})^{k}a_{1}^{k} $ in the fourth step, we implicitly expanded the expression $\nord{\hat{n}^k}$ in terms of the commuting mode operators of the different modes and used the fact that all modes except the first are in the vacuum state. Consequently, any term involving at least one mode operator from the other modes vanishes, since it contains an annihilation operator acting on vacuum. Similarly,
\begin{align}\label{eqtn 4 lem:ntotnum}
  \bra{\vec{\alpha}}
  (\hat n)_{k}\ket{\vec{\alpha}}
  &= \bra{\alpha\vec{e_1}}U^{\dagger}(\hat n)_{k}
  U \ket{\alpha\vec{e_1}}\nonumber \\
  &= \bra{\alpha\vec{e_1}}(\hat n)_{k}
  \ket{\alpha\vec{e_1}}\nonumber \\
  &= \bra{\alpha \vec{e_1}}(a_{1}^{\dagger}a_{1})_{k}\ket{\alpha \vec{e_1}},
\end{align}
where, as before, in the last line we have implicitly expanded $(\hat{n})_k$ and retained only the terms involving mode $1$. Direct computation shows that $(a_{1}^{\dagger}a_{1})_{k}=
{a_{1}^{\dagger}}^{k}a_{1}^{k}$:
\begin{align}
& (a_{1}^{\dagger})^{k}a_{1}^{k} =   \sum_{m=k}^{\infty} (m)\cdots(m-k+1)\ket{m}\bra{m} \nonumber \\
 & = \left[ \sum_{m_1 =1}^{\infty} m_1 \ket{m_1}\bra{m_1}\right]\left[  \sum_{m_2 =1}^{\infty} (m_2-1) \ket{m_2}\bra{m_2}  \right]\cdots \nonumber  \\
 & \cdots \left[  \sum_{m_k =1}^{\infty} (m_k-k+1) \ket{m_k}\bra{m_k}  \right] \nonumber \\
 &=  (a_{1}^{\dagger}a_{1})(a_{1}^{\dagger}a_{1}-1)\cdots (a_{1}^{\dagger}a_{1}-k+1) =  (a_{1}^{\dagger}a_{1})_{k},
\end{align}
where the kets and bras correspond to Fock states in mode $1$. It then follows that Eq.~\ref{eqtn 4 lem:ntotnum} evaluates to $\alpha^{2k}$.
\end{proof}

As a direct consequence of Lem.~\ref{lem:ntotnum}, the ordinary powers of $\hat{n}$ can be systematically recovered from the normally ordered powers of $\hat{n}$.

\begin{corollary}\label{cor:usual_moments}
For every nonnegative integer \(m\), the ordinary power \(\hat{n}^m\) can be expressed as a linear combination of the normally ordered powers \(\nord{\hat{n}^k}\) for \(k \leq m\). 
\end{corollary}

\begin{proof}
By the definition in Eq.~\ref{eq: (n)_k}, the falling factorial \((\hat{n})_m\) is a polynomial in \(\hat{n}\) of degree \(m\) with a leading term of \(\hat{n}^m\). Rearranging this expression yields \(\hat{n}^m = (\hat{n})_m + \hat{P}_{m-1}(\hat{n})\), where \(\hat{P}_{m-1}(\hat{n})\) is a polynomial in \(\hat{n}\) of degree at most \(m-1\). By recursively applying this same logic to the lower-degree terms in \(\hat{P}_{m-1}(\hat{n})\), the operator \(\hat{n}^m\) can be expanded entirely as a linear combination of falling factorials \((\hat{n})_k\) for \(k \leq m\). Substituting \((\hat{n})_k = \nord{\hat{n}^k}\) from Lem.~\ref{lem:ntotnum} establishes the operator identity. 
\end{proof}

\section{Determinate moment sequences}\label{app: Husimi_det}
Here we first show that the class of states with Husimi functions that
have a determinate moment sequence includes Gaussian states, states
with bounded photon number, and their finite mixtures, as well as
finite superpositions of pure states of these forms. We then show that
for this class of states, the measurement statistics for every probe
state $\gamma$ are well-behaved, in the sense that they are uniquely
determined by their photon-number moments.

\begin{prop}\label{prop:moment_determinate}
  Let $\rho$ be an
    $S$-mode quantum state that is Gaussian or has bounded photon
    number.  Then its Husimi function
  $Q(\vec{\alpha}, \vec{\alpha}^*)$ is bounded by
  \(Ce^{-\epsilon|\vec{\alpha}|^{2}}\) for some state-dependent
  positive constants \(C\) and \(\epsilon\).
\end{prop}

\begin{proof}
  We characterize the asymptotic tail behavior of
  $Q(\vec{\alpha}, \vec{\alpha}^*)$ for the specified classes of
  states. For a pure state $\ket{\psi}$ with a bounded total photon
  number $N$, the overlap \(\braket{\psi}{\vec{\alpha}}\) is a
  polynomial in the variables $\alpha_j$ of maximum degree $N$,
  weighted by the coherent state normalization factor
  $e^{-|\vec{\alpha}|^2/2}$. Consequently, its Husimi function takes
  the form
  $Q(\vec{\alpha}, \vec{\alpha}^*) = |P(\vec{\alpha})|^2
  e^{-|\vec{\alpha}|^2}$, where $P$ is a polynomial. For any $S$-mode
  Gaussian state, the Husimi function is a Gaussian
  \cite{LEONHARDT2}. Consequently states of bounded photon number and
  Gaussian states both satisfy that their Husimi functions are bounded
  by a function of the form \(C e^{-\epsilon|\vec{\alpha}|^{2}}\),
  where \(C\) and \(\epsilon\) are positive and depend on the
  state. 
\end{proof}

Let \(\cS_{g}\) be the family of positive semidefinite bounded
  operators whose Husimi function are bounded by
\(C e^{-\epsilon|\vec\alpha|^{2}}\) for some positive constants \(C\)
and \(\epsilon\). We refer to positive semidefinite bounded
  operators as positive operators. Such operators are non-negative
  multiples of density operator states. 

\begin{prop}
  The family \(\cS_{g}\) is closed under finite mixture,
  pure state superpositions, PLTs, squeezing, displacements and transformations
  of the form \( \rho \mapsto p\rho p^{\dagger}\) with \(p\) a
  polynomial of the mode creation operators.
\end{prop}

\begin{proof}
We tackle each claim in the order listed in the proposition statement.
  If \(\rho\) is a mixture of positive operators \(\rho_{i}\) where the
  Husimi function of \(\rho_{i}\) is bounded by
  \(C_{i}e^{-\epsilon_{i}|\vec{\alpha}|}\), then \(\rho\) is bounded
  by \(\max_{i}(C_{i})e^{-\min_{i}(\epsilon_{i})|\vec{\alpha}|^{2}}\).

  Suppose that \(\ketbra{\psi_{i}}\) for
  \(i=1,\ldots, k\) has Husimi function bounded by
  \(C_{i}e^{-\epsilon_{i}|\vec{\alpha}|^{2}}\).  Then
  \begin{align}
    \frac{1}{\pi^{S}}|\braket{\vec \alpha}{\psi_{i}}\braket{\psi_{j}}{\vec \alpha}|
    &=
      \sqrt{\frac{1}{\pi^{S}}\braket{\vec \alpha}{\psi_{i}}\braket{\psi_{i}}{\vec \alpha}}
      \sqrt{\frac{1}{\pi^{S}}\braket{\vec \alpha}{\psi_{j}}\braket{\psi_{j}}{\vec \alpha}}
      \nonumber \\
    &= \sqrt{ Q_{\psi_i}(\vec{\alpha},\vec{\alpha}^*) }\sqrt{  Q_{\psi_j}(\vec{\alpha},\vec{\alpha}^*) }     \nonumber \\
    & \leq
      \max_{i}(C_{i}) e^{-\min_{i}(\epsilon_{i})|\vec{\alpha}|^{2}}
  \end{align} from
  which it follows that the Husimi function
  of  \(\ketbra{\phi}\) with \(\ket{\phi}\) the superposition 
  \(\ket{\phi}=\sum_{i=1}^{k} w_i \ket{\psi_{i}}\) is bounded by
  \(k^{2}\max_{i}(C_{i}) e^{-\min_{i}(\epsilon_{i})|\vec{\alpha}|^{2}}\).

  Let \(Q(\vec{\alpha}, \vec{\alpha}^{*})\) be the Husimi function of \(\rho\in\cS_{g}\) bounded by
  \(Ce^{-\epsilon|\vec\alpha|^{2}}\) for some positive \(C\) and
  \(\epsilon\).  PLTs and squeezing transform the Husimi function
  according to an invertible linear map \(L\) to
  \(Q'(\vec{\alpha}, \vec{\alpha}^{*}) = Q(L\vec{\alpha}, (L
  \vec{\alpha})^{*})\leq Ce^{-\epsilon|L\vec\alpha|^{2}}\)
  \cite{leonhardt:qc1997a}. Let \(\lambda\) be the minimum singular
  value of \(L\), which is positive \cite{leonhardt:qc1997a}. Then
  \(|L\vec\alpha|^{2}\geq \lambda^{2}|\vec\alpha|^{2}\) so
  \(Q'(\vec{\alpha}, \vec{\alpha}^{*})\leq
  Ce^{-\epsilon\lambda^{2}|\vec\alpha|^{2}}\). Since
  \(\epsilon\lambda^{2}\) is positive, the transformed operator
  is also in \(\cS_{g}\).  A displacement by \(\vec{\beta}\)
  transforms the Husimi function of \(\rho\) to
  \begin{align}
    Q'(\vec{\alpha}, \vec{\alpha}^{*})
    &= Q(\vec{\alpha}+\vec{\beta}, (\vec{\alpha}+\vec{\beta})^{*})
      \notag\\
    &\leq Ce^{-\epsilon|\vec\alpha+\vec\beta|^{2}}
      \notag\\
    &\leq Ce^{2\epsilon|\vec\beta|^{2}}e^{-(\epsilon/2)|\vec\alpha|^{2}}.
  \end{align}
The last inequality follows from
    \(|\vec\alpha+\vec\beta|^{2}\geq |\vec\alpha|^{2}/2 -
    2|\vec\beta|^{2} \). To see this, first note that
    \(|\vec\alpha+\vec\beta|^{2}\geq
    |\vec\alpha|^{2}-2|\vec\alpha||\vec\beta|\). In general, for
    \(x,y\geq 0\) we have \(-2xy \geq -x^{2}/2 -2y^{2}\).  We apply
    this inequality to the term \(-2|\vec\alpha||\vec\beta|\) with
    \(x=|\vec\alpha|\) and \(y=|\vec\beta|\) for
    \(|\vec\alpha+\vec\beta|^{2}\geq |\vec\alpha|^{2}-
    |\vec\alpha|^{2}/2- 2|\vec\beta|^{2} = |\vec\alpha|^{2}/2 -
    2|\vec\beta|^{2} \).  
      
  For the last claim, it suffices to consider a single creation
  operator \(a_{i}^{\dagger}\).  If \(\rho\) is transformed to
  \(a_{i}^{\dagger}\rho a_{i}\), the Husimi function transforms to
  \(|\alpha_{i}|^{2}Q(\vec{\alpha}, \vec{\alpha}^{*})\).  We have that
  for positive \(\delta\), there exists \(C(\delta)>0\) such that
  \(|\alpha_{i}|^{2}\leq C(\delta)e^{\delta|\vec\alpha|^{2}}\).  We
  may choose \(\delta=\epsilon/2\) and \(C'=C C(\delta)\) to ensure
  \(Q'(\vec{\alpha}, \vec{\alpha}^{*})\leq
  C'e^{-(\epsilon/2)|\vec{\alpha}|^{2}}\). 
\end{proof}

\begin{prop}
  We have \(\cS_{g}\subseteq\cS\).
\end{prop}

\begin{proof}
  Let \(Q(\vec{\alpha}, \vec{\alpha}^{*})\) be the Husimi function of
  a state in \(\cS_{g}\). By definition, there exist positive \(C\)
  and \(\epsilon \) such that
  \(Q(\vec{\alpha}, \vec{\alpha}^{*})\leq
  Ce^{-\epsilon|\vec\alpha|^{2}}\).  We argue that the moment
  generating function converges in a neighborhood of \(0\), which is a
  sufficient criterion for the  sequence to be determinate
  \cite[Chap.~14,15]{schmudgen2017moment}.
  The moment generating function is given by
  \begin{align}
    M(\vec{\beta}, \vec{\beta}^*) = \int_{\mathbb{C}^S} e^{\vec{\beta} \cdot \vec{\alpha}^* + \vec{\beta}^* \cdot \vec{\alpha}} Q(\vec{\alpha}, \vec{\alpha}^*) \, d\vec{\alpha}\, d\vec{\alpha}^*,
  \end{align}
  where $\vec{\beta} \in \mathbb{C}^S$ is an arbitrary complex
  vector. Applying the upper bound
  $Q(\vec{\alpha}, \vec{\alpha}^*) \leq C e^{-\epsilon
    |\vec{\alpha}|^2}$, we see that the integrand is strictly bounded
  by a displaced multivariate Gaussian. Consequently, the integral
  converges absolutely for all $\vec{\beta} \in \mathbb{C}^S$. This
  global convergence implies that $M(\vec{\beta}, \vec{\beta}^*)$
  converges everywhere, satisfying the sufficient condition for moment
  determinacy and completing the proof.
\end{proof}

The notation and definitions required for the following proposition are established in Sec.~\ref{subsec descr}.
 \begin{prop}\label{prop:pnr_moment_determinate}
   Consider a measurement configuration with two counters.
  Suppose that the input state \(\rho\) is in \(\cS_{g}\). Then, for
   every probe state
   $\sigma = \ket{\gamma}\bra{\gamma} \otimes
   \ket{\vec{0}}\bra{\vec{0}}$ the joint photon-counting probabilities
   $p_{ij}$ are uniquely determined by the moments of the
   total-photon-number operators,
   $\langle (\hatnpa )^k (\hatnpb )^l \rangle$.
\end{prop}
\begin{proof}
  Let \(\vec{\alpha}\) and \(\vec{\beta}\) be the variables for the
  Husimi functions of the input and probe modes.  By definition,
    there are positive \(C\) and \(\epsilon\) such that the Husimi
  function of \(\rho\) is bounded by
  \(Ce^{-\epsilon |\vec{\alpha}|^{2}}\) for some positive constants
  \(C\) and \(\epsilon\). PLTs induce orthogonal
  coordinate rotations in phase space, which imples that the Gaussian
  bound applies uniformly over all PLT transformations of \(\rho\).
  Since the twirled state \(\rho_{t}\) is obtained by a Haar integral
  over such PLT transformations of \(\rho\), \(\rho_{t}\) also has a
  Husimi function bounded by the same Gaussian. Because the probe
  state is Gaussian, it satisfies a bound of the form
  \(C'e^{-\epsilon'|\vec{\beta}|^{2}}\). After twirling, the state on
  input and probe modes is \(\rho_{t}\otimes \sigma\), so its Husimi
  function factors as the product of the Husimi functions of
  \(\rho_{t}\) and \(\sigma\). The joint Husimi function is therefore
  bounded by the Gaussian
  \(\max(C,C')e^{-\min(\epsilon,\epsilon')(|\vec{\alpha}|^{2}+|\vec{\beta}|^{2})}\).
  Let \(C''=\max(C,C')\) and \(\epsilon''=\min(\epsilon,\epsilon')\).
  The coordinate rotation property of PLTs implies that the same bound
  also applies to the state \(\rho_{out}\) after the BS that is
  measured by the two counters.

  The measurement statistics
  $p_{jk} = \langle \Pi_j \otimes \Pi_k \rangle$ are obtained by
  projecting $\rho_{out}$ onto the total-photon-number eigenspaces of
  the two output ports. Crucially, total-photon-number measurements
  are completely insensitive to local PLTs acting within their
  respective output spaces. We may therefore independently apply a
    uniform Haar-measure twirl over $\mathrm{U}(S)$ on each port's
    output modes without altering the probabilities $p_{jk}$. The
    Gaussian bound on the joint Husimi function is again preserved.
  By the same Schur's lemma argument utilized in Thm.~\ref{thm
    Tau(rho)}, this dual twirling operation  maps $\rho_{out}$ to a state $\tilde{\rho}_{out}$ that
  is completely diagonal in the joint total-photon-number basis:
  \begin{align}
    \tilde{\rho}_{out} = \sum_{j,k=0}^\infty \frac{p_{jk}}{d(j)d(k)} \Pi_j \otimes \Pi_k,
  \end{align} 
  where $d(l)$ is the dimension of the \(k\)-photon subspace of
    \(S\) modes. For this
  diagonal state, the relationship between the photon-number
  probabilities and the phase-space variables simplifies significantly. Using the expressions for the probabilities of the total photon number of a multi-mode coherent state \cite{leonhardt:qc1997a}, we evaluate the Husimi function of $\tilde{\rho}_{out}$, which
  yields the explicit power series:
\begin{align}
&\tilde{Q}_{out}(\vec{\alpha}', \vec{\beta}')  = \nonumber\\
& \frac{1}{\pi^{2S}} e^{-(|\vec{\alpha}'|^2 + |\vec{\beta}'|^2)} \sum_{j,k=0}^\infty \frac{p_{jk}}{d(j)d(k)} \frac{|\vec{\alpha}'|^{2j}}{j!} \frac{|\vec{\beta}'|^{2k}}{k!}.
\end{align}
Applying the inherited Gaussian upper bound and multiplying through by $\pi^{2S} e^{|\vec{\alpha}'|^2 + |\vec{\beta}'|^2}$, we obtain the inequality
\begin{align}
  g(|\vec{\alpha}|^{2},|\vec{\beta}|^{2}) &= \sum_{j,k=0}^\infty \frac{p_{jk}}{d(j)d(k)} \frac{|\vec{\alpha}'|^{2j}}{j!} \frac{|\vec{\beta}'|^{2k}}{k!}
           \notag\\ &\leq \pi^{2S} C''e^{(1-\epsilon'')(|\vec{\alpha}'|^2 + |\vec{\beta}'|^2)}.
  \label{eq:pij bound 1}
\end{align}

The moment generating
  function for \(p_{jk}\) is
  \begin{align}
    h(u,v)=\sum_{j,k=0}^{\infty} p_{jk} e^{uj+vk}.
    \label{eq:def_huv}
  \end{align}
  The goal is to show that this is finite for some neighborhood of the
  origin.  For this purpose, note that $d(l) \leq S^l$, and since
  \(\int_{0}^{\infty} dx x^{l}e^{-x} = l!\), we have
  \begin{align}
    e^{ul} \leq e^{ul} \frac{S^l}{d(l)} \cdot \frac{ \int_{0}^{\infty} dx x^{l}e^{-x}}{l!} = \int_0^\infty dx \frac{(Se^{u}x)^{l}}{d(l)l!}e^{-x} .
  \end{align}
  Applying this inequality to the exponential terms in the
  series for $h(u,v)$ gives
  \begin{align}
    h(u,v) &\leq \sum_{j,k=0}^\infty p_{jk} \int_{x,y=0}^\infty dxdy \frac{(Se^{u}x)^{j}}{d(j)j!}e^{-x} \frac{(Se^{v}y)^{k}}{d(k) k!}e^{-y}  \nonumber \\
           & = \int_{x,y=0}^{\infty} dxdy e^{-(x+y)}g(Se^{u}x,Se^{v} y)
             \notag\\
           &\leq \int_{x,y=0}^{\infty}dxdy e^{-(x+y)} \pi^{2S}C'' e^{(1-\epsilon'')(Se^{u}x+Se^{v}y)}, 
  \end{align}
  which is finite for \(Se^{u}< 1/(1-\epsilon'')\) and
  \(Se^{v}<1/(1-\epsilon'')\). In the second line, the interchange of
  the sums over $j,k$ with the integrals over $x,y$ is justified by
  Tonelli's theorem \cite[Chap.~1]{rudin1987real}. In particular, each
  summand in $g(Se^{u}x,Se^{v} y)$ is a continuous, hence
  measurable, and non-negative function of
  $(x,y) \in \mathbb{R}_{\geq}^2$, so the double sum and the double
  integral may be exchanged, both sides being well defined as elements
  of $[0,\infty)$. The moment generating function therefore converges
  on a neighborhood of the origin and satisfies the condition for
  moment determinateness \cite[Chap.~14,15]{schmudgen2017moment}.

\end{proof}
 
\begin{corollary}\label{cor:pnr_moment_determinate}
  Consider a measurement configuration with one counter. Suppose
  that the input state \(\rho\) is in \(\cS_{g}\). Then, for every probe state $\sigma = \ket{\gamma}\bra{\gamma} \otimes \ket{\vec{0}}\bra{\vec{0}}$ the photon-counting probabilities $p_{i}$ are uniquely determined by the moments of the total-photon-number operator, $\langle (\hatnpa )^k  \rangle$.
\end{corollary}
\begin{proof}
 The moment generating function for one counter is obtained
    by substituting \(v=0\) in the generating function \(h(u,v)\) for
    two counters defined in Eq.~\ref{eq:def_huv}. It therefore also
    converges in a neighborhood of the origin, which implies moment determinateness.
\end{proof}

\section{Anti-normally ordered powers of $\hat{n}$}\label{anti_nord_n^k}
Here, we show that anti-normally ordered powers of $\hat{n}$ can be
expressed as linear combinations of ordinary powers of
$\hat{n}$. While not utilized in the main text, the lemma below
complements Lem.~\ref{lem:ntotnum} for normally-ordered powers of
\(\hat{n}\) and provides an explicit expression.

We first observe that anti-normal ordering, like normal ordering, commutes
with PLTs. In particular, for any PLT $\cU$ and any product of mode operators $M(\hat{\mathbf{a}}^\dagger, \hat{\mathbf{a}})$, one has  
\begin{align}
\cU^\dagger \anormord{M(\hat{\mathbf{a}}^\dagger, \hat{\mathbf{a}})} \cU 
= \anormord{\cU^\dagger M(\hat{\mathbf{a}}^\dagger, \hat{\mathbf{a}}) \cU}.
\end{align}
The proof of this fact parallels the one
given for normal ordering in Appendix~\ref{app nor_PLR}, and can be adapted directly to the
anti-normal case. 

\begin{lemma}
For every nonnegative integer \(k\), the anti-normally ordered power of \(\hat{n}\) is given by
  \begin{align}
\anormord{\hat{n}^{k}} &=(\hat n+(S-1) + k)_{k}.
  \end{align}
\end{lemma}
\begin{proof}
We prove the lemma by induction on the number of modes \(S\). The base case $S=1$ can be verified directly by calculation:
  \begin{align}
& a_{1}^{k} (a_{1}^{\dagger})^{k}=   \sum_{m=0}^{\infty} (m+1)\cdots(m+k)\ket{m}\bra{m} \nonumber \\
 & = \left[ \sum_{m_1 =0}^{\infty} (m_1+1) \ket{m_1}\bra{m_1}\right]\left[  \sum_{m_2 =0}^{\infty} (m_2+2) \ket{m_2}\bra{m_2}  \right] \nonumber \\
 & \cdots \left[  \sum_{m_k =0}^{\infty} (m_k+k) \ket{m_k}\bra{m_k}  \right] \nonumber \\
 &=  (a_{1}^{\dagger}a_{1}+1)(a_{1}^{\dagger}a_{1}+2)\cdots (a_{1}^{\dagger}a_{1}+k) =  (a_{1}^{\dagger}a_{1} +k)_{k}.
\end{align}
 For the general case $S>1$, let $\hat{n}_{-}$ denote the total number
operator on the first $S-1$ modes, and let $\hat{n}_{S}$ be the number operator on the last mode. 
Assuming the formula holds for the first $S-1$ modes, by commutativity of operators on different modes, we obtain
\begin{align}\label{lem_antiord_eq1}
     \anormord{\hat{n}^{k}}
          &=    
          \anormord{\sum_{j=0}^{k}\binom{k}{j} \hat{n}_{-}^{j} \hat{n}_{S}^{k-j}}\nonumber\\
          &= 
          \sum_{j=0}^{k}\binom{k}{j}\anormord{\hat{n}_{-}^{j}}\;
            \anormord{\hat{n}_{S}^{k-j}}\nonumber\\
          &=
          \sum_{j=0}^{k}\binom{k}{j}(\hat n_{-}+S-2+j)_{j}
            (\hat n_{S}+(k-j))_{k-j}.
\end{align}

We want to show that the last line above is equal to $(\hat{n}+S-1+k)_k$. 
Since each term in the sum contains an equal number of annihilation and creation operators for each mode, 
$\anormord{\hat{n}^{k}}$ is diagonal in the Fock basis. 
Hence, it suffices to verify that for any Fock state $\ket{\vec{m}}$, with total photon number 
$m = m_1 + \cdots + m_S$, $\bra{\vec{m}} \anormord{\hat{n}^{k}} \ket{\vec{m}}$ equals
 \begin{align}
\bra{\vec{m}} (\hat{n}+S-1+k)_k \ket{\vec{m}}  =(m+S-1+k)_{k}.
 \end{align}
   Denoting $m_1+\hdots+m_{S-1} = m'$, we obtain
   \begin{align}
  & \bra{\vec{m}} \anormord{\hat{n}^{k}} \ket{\vec{m}} = \nonumber \\
   & =  \sum_{j=0}^{k}\binom{k}{j}  \bra{\vec{m}} (\hat n_{-}+S-2+j)_{j}
            (\hat n_{S}+(k-j))_{k-j}  \ket{\vec{m}} \nonumber \\
            & = \sum_{j=0}^{k}\binom{k}{j}  (m'+S-2+j)_{j}(m_S+ k-j)_{k-j} \nonumber \\
            & = \sum_{j=0}^{k}\binom{k}{j} j! \binom{m'+S-2+j}{j} (k-j)! \binom{m_S+ k-j}{k-j} \nonumber \\
            & = k! \sum_{j=0}^k \binom{m'+S-2+j}{j} \binom{m_S+ k-j}{k-j} \nonumber \\
            & = k! \binom{m'+m_S+S-1+k}{k} =  (m+S-1+k)_{k}.
   \end{align}
where we have used the identity $(n)_j  = j! \binom{n}{j}  $ repeatedly.
 \end{proof}

\end{document}